\documentclass[twocolumn]{aastex62}

\usepackage{amsmath}
\usepackage{graphicx}
\usepackage{natbib}
\usepackage{txfonts}
\usepackage[referable]{threeparttablex}
\usepackage{tensor}
\usepackage{mathtools}
\usepackage{mathrsfs}
\usepackage{multirow}
\usepackage{bm}
\usepackage{booktabs}
\usepackage{lineno}

\shorttitle{ nflow-$z$ }
\shortauthors{ Ren \& Chan, et al  }

\newcommand{\comment}[1]{}

\newcommand{\beq}{\begin{equation}}
\newcommand{\eeq}{\end{equation}}
\newcommand{\beqa}{\begin{eqnarray}}
\newcommand{\eeqa}{\end{eqnarray}}

\newcommand{\snmad}{\sigma_{\rm NMAD} }
\newcommand{\change}[1]{\textcolor{black}{#1}}

\begin{document}

\title{ Photo-$z$ Estimation with Normalizing Flow }

\correspondingauthor{Kwan Chuen Chan}
\email{chankc@mail.sysu.edu.cn}

\author{Yiming Ren}
\affiliation{ School of Physics and Astronomy, Sun Yat-Sen University, 2 Daxue Road, Tangjia, Zhuhai 519082, China }
\affiliation{ CSST Science Center for the Guangdong-Hongkong-Macau Greater Bay Area, SYSU, Zhuhai 519082, China }

\author{Kwan Chuen Chan}
\affiliation{ School of Physics and Astronomy, Sun Yat-Sen University, 2 Daxue Road, Tangjia, Zhuhai 519082, China }
\affiliation{ CSST Science Center for the Guangdong-Hongkong-Macau Greater Bay Area, SYSU, Zhuhai 519082, China }

\author{Le Zhang}
\affiliation{ School of Physics and Astronomy, Sun Yat-Sen University, 2 Daxue Road, Tangjia, Zhuhai 519082, China }
\affiliation{ CSST Science Center for the Guangdong-Hongkong-Macau Greater Bay Area, SYSU, Zhuhai 519082, China }

\author{Yin Li}
\affiliation{ Department of Mathematics and Theory, Peng Cheng Laboratory, Shenzhen, Guangdong 518066, China }

\author{Haolin Zhang}
\affiliation{ School of Physics and Astronomy, Sun Yat-Sen University, 2 Daxue Road, Tangjia, Zhuhai 519082, China }
\affiliation{ CSST Science Center for the Guangdong-Hongkong-Macau Greater Bay Area, SYSU, Zhuhai 519082, China }

\author{Ruiyu Song}
\affiliation{ School of Physics and Astronomy, Sun Yat-Sen University, 2 Daxue Road, Tangjia, Zhuhai 519082, China }
\affiliation{ CSST Science Center for the Guangdong-Hongkong-Macau Greater Bay Area, SYSU, Zhuhai 519082, China }

\author{Yan Gong}
\affiliation{ National Astronomical Observatories, Chinese Academy of Sciences, 20A Datun Road, Beĳing 100012, China }
\affiliation{ School of Astronomy and Space Sciences, University of Chinese Academy of Sciences (UCAS), Yuquan Road No.19A Beĳing 100049, China }
\affiliation{ Science Center for China Space Station Telescope, National Astronomical Observatories, Chinese Academy of Sciences,  \\  20A Datun Road, Beĳing 100101, China }

\author{ Xian-Min Meng }
\affiliation{ National Astronomical Observatories, Chinese Academy of Sciences, 20A Datun Road, Beĳing 100012, China }

\author{Xingchen Zhou}
\affiliation{ National Astronomical Observatories, Chinese Academy of Sciences, 20A Datun Road, Beĳing 100012, China }


\begin{abstract}

 Accurate photometric redshift (photo-$z$) estimation is a key challenge in cosmology, as uncertainties in photo-$z$ directly limit the scientific return of large-scale structure and weak lensing studies, especially in upcoming Stage IV surveys. The problem is particularly severe for faint galaxies with sparse spectroscopic training data.    In this work, we introduce nflow-$z$, a novel photo-$z$ estimation method using the powerful machine learning technique of normalizing flow.   nflow-$z$ explicitly models the redshift probability distribution conditioned on the observables such as fluxes and colors.   We build two nflow-$z$ implementations, dubbed cINN and cNSF, and compare their performance. We demonstrate the effectiveness of nflow-$z$ on several datasets, including a CSST mock, the COSMOS2020 catalog, and samples from DES Y1, SDSS, and DESCaLS.  Our evaluation against state-of-the-art algorithms shows that nflow-$z$  performs favorably. For instance, cNSF surpasses Random Forest, Multi-Layer Perceptron, and Convolutional Neutral Network on the CSST mock test. We also achieve a ~30\% improvement over official results for the faint DESCaLS sample and outperform conditional Generative Adversarial Network and Mixture Density Network methods on the DES Y1 dataset test.  Furthermore, nflow-$z$ is computationally efficient, requiring only a fraction of the computing time of some of the competing algorithms.   Our algorithm is particularly effective for the faint sample with sparse training data, making it highly suitable for upcoming Stage IV surveys. 

\end{abstract}



\section{ Introduction}

Wide-area imaging surveys provide powerful cosmological probes to constrain cosmology. Weak gravitational lensing is a leading application  \citep{BartelmannSchneider2001,Heymans_etal2013,Hildebrandt_etal2017,Troxel_eltal2018,Asgari_etal2021,Hikage_etal2019,Amon_etal2022,Secco_etal2022,Li_etal2023,Dalal_etal2023, Wright_etal2025}.  Angular clustering emerges as another indispensable cosmological probe, and it is often bundled with weak lensing to form the 3x2 point analysis to enhance the cosmological constraining power  \citep{DESY1_3x2pt, DESY3_3x2pt, Heymans_etal2021, Miyatake_etal2023,Sugiyama_etal2023}.  As a standalone probe, it can be used to measure the transverse Baryon Acoustic Oscillations \citep{Padmanabhan_etal2007, Seo_etal2012, Abbott:2017wcz, DES:2021esc, Chan_xip2022, Song_etal2024,DESY6_BAO}.  Ongoing or forthcoming Stage IV surveys will provide enormous amount of photometric data with significant increase in depth, leading to substantially more exquisite cosmological findings. Among them, the flagship imaging surveys include Rubin Observatory Legacy Survey of Space and Time (LSST) \citep{LSST_2019,LSST_DESC_2018}, Euclid \citep{Euclid_2011}, Chinese Space Station Survey Telescope (CSST) \citep{Zhan_2011,Gong_etal2019,CSST_intro2025}, and Roman Space Telescope (WFIRST) \citep{WFIRST_2015,WFIRST_2021}.

To achieve the impressive cosmological results from imaging surveys, accurate photometric redshift (photo-$z$) estimation is a prerequisite. photo-$z$s are often derived from the photometry information measured from a few broadband filters. Two primary approaches for photo-$z$ estimation are template fitting methods and data-driven training techniques (see \citet{Salvato_etal2019,NewmanGruen_2022} for a review).

The template-fitting method (e.g., \citet{Arnouts_1999, Bolzonella_etal2000, Benitez_2000, Ilbert_etal2006}) derives photo-$z$s by fitting observed galaxy colors or magnitudes to spectral energy distributions (SEDs) with the photo-$z$  treated as a free parameter in the model.  The fitting process can incorporate prior information. However, the accuracy of this approach depends critically on the completeness and representativeness of the adopted SED templates and the reliability of the priors.

Training-based approaches typically employ machine learning techniques.  Various algorithms have been explored, and an incomplete list includes Multi-Layer Perceptron \citep{CollisterLahav_2004, Sadeh:2015lsa}, Nearest Neighbor Fitting \citep{DeVicente:2015kyp}, Random Forest \citep{Carliles_etal2010,ZhouNewman_etal2021, Lu_etal2024}, (Bootsted) Decision Tree \citep{Gerdes_etal2010, CarrascoKind_Brunner2013}, Sparse Gaussian Process \citep{Almosallam_etal2016}, Self-Organizing Map \citep{KindBrunner_2014,Masters_etal2015,Buchs_etal2019,Wright_etal2020}, Convolutional Neural Network \citep{Hoyle_2016,DIsantoPolsterer2018,Schuldt_etal2021,LiNapolitano_etal2022, Zhou_etal2022}, Bayesian Neural Network \citep{Zhou_etal2022RAA}, Recurrent Neural Network \citep{Luo_etal2024}, and conditional Generative Adversarial Networks (cGAN) \citep{Garcia-Fernandez2025}. These methods require spectroscopic redshift (spec-$z$) or sometimes high quality photo-$z$ data for training, meaning that their accuracy is contingent on both the size and representativeness of the spec-$z$ sample.

In general, if there is little  spec-$z$ data available, especially at high redshifts, then we have to rely on template fitting using locally observed or synthetic SEDs. On the other hand, if there are sufficient spec-$z$ training data, the training methods usually deliver higher quality results. Currently, the primary photo-$z$ estimates in all the major large-scale wide field imaging surveys are based on training-based methods.  However, there is still room for improvement, especially in the faint magnitude ends where there are large measurement errors, yet with sparse training spec-$z$ data.

 In addition, clustering-based methods provide an independent route to calibrate the true redshift distribution of a photo-$z$ sample. It can be categorized into clustering-$z$ (CZ) and self-calibration (SC). CZ, based on cross-correlations with overlapping spectroscopic samples \citep{Newman_2008, McQuinnWhite_2013, Menard_etal2013, Schmidt_etal2013, vandenBusch_etal2020} has been has been widely used in current surveys, while SC relies only on photometric auto- and cross-correlations \citep{Schneider_etal2006, Zhang_etal2010, Zhang_etal2017, Peng_etal2022, XU2023}. Since spectroscopic coverage becomes sparse at high redshift, CZ calibration is limited there, and SC is expected to play an increasingly important role in future surveys. Combining CZ with SC is an effective means to extend the clustering-based method to high redshift \citep{Zheng_etal2024}.

 In this work, we apply a powerful machine learning method, normalizing flow (NF) \citep{Kobyzev_etal2019,Papamakarios_etal2019} to estimate photo-$z$, and call this photo-$z$ estimation framework with NF, nflow-$z$. NF learns the coordinate transformation to some simple base distribution to approximate the (complex) target distribution.  The NF networks are invertible by construction. In particular, the invertible nature of this type of network was emphasized in \citet{Ardizzone_etal2018,Ardizzone_etal2019}.  Degeneracy in the redshift-photometry relation is a key factor limiting the accuracy of the photo-$z$ estimation. This problem is at best tackled implicitly in most of the existing methods.  By addressing the degeneracy issue directly, nflow-$z$ holds the promise to improve the photo-$z$ estimation accuracy. Moreover, nflow-$z$ directly models the probability distribution of the data, and it is expected to yield an accurate error estimate.  \change{ We note that  \cite{Crenshaw_etal2024} use the NF to estimate the  joint probability distribution between redshift and magnitudes in order to forward model the photometric galaxy catalog. The redshift estimation is then obtained by marginalizing over the extra degrees of freedom. Instead, we apply the conditional network to estimate photo-$z$ by treating the photometry information as a condition. This simple network structure enables us to focus on the redshift information and eliminate the extra degrees of freedom effectively.   Our network structure is similar to \cite{Sun_etal2023}, which briefly mentions the NF method and applies it to estimate photo-$z$ on the HSC-SSP survey PDR3 data \citep{Aihara_etal2022}. }  Here, we study the performance of nflow-$z$ in detail and we contrast the results for two different NF architectures. We test nflow-$z$ on a number of datasets and compare them against other state-of-the-art algorithms.  Plus, a primary motivation of this work is to prepare a photo-$z$ estimation pipeline for CSST.

This paper is organized as follows. In Sec.~\ref{sec:NF_ideas_implementations}, we review the general idea of NF, present the nflow-$z$ framework, and describe two principal architectures to implement nflow-$z$. We test nflow-$z$ with different datasets and compare our test results against those from other methods in Sec.~\ref{sec:algorithm_testing}. We conclude in Sec.~\ref{sec:conclusions}.

\section{ Normalizing flow and its implementations  }
\label{sec:NF_ideas_implementations}

\subsection { A brief review on NF }

NF \citep{Tabak:2013cnz, RezendeMohamed2015} is a powerful machine learning method capable of estimating a complex high-dimensional probability distribution. Here, we briefly describe its key ideas and refer the reader to \citet{Prince2023,Kobyzev_etal2019,Papamakarios_etal2019} for more details.

The central idea of NF is to learn a coordinate transformation $ g $  that maps a complex target distribution $p_X(x)$  to some tractable base distribution $p_Y( y )$.  Mathematically, with the bijective differentiable transformation $x = g(y;c) $, we have 
\begin{align}
p_X( x;c ) d x = p_X( x;c ) \Big| \frac{d x }{ d y}  \Big| d y  = p_X \big( g(y;c) ;c\big) \Big| \frac{d g }{ d y}  \Big| d y, 
\end{align}
where $p_X $ describes the probability distribution in physical space  and $ \big| d g / d y  \big| $ is the Jacobian determinant. In this work, the physical space is taken to be the redshift space, i.e.,~$x=z$.  Our notation reflects that our target distribution is in the form of a 1-dimensional conditional distribution with the condition denoted by $c$. In our case, the condition represents the observables such as fluxes and colors. We then demand that the base distribution in latent space, $p_Y$ to satisfy \footnote{In the machine learning literature, the latent space variable is often denoted as $z$. Since redshift is the key subject in this work, we reserve the notation $z$ for redshift. } 
\begin{align}
\label{eq:NF_latentdist}
p_Y(y) = p_X( g(y;c);c ) \Big| \frac{d g }{ d y}  \Big|.
\end{align}
In words, the target distribution is normalized to the  base distribution by the coordinate transformation (flow). The base distribution is chosen to be some tractable probability distribution, and the standard normal (or Gaussian) distribution, which is sometimes called noise in the machine learning literature, is commonly adopted.  We stick to the normal distribution in this work.  Transformation $g$ is also a function of the network parameters, which are learned in the training process, so that Eq.~\eqref{eq:NF_latentdist} is satisfied to a high accuracy.

We train the network parameters by maximizing the likelihood of the data 
\begin{align}
\label{eq:likelihood_transform}
p_X(x;c) =   p_Y( f(x;c))  \Big| \frac{ d{f} }{d x }  \Big|, 
\end{align}
where  $y =  g^{-1}(x) \equiv f(x) $.  Note that, in this case, we need to evaluate $f$, while to generate samples, we use $g$ instead: 
\begin{align}
y \sim p_Y(y), \quad x = g(y). 
\end{align}

The key to successful probability distribution modeling is to be able to learn a general bijective  differentiable transformation. However, for the method to be practical, the model needs to perform both the inference and sampling efficiently. This implies that the transformation $g$, its inverse $f$, and the Jacobian determinant must be efficiently calculated.  

\change{An approach to realize an invertible differentiable function of sufficient complexity is through the composition of simpler invertible differentiable functions. }
More precisely, if
\begin{align}
g = g_N \circ g_{N-1} \circ \dots \circ g_1,
\end{align}
with $g_i$ being an invertible function,  then its inverse is also invertible and is given by 
\begin{align}
f = f_1 \circ f_{2} \circ \dots \circ f_N, 
\end{align}
where $f_i = g_i^{-1} $. With the help of the chain rule, the corresponding Jacobian determinant can be conveniently written as 
\beq
\bigg| \frac{d f }{d x}  \bigg| = \bigg| \frac{d f_1 }{d f_2}  \bigg| \bigg| \frac{d f_2 }{d f_3}  \bigg| \dots \bigg| \frac{d f_N }{d x}   \bigg|.
\eeq
This composition is implemented using layers of networks in deep learning.  Each layer represents a differentiable invertible function. We discuss two types of architecture to construct $f_i$ in Sec.~\ref{sec:NF_implementations}.

Both NF and the Generative Adversarial Network (GAN, \citet{Goodfellow_etal2014}) are well-known members of the generative modeling family in deep learning. Compared to GAN, NF has the advantage of exact probability distribution modeling and is relatively easy to train. In the literature, NF is often perceived to deliver less dazzling results than GAN, but this is often framed in the context of high-dimensional problems, such as image generation. In Sec.~\ref{sec:cGAN_confrontation}, we demonstrate that our NF results are superior to the GAN ones.

\begin{figure}[!tb]
\centering
\includegraphics[width=\linewidth]{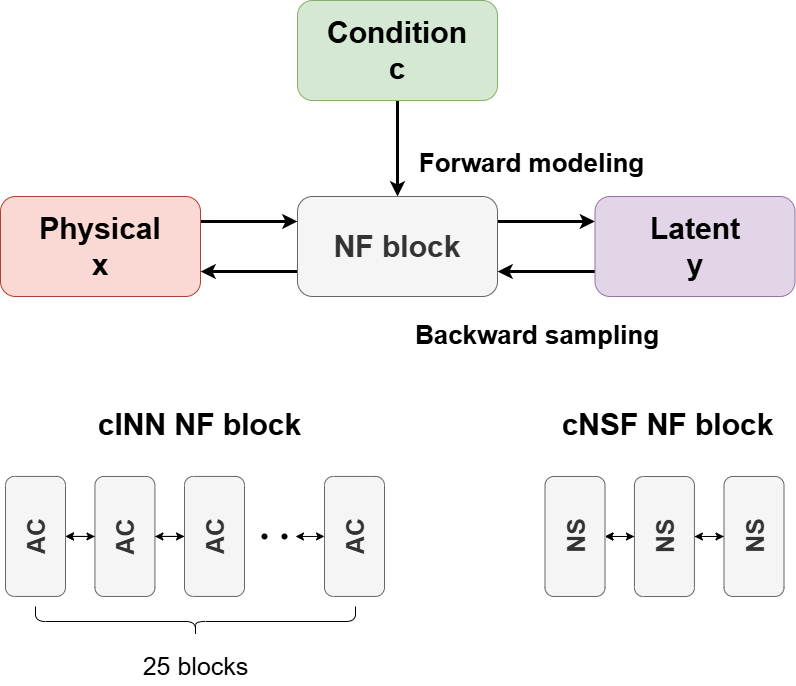}
\caption{  The general framework of the nflow-$z$  network  is shown in the upper part of the figure. The schematic structure of the two implementations of the NF block, cINN and cNSF, is shown in the lower part of the figure.  }
\label{fig:general_framework}
\end{figure}

\subsection{ Implementation of nflow-$z$}
\label{sec:NF_implementations}

An illustration of the nflow-$z$ network structure is shown in Fig.~\ref{fig:general_framework}. The physical space $X$ is a 1D redshift space, and the latent space is also 1D. The observables, including fluxes, colors, and their error bars, play the role of conditions. The conditions are injected into some neural networks, whose output then serves as the transformation parameters for the function $f_i$.  The direction of the arrows indicates the function evaluation required: $f$ for forward modeling and $g$ for backward sampling.

Differences in architecture implementation lie in the central NF block. Many efforts have been devoted to constructing an expressive yet easy-to-compute function $f_i$. Here, we consider two widely used implementations of NF, the Conditional Invertible Neural Network (cINN, \citet{Ardizzone_etal2019}) and Conditional Neural Spline Flow (cNSF, \citet{Durkan_etal2019,Dolatabadi_etal2020}). Roughly speaking, in cINN, $f_i $ is implemented using affine transformation and many layers (25 here) are used, while in cNSF, each layer is realized by a more sophisticated spline transformation and thus \change{fewer} layers (3 here) are necessary.

\subsubsection{cINN}

cINN is proposed in \citet{Ardizzone_etal2019}, which is built upon the previous version, INN  \citep{Ardizzone_etal2018}. The software {\tt FrEIA } \citep{freia} offers a flexible framework to construct such a neural network.  This architecture has been used in various inverse problems in astronomy, including stellar parameter determination \citep{Ksoll_etal2020, Kang_etal2022}, cosmic ray source inference \citep{Bister_etal2022}, exoplanet characterization \citep{Haldemann_etal2023}, and radio sky map making \citep{Zhang_etal2023}.

Following the original cINN \citep{Ardizzone_etal2019}, we \change{implement} the NF block using the GLOW coupling block \citep{Dinh_etal2016,KingmaDhariwal_2018}. In the GLOW framework, each data vector is split into two parts, \change{and} a set of affine coupling transformations is applied to them. The merit of this transformation is that $g$, $f$, and their associated Jacobian determinant can be computed efficiently. The operation, together with the permutation among dimensions, enables different dimensions of the data vector to couple with each other.  


\change{  The GLOW  affine coupling transformations are based on the Real NVP (non-volume preserving) transformation developed by \citet{Dinh_etal2016}. In this algorithm, to allow the components of the input array to be coupled, the input data array is split into two parts as  $(u_1,u_2)$, and }   the forward affine transformation from $(u_1,u_2)$  to  $(v_1,v_2)$  is then given by
\begin{align}
\label{eq:affine_coupling_block1}
v_1 &= u_1 \exp[ s_2( u_2) ] + t_2(u_2),  \\
\label{eq:affine_coupling_block2}
v_2 &= u_2 \exp[ s_1( v_1) ] + t_1(v_1), 
\end{align}
where $s_i$ and $t_i$ are arbitrary functions of the condition $c$, implemented using neural networks. The structure of the transformation is so constructed that the inverse transformation can be easily written down as well:
\begin{align}
u_2 &= ( v_2 - t_1(v_1) ) \exp[ - s_1( v_1) ] , \\
u_1 &= ( v_1 - t_2(u_2) ) \exp[ - s_2( u_2) ] .
\end{align}
In particular, the general functions $s_i$ and $t_i$ do not need to be inverted.  The log-determinant of the transformation is simply equal to $ s_2(u_2) + s_1 (v_1) $. This can be shown by performing the transformation (Eqs.~\eqref{eq:affine_coupling_block1} and \eqref{eq:affine_coupling_block2})  in two steps:  $ (u_1,u_2)\rightarrow (v_1,u_2) \rightarrow (v_1,v_2)$.  
\change{ However, our input data array is 1D (a redshift); to directly match this algorithm, we duplicate it to form a 2D array. }  We point out that the {\tt FrEIA } implementation still works for a 1D input without duplication, but we find that its performance is not as good as the 2D case. Thus, we opt to use the 2D setting, at the expense of more computation time.

Our cINN NF block consists of 25 layers, with each layer being  a GLOW affine coupling block. The scale function $s_i$ and shift function $t_i$  in each block are modeled by a two-layer  fully connected neural network with 512 hidden units  and ReLU activation functions. The conditional inputs are injected into every block through the {\tt FrEIA}  ConditionNode mechanism.

\subsubsection{cNSF}

We consider an alternative implementation using the rational spline bijections of linear order \citep{Durkan_etal2019,Dolatabadi_etal2020} implemented in {\tt pyro} \citep{Pyro_2019, Pyro_2019composable}.  We call this architecture cNSF. In place of the affine transformation in cINN,  $ f $ is realized by the piecewise rational spline.  To model the spline transform, the interval within some bounds $[-B, B]$ is divided into $K-1$ bins by $K$ points, known as knots. Within  each bin, the transformation function is given by a rational spline of linear order function, whose numerator and denominator are linear functions. The function in the whole interval $[-B,B]$ must be monotonically increasing. The transformation parameters include the position of the nodes and the derivatives at the knots. In particular, in the formulation of \citet{Dolatabadi_etal2020}, the piecewise rational linear spline is differentiable at the knots as well.  As in cINN, in the {\tt pyro} implementation, the conditions are incorporated by feeding them into a shallow neural network, whose output are the parameters for the transformation $f$. We refer the readers to the original papers \citep{Durkan_etal2019,Dolatabadi_etal2020} for more details on the transformation.

The advantage of the rational spline transformation is that it provides a more flexible parametrization than the affine transformation due to the division into bins by the knots and the usage of the rational linear function. Yet with these complexities,  the inversion and derivative can still be written in closed form and hence can be computed efficiently.

Our cNSF block consists of three layers, and each is implemented by a linear rational spline flow block, within which we have used eight segments for the spline transform. In each block, the conditions are injected into an Multi-Layer Perceptron network of size $ 32 \times 32 \times 32 $, which returns the transformation parameters for the spline flow transformation.

\subsection{Setting of the network parameters}

Here, we mention some technical details of the implementations and describe the setting of the network parameters. Unless otherwise specified, they apply to both implementations.

To put all the data, both the redshift and the condition data, such as fluxes, on a similar scale, we pre-process the data by the standard scaler transformation $ ( d - \mu )/ \sigma$ on the data vector $d$, where $\mu$  and  $\sigma $ are the mean and the standard deviation of $d$, respectively. The full dataset is split into training, validation, and test parts with the proportion of 8:1:1.  In both implementations, the training data are further divided into batches with  256 samples in each batch.

Recall that Eq.~\eqref{eq:likelihood_transform} enables us to transform the likelihood of the data to some tractable distribution $p_Y$, which is taken to be the normal distribution. However, the transformation to reach this final distribution is unknown; at least it has to be solved for  iteratively. To proceed, we assume that we can treat $p_Y$ as normal, and this approximation holds best when the likelihood is maximized for a given NF model. The exactness of this approximation depends on the capacity of the NF model.   It is convenient to consider the loss function in the form of the negative of the log-likelihood, and we have
\beq
- \log p_X(x;c) =  \frac{ f^2(x;c) }{2} - \log \Big| \frac{ d{f} }{d x }  \Big| + {\rm const}. 
\eeq
The desired loss function is obtained by averaging over the individual loss function of the samples (and ignoring the irrelevant constant) as
\beq
\label{eq:loss_function}
\mathcal{L} =  \frac{1}{N}\sum_i \Big(   \frac{ f_i^2(x_i;c_i) }{2} - \log \Big| \frac{ d{f_i} }{d x_i }  \Big| \Big), 
\eeq
where $N$ is the total number of samples. For the cINN case with duplication, $f_i $ is a 2D array, and Eq.~\eqref{eq:loss_function} is generalized straightforwardly.

 Adam optimizer \citep{KingmaBa2015} is used to train the network by minimizing $\mathcal{L}$. The initial learning rate is set to $10^{-3}$ and is subsequently reduced by a factor of $10^{-1}$ after 150 and 250 epochs, respectively.

 It is crucial not to overfit the network; otherwise, the trained model becomes sensitive to the fluctuations in the training data and thus does not generalize well to new data. We use the validation data to check for overfitting. In the event of overfitting, the loss function for the training data keeps on decreasing while the validation loss starts to rise.  We evaluate the validation loss at the end of each epoch by averaging over the results from all the batches.  The training is terminated if the validation loss stays higher than the lowest recorded validation loss for 40 consecutive evaluations. Moreover, to prevent excessive computation, we set the maximum number of training epochs to 400. In either case, the model corresponding to the lowest validation loss is selected.

NF naturally produces the probability distribution function (abbreviated as PDF) for the redshift estimate.  After training, we can make predictions by inserting the observables as conditions and generating samples from the latent space.  To get a smooth estimate of the PDF, 20,000 samples are generated for each object. From the PDF, different point estimates can be calculated, including the mean, the median, and the mode of the distribution.  We find that the median generally leads to the most accurate results, and so it is adopted as the default point estimate below. For the error estimate, we adopt the half width of the 68th percentile about the median rather than the standard deviation, as the former gives a more reliable error estimate.

We run the codes on the NVIDIA GeForce RTX 3080 GPU. Using the CSST mock dataset as an example, the cINN model typically finishes training at around 350 epochs, while the cNSF  requires about 200 epochs. The total time, training plus testing, is quite modest,  approximately 25 minutes for cINN and 6 minutes for cNSF.

\section{ Algorithm testing with datasets}
\label{sec:algorithm_testing}

In this section, we test nflow-$z$ using different datasets and contrast the results with other algorithms.  We conduct tests on four different datasets: a CSST mock, COSMOS2020 catalogs, an SDSS data sample, and a DES Y1 sample, with special emphasis on the CSST mock. These diverse datasets differ in the redshift span, magnitude ranges, and size of the training sample, and hence, these help to test the algorithm under different situations.  When comparing with other algorithms, if possible, we use  the results obtained by others on the same dataset.  This avoids misusing others' algorithms or running their code with un-optimized parameters.


\begin{figure*}[!ht]
\centering
\begin{minipage}{0.45\textwidth}
\centering
\includegraphics[width=\textwidth]{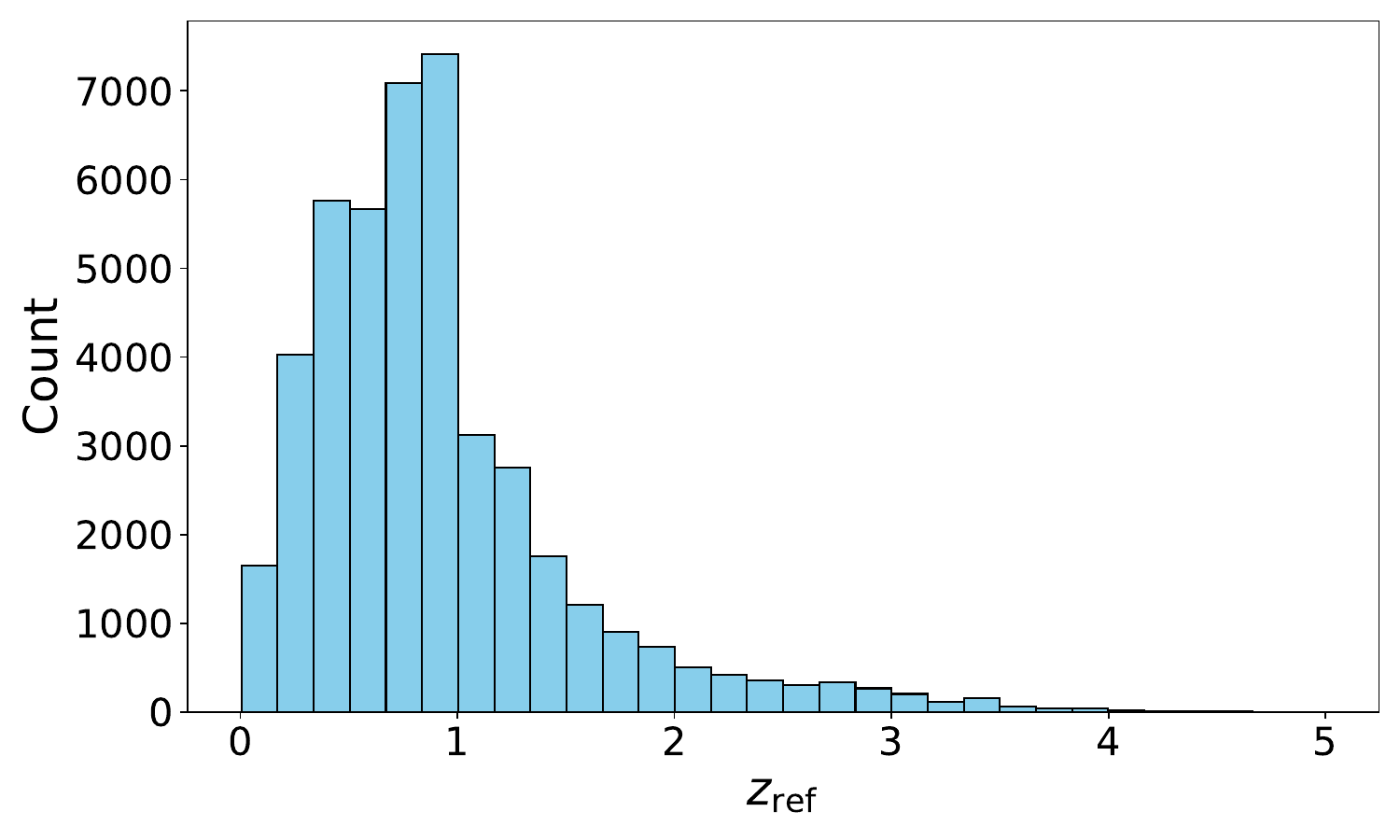}
\end{minipage}%
\hspace{0.5cm}
\begin{minipage}{0.45\textwidth}
\centering
\includegraphics[width=\textwidth]{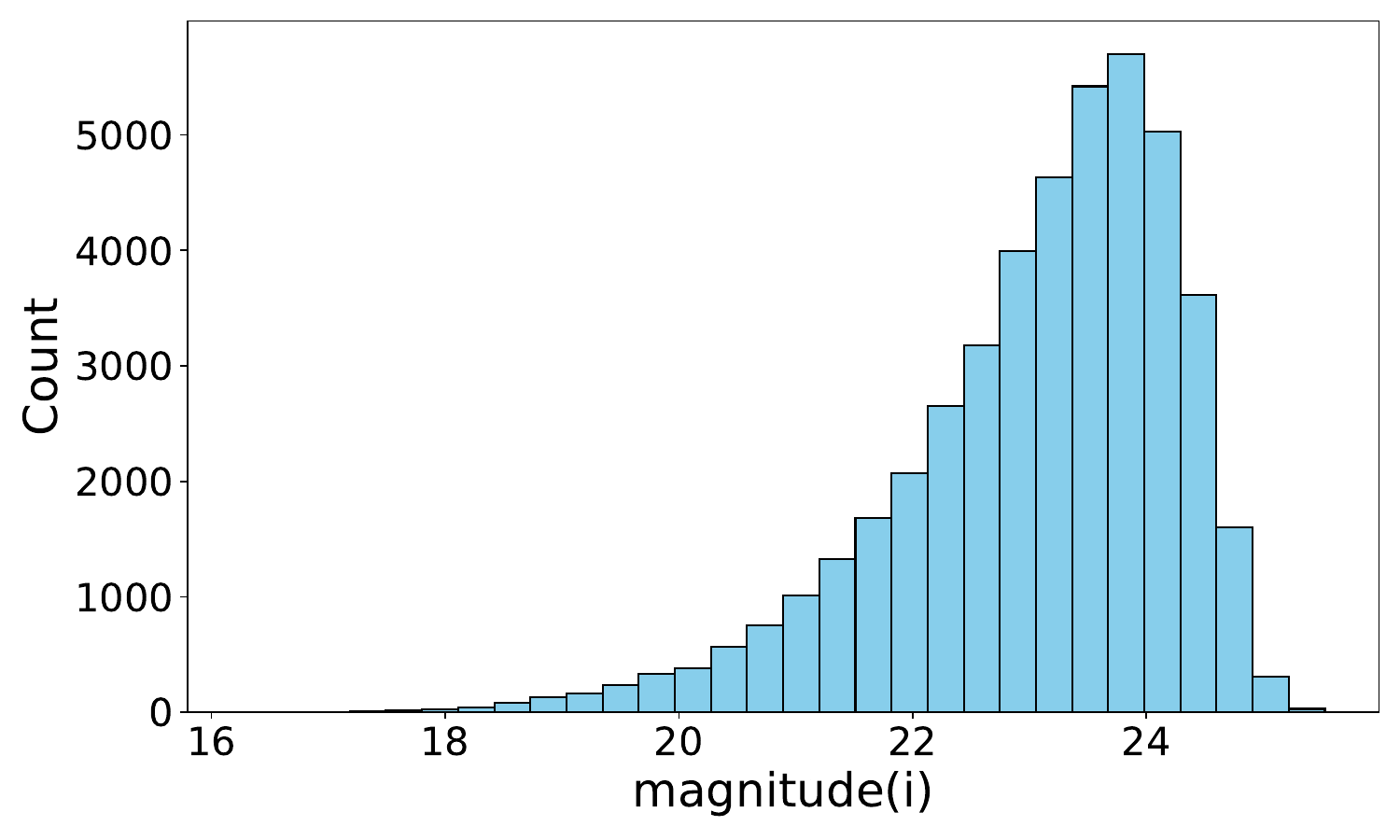}
\end{minipage}
\caption{  Redshift (left) and $i$-band magnitude (right) distribution of the CSST mock catalog. }
\label{fig:z_magi_dist_CSST}
\end{figure*}

\subsection{ Testing on the CSST mock catalog  }

\subsubsection{ CSST mock catalog overview } 

The first dataset that we consider is a CSST mock catalog. The CSST survey is a stage IV galaxy survey  \citep{Zhan_2011,Cao_etal2018,Gong_etal2019, CSST_intro2025}, conducted on a two-meter telescope onboard a satellite orbiting in tandem with the China Space Station. The satellite is expected to be launched in 2027, and is scheduled to operate over a ten-year period  covering 17,500 ${\rm deg}^2 $.  Among its tasks is the wide field cosmological survey, which is realized by a slitless spectroscopy device and a multi-band photometric instrument. In the wide-field photometric survey, there are seven broadband filters: $NUV$, $u$, $g$, $r$, $i$, $z$, and $y$. Their wavelengths range from 250 nm to 1000 nm and  5-$\sigma$ limiting magnitudes are 25.4, 25.4, 26.3, 26.0, 25.9, 25.2, and 24.4, respectively. Further details on the filter properties can be found in \cite{Cao_etal2018}. A primary goal of this work is to develop an effective photo-$z$ estimation code for the CSST.

We test our method on a CSST mock catalog, which was first used in \citet{Zhou_etal2021}. This catalog includes realistic mock images consistent with the CSST observing conditions, and has subsequently been used in a number of other CSST photo-$z$ preparation studies \citep{Zhou_etal2022,Lu_etal2024, Luo_etal2024}. Since we only use the flux information rather than the full image, we limit ourselves to describing only the generation of the flux data.  

This catalog is built upon the COSMOS2015 catalog \citep{Laigle_etal2016}.  On galaxies with reliable photometry measurements in  COSMOS2015 catalog, SED fitting is performed to determine their SED properties. The SED fitting is done through {\tt LePhare}  \citep{Arnouts_etal2002,Ilbert_etal2006} with the photo-$z$ fixed to be the fiducial value from COSMOS2015. The SED templates used are modified from the original 31 templates from {\tt LePhare} following the prescription in \citet{Cao_etal2018}.  First, the SEDs are extended from about 90 \r{A} to 900 \r{A} in order to fit galaxies with $z>2$. Second, the elliptical and spiral galaxy templates are interpolated so that the final number of SED templates totals 37. Furthermore, the emission lines Ly$\alpha$, H$\alpha$, H$\beta$, [OII], [OIII] are included in the continuum spectra by {\tt LePhare}.

Armed with the best fit SED, $S$, the expected \change{total photon count} in a band $j$ can be calculated as 
\beq 
F_j = t_{j} A_j N_j  \int d\lambda \frac{ \lambda }{ hc}  \, S(\lambda) \tau_j(\lambda), 
\eeq
where $\tau_j$ is total system throughput including  the intrinsic filter transmission, detector quantum efficiency, and total mirror efficiency, respectively \citep{Cao_etal2018,Zhou_etal2022}, while $h$ and $c$ are Planck constant and light speed, respectively.  The prefactors $t_{j}$, $A_j$, and $ N_j $ denote the exposure time, the effective telescope aperture area, and the number of exposures for this band, respectively.   The Poisson noise and background noise are introduced in the flux modeling.  The latter incorporates the effects of dark current, read-out noise, and sky background (see \citet{Zhou_etal2022} for more details).


The full catalog consists of 99,095 galaxies. Following \citet{Zhou_etal2022}, we select high-quality galaxies with signal-to-noise ratio larger than 10 in $i$ or $g$ bands. After this quality cut, we end up with  44,991 galaxy samples, and we use these galaxies in the following analyses.  In Fig.~\ref{fig:z_magi_dist_CSST}, we plot the redshift and magnitude distribution of this sample.

Our input condition data include fluxes and their errors in seven bands, and the six colors computed from them.  However, some fluxes are negative due to error fluctuations, and direct computation of the magnitude and hence color is not possible. Instead, we use the asinh magnitude \citep{Lupton_etal1999}, which is well-defined even for negative flux.

\begin{figure}[!tb]
\centering
\includegraphics[width=\linewidth]{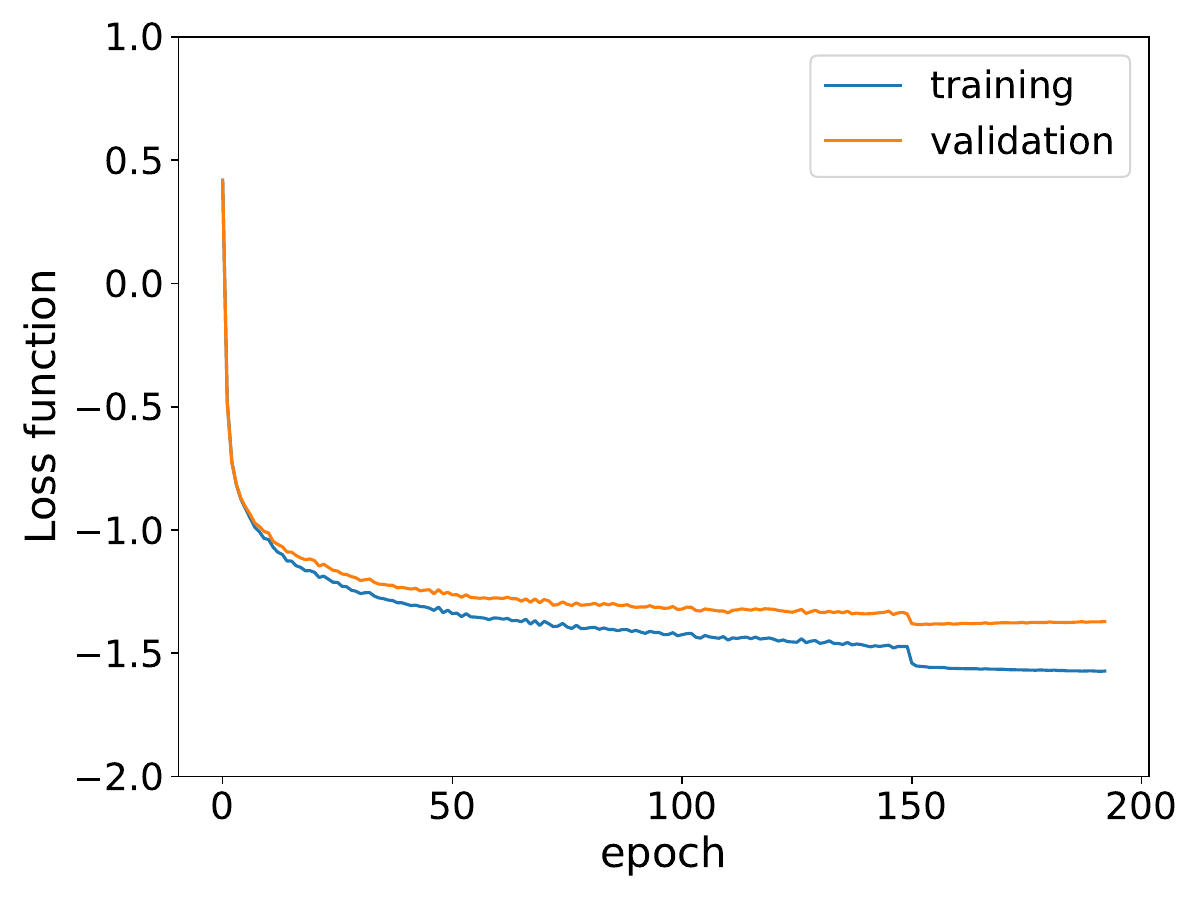}
\caption{ Loss function for the training (blue) and validation (orange) samples as a function of the epoch. The reduction in learning rate at epoch 150 causes the sudden drops in the loss functions. The training is stopped when the average validation loss starts to rise at epoch 191.     }
\label{fig:loss_function}
\end{figure}

We plot the loss function (Eq.~\eqref{eq:loss_function}) for the training and validation samples from a typical run in Fig.~\ref{fig:loss_function}.  The results shown are from cNSF, and cINN shares similar trend.   At the beginning, both loss functions decrease rapidly as epoch increases, and the reduction becomes mild after $  \sim 100 $ epochs. The sudden drop at epoch 150 corresponds to the reduction of the learning rate (by a factor of $10^{-1}$). After that, the validation loss merely drops, although the training loss still keeps on decreasing. The training is terminated at epoch 191 when the average validation loss starts to rise.

\begin{figure*}[!htb]
\centering
\includegraphics[width=0.8\linewidth]{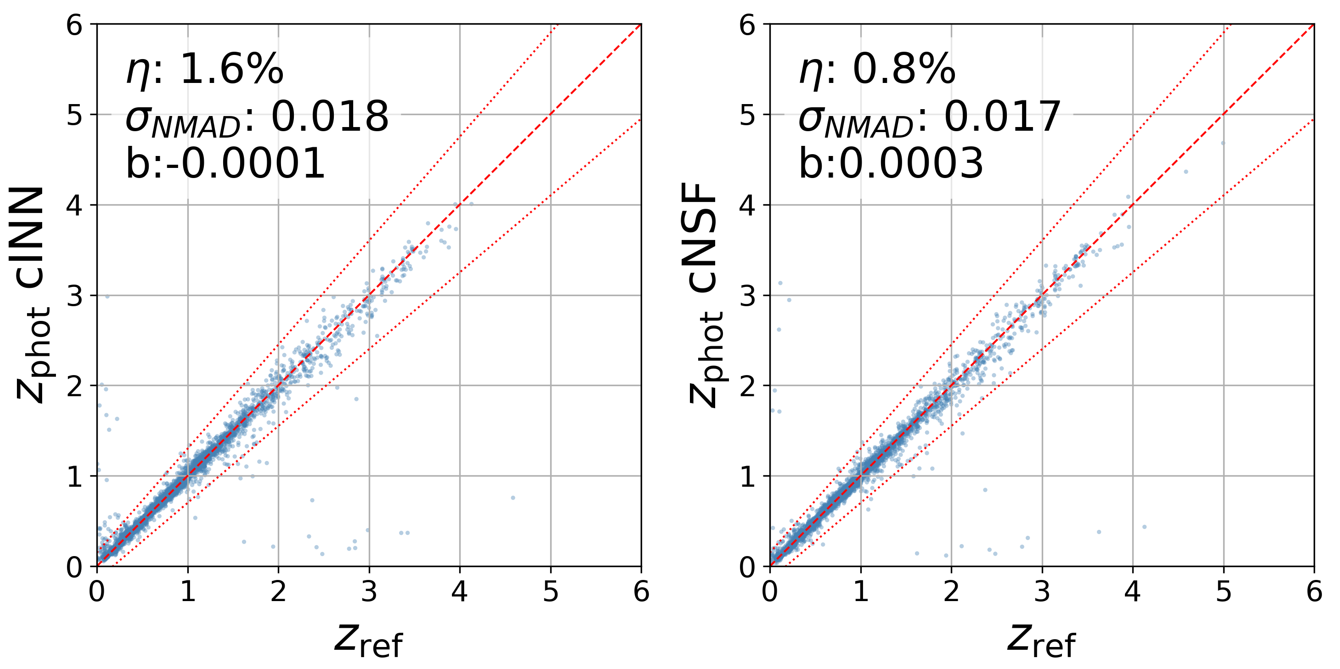}
\caption{ Scatter plot between the spectroscopic redshift, $z_{\rm spec} $ and the photo-$z$, $z_{\rm phot} $ estimated by cINN (left panel) and cNSF (right panel), respectively. The diagonal line indicates perfect prediction, and the accompanying dotted lines depict the boundaries for the outlier fraction.   }
\label{fig:zp_zs_CSST}
\end{figure*}

\subsubsection{ Quantification of the photo-$z$ accuracy}
\label{sec:CSST_quantification}

We quantify the accuracy of the results using the following metrics, which are commonly used in photo-$z$ estimation analyses: 
\begin{itemize}
\item Bias, $b$: the median of the distribution of  $\Delta z \equiv z_{\rm phot} - z_{\rm ref} $, where $z_{\rm phot}$ and $z_{\rm ref} $ denote the estimated photo-$z$ and the reference redshift.  The latter is either the spec-$z$ or high-quality photo-$z$. 
\item Normalized median absolute deviation, $\snmad$: $1.48 \times \mathrm{median}\Big(   |  \Delta z - \mathrm{median}(\Delta z ) | / (1 + z_{\rm ref} ) \Big) $ \citep{Brammer_etal2008}, which quantifies the spread of $ \Delta z / (1 + z_{\rm ref} )$ and it is not sensitive to extreme outliers. 
\item Outlier fraction, $\eta$: the fraction of galaxies with $|\Delta z |  / (1 +z_{\rm ref} ) > 0.15 $. This encodes the significance of catastrophic photo-$z$ error by measuring the fraction of galaxies with substantial  deviation from the reference redshift. 
\end{itemize}


In Fig.~\ref{fig:zp_zs_CSST} we show the scatter plot between the reference redshift and the photo-$z$ estimated by cINN and cNSF.  The metrics $b$, $\snmad$, and $\eta$ are also printed.  We see that cNSF gives lower  $\eta $ and $\snmad$, but  cINN has a smaller bias.

As we mentioned, a number of other machine-learning methods have been applied to this dataset. Here we contrast ours against other available results: 
\begin{itemize}
    \item cINN: $\snmad= 0.018\pm 0.0004$,   $\eta=1.60\pm 0.16\%$ (this work)
    \item cNSF: $\snmad= 0.017\pm0.002$,   $\eta=0.83\pm0.07\%$ (this work)
    \item Random forest: $\snmad =0.025$, $\eta=2.0\%$ \citep{Lu_etal2024}
    \item Multi-Layer Perceptron (MLP): $\snmad =0.023$, $\eta=1.4\%$ \citep{Zhou_etal2022} 
    \item Convolutional Neural Network (CNN):  $\snmad =0.025$, $\eta=1.2\%$ \citep{Zhou_etal2022} 
    \item Hybrid Transfer Network:  $\snmad =0.020$, $\eta=0.90\%$ \citep{Zhou_etal2022} 
    \item Recurrent Neural Network:  $\snmad =0.027$, $\eta=1.6\%$   \citep{Luo_etal2024}
\end{itemize}
To quantify the level of fluctuations, for cINN and cNSF, we have shown the mean and the associated standard deviation from 10 runs.  Among these methods, the most sophisticated network is the Hybrid Transfer Network, which combines the Bayesian CNN and Bayesian MLP so that it takes both images and flux data as input. It is also the best performer among the non-NF methods.   We find that the performance of cINN is comparable to Hybrid Transfer Network, with $\snmad$ being  marginally better but  $\eta$ less appealing. Remarkably, cNSF achieves the best results across these two metrics.

It is worthwhile commenting on the key properties of nflow-$z$ here.  \change{ First}, the redshift-color degeneracy is a key physical factor limiting the accuracy of the photo-$z$ estimation.  \change{ nflow-$z$ explicitly accounts for it by assigning a latent variable to each training galaxy sample.  To elaborate this point further, suppose that there is a degenerate point in the color space, so that many galaxies with different redshifts are collapsed to this point.  In NF, each galaxy training sample ($z$) is assigned a unique redshift distribution corresponding to  a latent variable $y$.  This degenerate point in color space is effectively lifted into some higher dimensional surface in the latent space. By resolving the degeneracy point, its impact on the photo-$z$ estimation is contained. }   This issue is not addressed in other methods, at least not explicitly.  \change{Second}, the nflow-$z$ network structure is relatively simple, and so the redshift information is highly concentrated, while \change{ the redshift information } can be diluted in more complicated models due to an excessively large number of degrees of freedom.  \change{  This comment is especially in reference to the finding that nflow-$z$ outperforms CNN. In principle, all the information, including the flux and color information, is contained in the images. Ideally, the algorithms like CNN that take the whole image as input should deliver better results than nflow-$z$, which uses only the distilled information from the image. However, it is conceivable that the degrees of freedom in the algorithm are so massive that the redshift information may not be effectively extracted. On the other hand, the usage of the distilled information can help the AI algorithm concentrate on the most physically relevant information.  The redshift information in nflow-$z$ is highly concentrated in the sense that the latent variables encode only the redshift distribution information.  }   Practically, a simple architecture structure  means that it takes much less time to run and is easy to tune its parameters to optimize the results.

\begin{figure}[!htb]
\centering
\includegraphics[width=\linewidth]{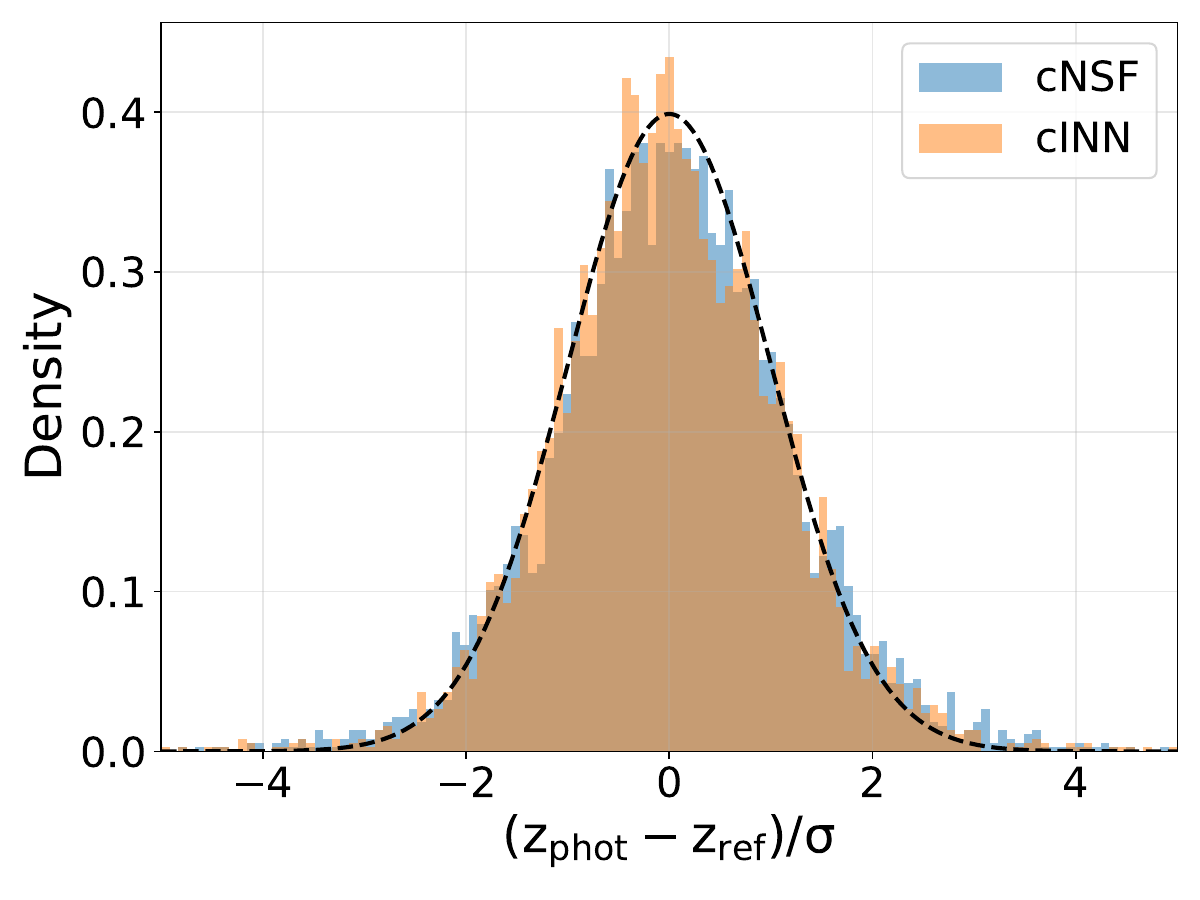}
\caption{ Distribution of the normalized deviation $\epsilon$ for both cINN (orange) and cNSF (blue). The standard normal (black dashed) is over-plotted for comparison.     }
\label{fig:zp_zs_sigma_CSST}
\end{figure}

\begin{figure}[!tb]
\centering
\includegraphics[width=\linewidth]{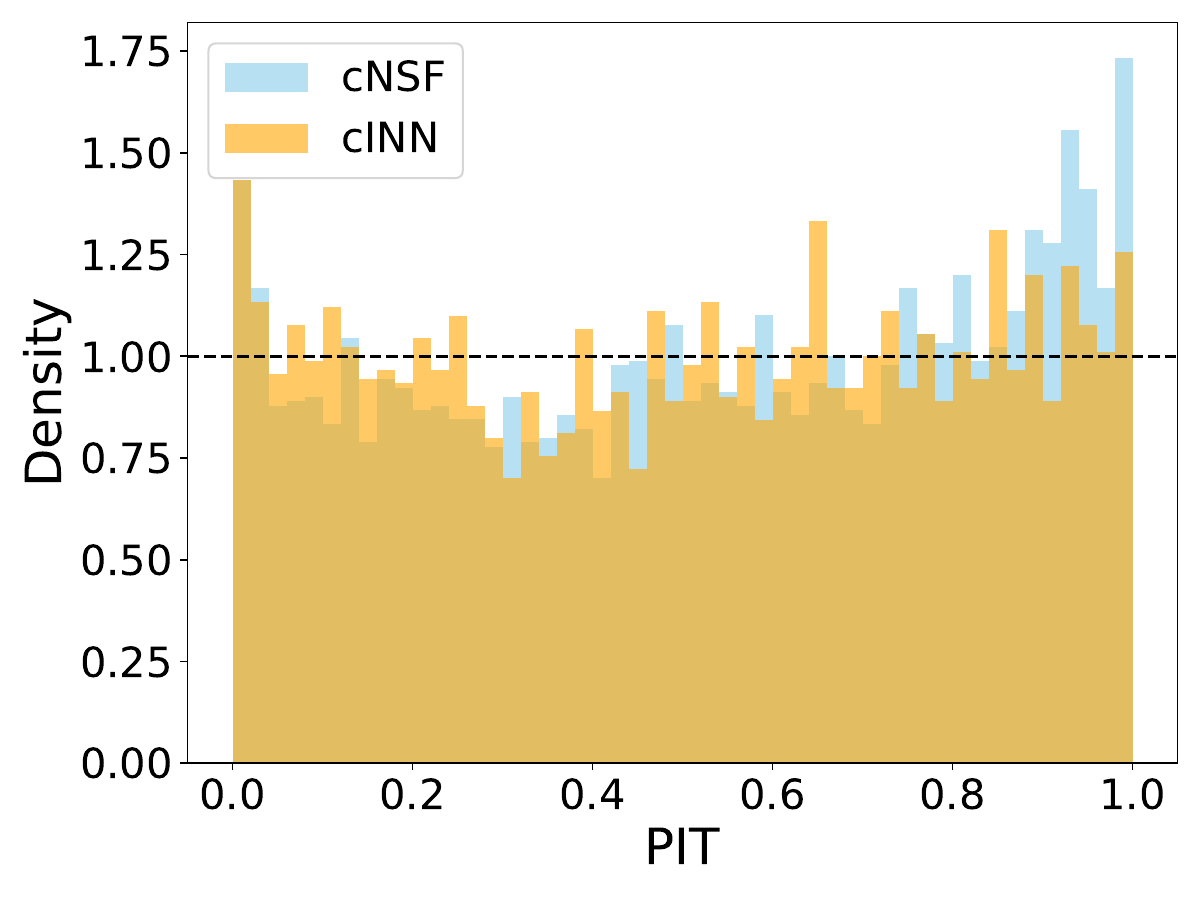}
\caption{ PIT transform for the photo-$z$ PDF obtained from cINN (orange) and cNSF (blue). If the PDF is exact,  then the PIT transform yields a uniform distribution, which is indicated as a dashed line.   }
\label{fig:PIT_CSST}
\end{figure}

\subsubsection{Accuracy of the error bar and PDF estimation}

We now check the accuracy of the error estimation using the normalized deviation 
\beq
\epsilon \equiv \frac{ \Delta z }{ \sigma}, 
\eeq
where $\sigma $ is the error estimate of the photo-$z$. In the ideal case, we anticipate the error bar to give an accurate estimate of the deviation $\Delta z$ and $\epsilon$ follows a standard normal distribution.  In Fig.~\ref{fig:zp_zs_sigma_CSST}, we show $\epsilon $ for both cINN and cNSF.  Overall, both distributions are in good agreement with the standard normal. Closer inspection reveals that the cNSF result is slightly wider than the standard normal, suggesting that the error bars are slightly underestimated.

Probability integral transform (PIT, \citep{Bordoloi_etal2010}) offers a model-independent way to verify if the PDF derived from the model is consistent with the data. Mathematically, ${\rm PIT}$  is defined as 
\beq
{\rm PIT}(z_{\rm ref} ) = \int_0^{z_{\rm ref}} P(z') dz',   
\eeq
where $P$ is the estimated PDF.  The idea behind is that if the probability distribution is correct, then the probability that a galaxy appears should be uniform when the distance is measured in terms of the cumulative probability.  We plot the PIT for cINN and cNSF in Fig.~\ref{fig:PIT_CSST}.  Overall, both PITs are close to the uniform distribution. In greater detail, the cINN PIT is more consistent with the uniform distribution, while the cNSF PIT peaks at both ends with a  mild trough at the center.  The cNSF PIT pattern indicates that the derived PDF tends to underestimate the width of the distribution. This trend is consistent with the implication from  Fig.~\ref{fig:zp_zs_sigma_CSST} that cNSF slightly underestimates the error bar.

\begin{figure*}[!ht]
\centering
\begin{minipage}{0.95\textwidth}
\centering
\includegraphics[width=\linewidth]{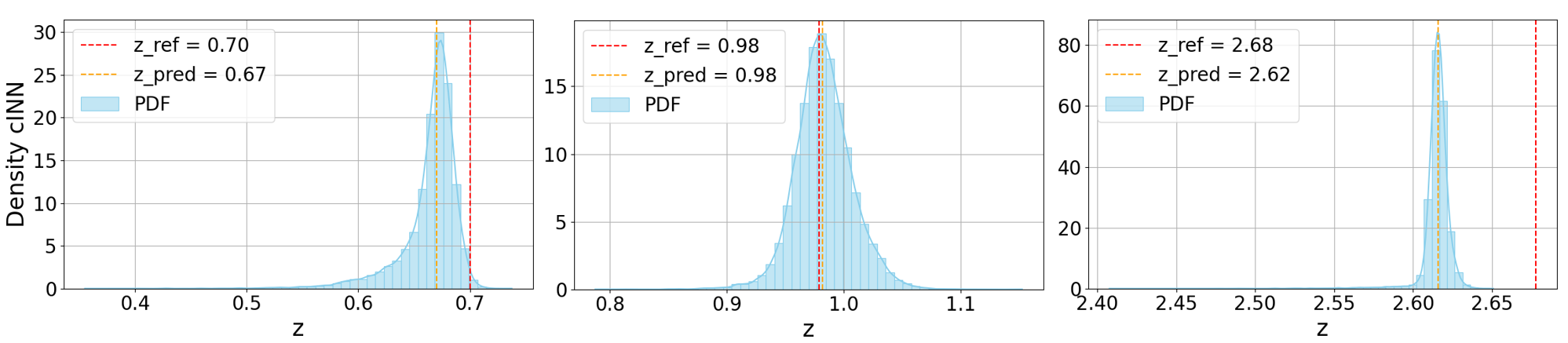}
\end{minipage}%
\vspace{0.5cm}
\begin{minipage}{0.95\textwidth}
\centering
\includegraphics[width=\linewidth]{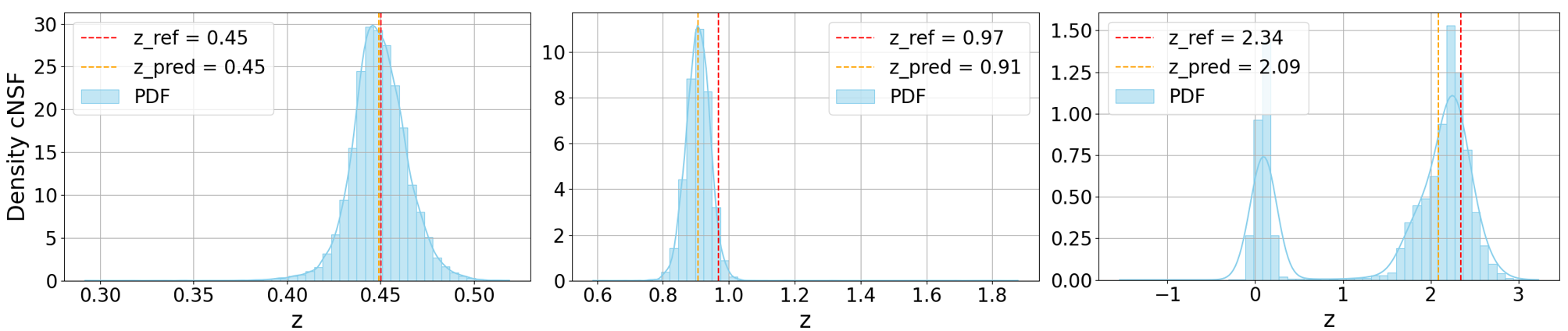}
\end{minipage}
 	\caption{ Some representative PDFs (histograms) obtained from cINN (upper panel) and cNSF (lower panel). The true spec-$z$ and the NF predicted point estimate are indicated as red and orange dashed lines, respectively. Both cINN and cNSF predominantly show only a single peak when the prediction is in good agreement with the true value; however, they may show a double-peak feature when the difference is large. }
   \label{fig:cINN_cNSF_PDF}
\end{figure*}

We now examine the characteristics of the PDFs, in particular, if the distribution  exhibits multi-modal behavior. For convenience, we first define  $\delta  = |\Delta z| / (1 + z_{\rm ref}) $ and consider its values in a few different ranges: [0, 0.01], [0.01, 0.02], [0.02, 0.05], [0.05, 0.1], [0.1, 0.15], and [0.15, $\infty$].  
We show the PDFs for a few representative examples in Fig.~\ref{fig:cINN_cNSF_PDF}. Among the cases examined, the cINN PDF exhibits only a single peak for $\delta \lesssim 0.05 $ and it has a small probability ($\sim 5\%$) showing a double-peak feature for $\delta \gtrsim  0.05$.  For cNSF, the PDF is also  single-peak only for $\delta \lesssim 0.05$, but it has a more appreciable chance ($\sim 20\%$) of exhibiting a double-peak feature for $ \delta \gtrsim 0.05 $.   


\begin{figure*}[!htb]
\centering
\includegraphics[width=\linewidth]{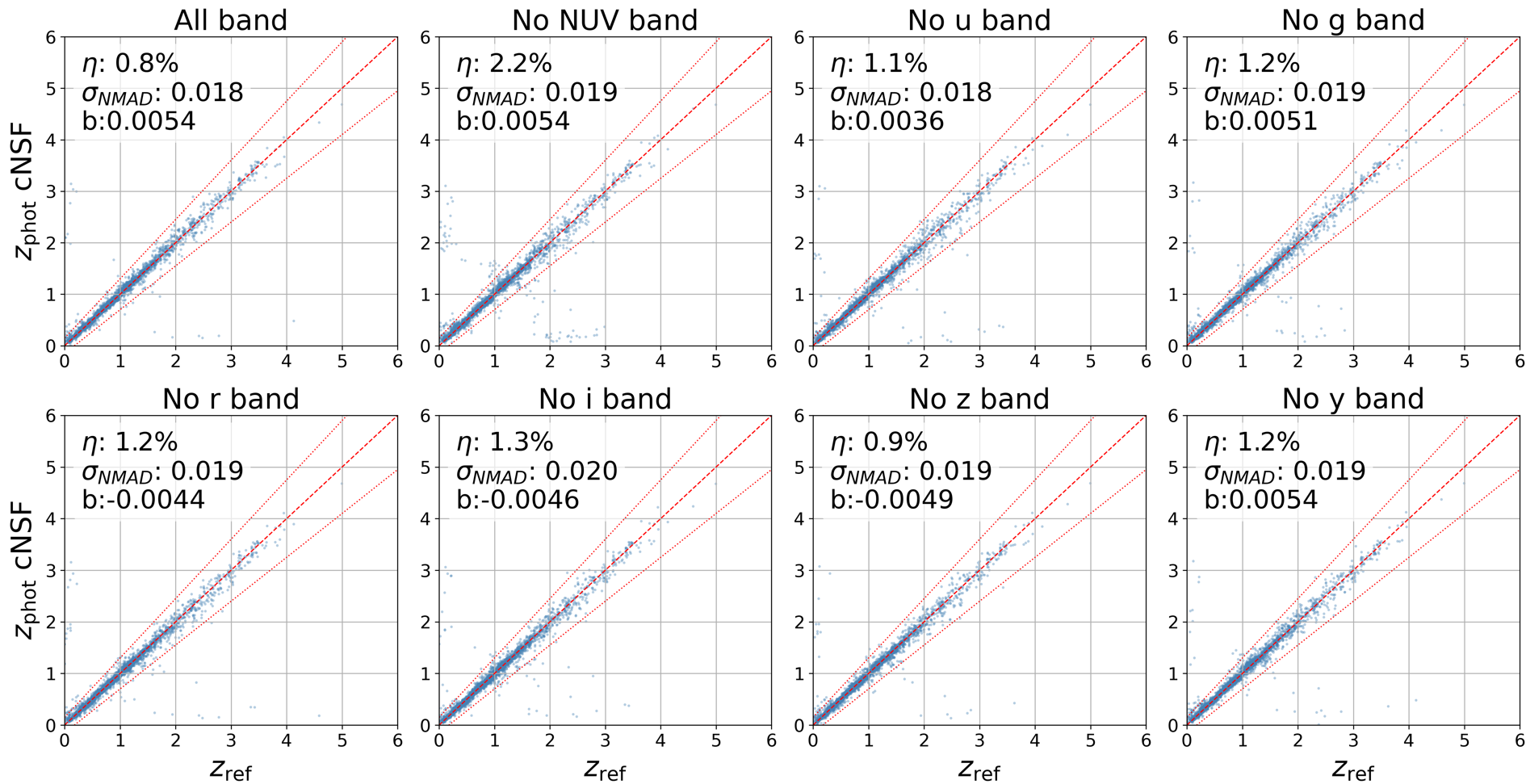}
\caption{ Impact of removing one of the photometric bands. Shown are the scatter plots between spec-$z$ and photo-$z$ estimated with full band information or one of the bands missing. The photo-$z$s are derived from cNSF.    }
\label{fig:cNSF_no-one-band}
\end{figure*}

\subsubsection{ The importance of the individual bands }

It is instructive to check the importance of various bands in the performance of photo-$z$ estimation. To do so, we consider the scenario when one of the photo-$z$ band data is missing.

In Fig.~\ref{fig:cNSF_no-one-band}, we show the scatter plot between the spec-$z$ and cNSF photo-$z$ using photometric data with full band data or one of the bands missing.  Note that for the missing band case, we use the data with one band missing in both training and testing.     The trend for cINN is broadly similar. We find that missing a single band does not significantly influence the performance. Among the bands, while missing $i$, $g$, or $r$ gives a noticeable increase in $ \eta$ and $\snmad$, missing $NUV$ leads to the most significant impact, with $\snmad $ and especially $\eta$ boosted most substantially. \change{  We note that the $NUV$ band is particularly effective in distinguishing the Lyman break from the Balmer break,  reducing the catastrophic error caused by the confusion between these breaks.  }

There is  a similar study on the importance of different bands in CSST photo-$z$ estimation by \citet{Cao_etal2018}.  Table 3 there shows the impact of removing one of the CSST bands on the photo-$z$ estimation using the SED fitting method. They find that  $r$, $g$, and $i$ are the most important ones as they lead to the most sizable increase in error when either of them is missing, while missing $NUV$ causes only a modest increase in the metrics.  There are a couple of notable differences between their study and ours. First, they use an old CSST mock, which is less realistic than the present one. Second, their conclusion is based on SED fitting with {\tt LePhare} only.

To investigate the cause of the discrepancy, we first perform SED fitting on this old catalog using {\tt LePhare}. While \citet{Cao_etal2018} refine the COSMOS SED templates by performing SED interpolation, here, for simplicity, we directly use the default COSMOS SED templates. Although our resultant   $\eta$  and  $\snmad $ are larger than the values shown in \citet{Cao_etal2018}  by a factor of a few, we also find a similar pattern that in the case of missing $g$, $r$, or $i$ band data, the photo-$z$ estimation suffers  the largest reduction in accuracy. We go on to apply cNSF on this dataset, and we find that the photo-$z$ precision metrics are smaller than theirs by an order of magnitude or so. The excellent performance of our algorithm is partly due to the fact that the mock is relatively simple.  For cNSF, we also find similar results that missing $NUV$ causes the most significant reduction in the photo-$z$ estimation accuracy, and missing $g$, $r$, or $i$ band data contribute to a mild decrease in accuracy only.

Our results suggest that the importance of the band depends on the level of accuracy in question. We can understand this in terms of the concepts of coarse and fine features in the data. For the low level of accuracy attained by SED fitting, the coarse features are mostly efficiently captured by $g$, $r$, or $i$. However, for the high level algorithms such as cNSF, even missing one of the bands, the coarse features can still be captured well, and the differentiating features are the fine ones, which is the $NUV$ in this case.

SED fitting is often used to guide survey design for its simplicity, e.g.,~in the case of CSST, it is used to find out the optimal photometric band configuration. Our results suggest that this may only lead to a partial result, and the conclusion so drawn may not be optimal.

\begin{table*}[!tb]
  \caption{ Impact of under-representative training sample and its mitigation by resampling, relative to the fiducial full sample. The resampling process increases the weight of the under-representative part. Both cINN and cNSF results are shown, in the form of mean and standard deviation estimated over ten runs.  }
\label{tab:unrepresentative_sample}
  \begin{tabular}{|l|c|c|c||c|c|c||}
    \hline
    \multirow{2}{*}{} &
      \multicolumn{3}{c}{  cINN } &
      \multicolumn{3}{c}{  cNSF }  \\
      \hline 
        &  Full  &  Under-representative &  Resampled  &   Full  &  Under-representative &  Resampled   \\
        \hline
 $\eta$              &  $1.60 \pm 0.16 \% $  &  $8.64 \pm 0.83$ \%  &  $3.37 \pm 0.018 \% $      &  $0.83 \pm0.07\%$    & $ 7.01\pm 0.62\% $   &  $3.58\pm 0.43 $\%   \\
 $\sigma_{\rm NMAD}$    &  $0.018 \pm 0.0004 $  & $ 0.027 \pm 0.0015 $   &  $0.022 \pm 0.0004  $   &  $0.017\pm0.0002$     &  $0.022 \pm 0.0004 $   & $0.023 \pm 0.0005 $   \\
 $b$                &  $-0.0001 \pm 0.0004$  &  $ -0.0024 \pm 0.0028 $   & $-0.0004 \pm 0.0005 $  & $0.0003\pm0.001$  &  $-0.0012 \pm 0.0014$  & $0.0002 \pm 0.0006  $  \\ \hline 
  \end{tabular}
\end{table*}

\subsubsection{ Unrepresentative training sample and its mitigation }

So far, we have considered the ideal scenario that the training dataset is a faithful representation of the final testing dataset. In reality, we may face the situation that the training data are not representative of the actual dataset to which it is applied. In particular, the training sample is likely to be under-representative at the faint end of the sample.

To investigate the issue of unrepresentative training sample, we consider a toy sample that is complete up to an $i$-band magnitude of 22, beyond which only a fraction of galaxies of the full sample are available.  We choose the fraction to be $\exp{ [ - ( i - 22 )/ \sigma ]}  $ with $\sigma = 0.5$.

In Table \ref{tab:unrepresentative_sample}, we compare the results obtained with the fiducial full sample with the under-representative case.  In the latter case, the training is carried out on the unrepresentative sample, while testing is performed on the full sample. We find that for both algorithms, the outlier fraction $\eta$ is substantially inflated, and $ \snmad $ also increases, albeit less significantly. 

The accuracy deteriorates because the under-representative part constitutes a reduced weighting in the estimation of the full conditional probability distribution.  To alleviate the situation, we consider boosting  the weight of the under-representative part by resampling it. We do so by resampling the under-representative part with replacement until the distribution of galaxies is the same as the full sample.

Here, we randomly resample the under-representative part. If the under-representative sample is biased in color-magnitude space, we can obtain color-dependent weighting using methods like kNN to get a representative sample \citep{Lima_etal2008,Sadeh:2015lsa}.  In case of a very low galaxy sample in the faint end, there is a possibility of generating the faint sample using the galaxy formation model to supplement (e.g.,~\citet{Ramachandra_etal2022,Moskowitz_etal2024}).

Table \ref{tab:unrepresentative_sample} shows that this simple up-weighting indeed improves the overall results, and it is particularly significant for $\eta$.  This again indicates that $ \eta $ is sensitive to the sample size of the training data. However, we note that there is an exceptional case where $\snmad$ actually increases slightly when resampling is applied in cNSF.  This is not easy to understand, given that $\eta$ does decrease.  We note that even in the under-representative case, cNSF already gives a tight $\snmad$, so any error introduced may degrade the results. It is conceivable that the resampled galaxies are less representative (noisy), and thus can compromise all galaxies when they are resampled.  Anyway, even in this case, the rise in the accuracy in $\eta$ is quite likely to compensate for the drop in $\snmad$, so it is worthwhile to perform resampling.

\subsection{Test on the COSMOS2020 catalog }

\begin{figure*}[!htb]
\centering
\includegraphics[width=\linewidth]{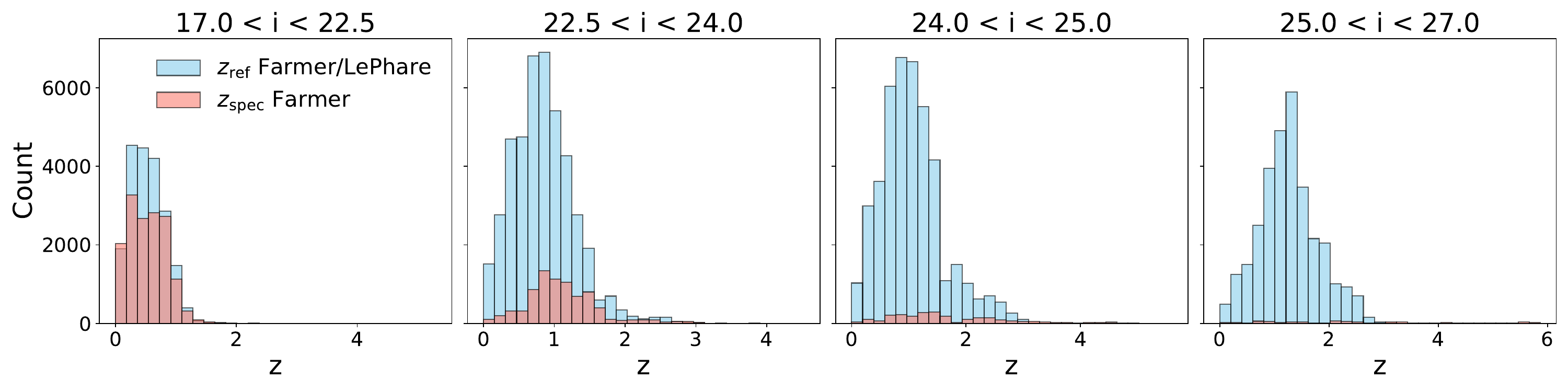}
\caption{ Redshift distribution histogram of the COSMOS2020 galaxy sample (blue) and the spec-$z$ catalog (red) in the COSMOS field.  The samples are divided into four $i$-magnitude bins.  The reference redshifts for these samples are the high-quality COSMOS2020 photo-$z$ and spec-$z$, respectively.  } 
\label{fig:COSMOS2020_mag_redshift}
\end{figure*}

\begin{figure*}[!htb]
\centering
\includegraphics[width=\linewidth]{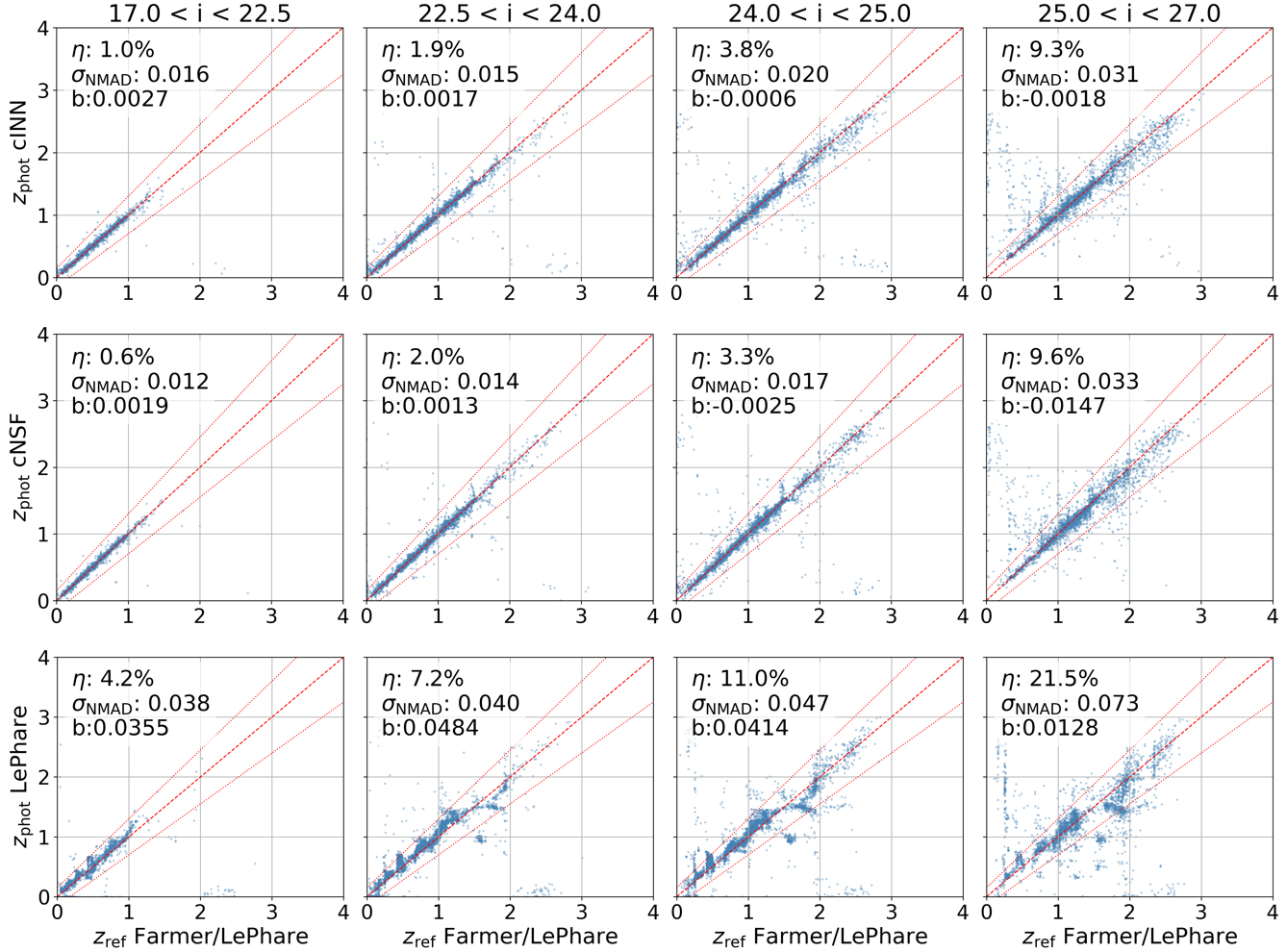}
\caption{ Scatter plot between the photo-$z$ estimated using the seven CSST-like band data versus the fiducial reference redshift from COSMOS2020. The results from cINN, cNSF, and {\tt LePhare} are compared (top to bottom). The samples are divided into four $i$-magnitude groups (left to right).  }
\label{fig:zp_zs_COSMOS2020_Farmer_LP}
\end{figure*}

\begin{table*}
  \caption{ Comparison of $b$, $\snmad$, and $\eta$ results obtained from cINN, cNSF, and {\tt LePhare} for different galaxy samples built from COSMOS2020 catalog. In the first four big rows,  the reference redshifts are high-quality photo-$z$s ({\tt LePhare} or {\tt EAZY}) based on the photometry from FARMER or CLASSIC, while the last big row shows the results for FARMER sample with spec-$z$.  The results for four $i$-magnitude bins are shown (from left to right).  The last column shows the galaxy sample used, with the number of galaxies shown in brackets.  To guide the eyes, the best metric in each sub-column is in bold. }
\label{tab:b_sigma_eta_COSMOS2020}
  \begin{tabular}{|l|c|c|c||c|c|c||c|c|c||c|c|c|| c|}
    \hline
    \multirow{2}{*}{} &
      \multicolumn{3}{c}{  $ 17<i<22.5$} &
      \multicolumn{3}{c}{ $22.5<i<24 $} &
       \multicolumn{3}{c}{ $24<i<25 $} &
       \multicolumn{3}{c}{ $25<i<27 $}  &
          \multicolumn{1}{c}{ }     \\
      \hline 
    & $b$ & $\sigma_{\rm NMAD}$ & $\eta$ &  $b$ & $\sigma_{\rm NMAD}$ & $\eta$ & $b$ & $\sigma_{\rm NMAD}$ & $\eta$ & $b$ & $\sigma_{\rm NMAD}$ & $\eta$  &  galaxy sample  \\ \hline       
cINN    & $\bold{0.001}$ & $0.015$ & $1.0\%$  &  $\bold{0.000}$ & $0.015$ & $2.0\%$                             & $\bold{0.000}$  & $0.020$ & $4.0\%$ & $\bold{-0.005}$ & $0.035$ & $9.7\%$  &  FARMER \\ 
cNSF    & $0.002$ & $\bold{0.012}$ & $\bold{0.6}\%$ &  $0.001$ & $\bold{0.014}$ & $\bold{1.9\%}$                 & $-0.002$       & $\bold{0.017}$ & $\bold{3.3\%}$ & $-0.015$ & $\bold{0.033}$ & $\bold{9.6\%}$  & {\tt LePhare} \\
{\tt LePhare }   & $0.036$ & $0.038$ & $4.2\%$ &  $0.048$ & $0.040$ & $7.2\%$                                    & $0.041$       & $0.047$ & $11.0\%$ & $0.013$ & $0.073$ & $21.5\%$ &  (137,892)    \\ \hline 
cINN    & $\bold{0.000}$ & $\bold{0.013}$ & $\bold{1.7\%}$ &  $\bold{-0.001}$ & $0.016$ & $\bold{2.7\%}$         & $\bold{0.000}$       & $\bold{0.022}$ & $8.0\%$ & $-0.004$ & $0.051$ & $21.2\%$  &  FARMER \\ 
cNSF    & $\bold{0.000}$ & $\bold{0.013}$ & $1.9\%$ &  $\bold{-0.001}$ & $\bold{0.014}$ & $2.9\%$                & $0.004$      & $0.023$ & $\bold{6.6\%}$ & $\bold{0.001}$ & $\bold{0.043}$ & $\bold{20.3\%}$  & {\tt EAZY} \\
{\tt LePhare}    & $0.032$ & $0.042$ & $6.7\%$ &  $0.049$ & $0.041$ & $7.8\%$                                    & $0.041$      & $0.054$ & $13.7\%$ & $-0.006$ & $0.096$ & $29.3\%$ & (161,297)   \\ \hline 
cINN    & $0.001$ & $0.012$ & $1.8\%$ &  $-0.004$ & $0.019$ & $3.9\%$                                            & $\bold{0.001}$       & $0.027$ & $\bold{7.1\%}$ & $-0.006$ & $0.051$ & $\bold{13.0\%}$  &  CLASSIC \\ 
cNSF    & $\bold{0.000}$ & $\bold{0.011}$ & $\bold{0.8\%}$ &  $\bold{0.000}$ & $\bold{0.017}$ & $\bold{3.7\%}$   & $0.002$ & $\bold{0.025}$ & $7.6\%$ & $\bold{-0.002}$ & $\bold{0.045}$ & $14.7\%$  & {\tt LePhare} \\
{\tt LePhare }    & $0.026$ & $0.043$ & $4.5\%$ &  $0.047$ & $0.057$ & $9.1\%$                                   & $0.038$ & $0.063$ & $14.3\%$ & $0.038$ & $0.092$ & $26.1\%$ & (145,779)  \\ \hline 
cINN    & $\bold{0.000}$ & $0.015$ & $3.3\%$ &  $\bold{0.000}$ & $0.022$ & $8.7\%$                               & $\bold{0.000}$ & $0.038$ & $\bold{17.0\%}$ & $-0.008$ & $0.096$ & $\bold{32.7\%}$  &  CLASSIC \\ 
cNSF    & $-0.002$ & $\bold{0.013}$ & $\bold{2.8\%}$ &  $\bold{0.000}$ & $\bold{0.020}$ & $\bold{7.2\%}$         & $0.001$ & $\bold{0.037}$ & $17.9\%$ & $0.006$ & $\bold{0.092}$ & $33.1\%$  & {\tt EAZY} \\
{\tt LePhare }    & $0.031$ & $0.049$ & $9.8\%$ &  $0.060$ & $0.056$ & $13.9\%$                                  & $0.053$ & $0.078$ & $24.2\%$ & $\bold{0.004}$ & $0.153$ & $40.7\%$ & (240,746) \\ \hline 
cINN    & $0.001$  & $0.014$ & $3.1\%$ &  $\bold{0.001}$  & $0.017$ &  $3.5\%$                                   & $0.021$   &   $0.062$ & $15.8\%$   &   $\bold{0.000}$   &    $0.158$   &    $40.7\%$    &  FARMAER  \\ 
cNSF    & $\bold{0.000}$  & $\bold{0.012}$ & $\bold{2.4\%}$ &  $\bold{0.001}$  & $\bold{0.012}$ & $\bold{2.5\%}$ & $\bold{-0.001}$   &   $\bold{0.027}$ & $\bold{6.0\%}$   &   $-0.005$   & $\bold{0.032}$      &    $\bold{28.4\%}$     &  spec-$z$  \\
{\tt LePhare } & $0.029$  & $0.044$ & $12.1\%$ &  $0.057$  & $0.046$ & $13.0\%$                                  & $0.075$   &   $0.060$ & $20.6\%$   &   $0.160$   & $0.151$      &    $42.9\%$     &  (26,379)   \\  \hline
\hline
  \end{tabular}
\end{table*}


COSMOS2020 catalog \citep{Weaver_etal2022} is the latest galaxy catalog compilation in the 2 deg$^2$ COSMOS field. This field has been observed by numerous surveys across wavelengths in the electromagnetic spectrum. Many of these measurements in the UV, optical, and IR regimes have been subsumed in the COSMOS2020 catalog.

This catalog provides photometry measured using two different pipelines, CLASSIC and FARMER, based on flux measurements in the bands $izJYHK_{\rm s}$.   The CLASSIC photometry is processed with the traditional pipeline, which has been used in previous releases of the COSMOS catalog \citep{Capak_etal2007,Ilbert_etal2009,Laigle_etal2016}. In this pipeline, the image is first homogenized to a target point spread function, and fluxes are then extracted within circular apertures.  On the other hand, the FARMER photometry is measured with the profile-fitting code,{\tt  TRACTOR} \citep{Lang_etal2016}, which derives a parametric model from the images assuming some morphological models. The FARMER photometry is shown to be more robust for faint galaxies.

The redshifts of galaxies in the catalog are given by high-quality photo-$z$s, derived from two template fitting codes, {\tt LePhare} \citep{Arnouts_etal2002,Ilbert_etal2006} and {\tt EAZY} \citep{Brammer_etal2008}, respectively.   The broad, medium, and narrow band data in the UV, optical, and IR frequency range are employed, totaling about 30 bands.  The precise number of galaxies in the catalog depends on the photometry and photo-$z$ estimator used. For example, for FARMER and {\tt LePhare}, there are 137,892 galaxies, and for  CLASSIC and {\tt EAZY}, there are 240,746 galaxies.  Besides the high-quality photo-$z$ catalogs, a spec-$z$ catalog in the COSMOS field is also available \citep{Khostovan_etal2025}.  In this catalog, there are 26,379 galaxies for FARMER photometry.  As an illustration, in Fig.~\ref{fig:COSMOS2020_mag_redshift}, we show the  reshift histograms for the FARMER-{\tt LePhare} sample  and the FARMER-spec-$z$ catalog.


We illustrate the performance of both cINN and cNSF via a scatter plot in Fig.~\ref{fig:zp_zs_COSMOS2020_Farmer_LP}. To be in line with the CSST mock setup, we only consider the photometric data from the CSST-like photometric bands, i.e.,~$NUVugrizy$. The reference redshift is the high-quality photo-$z$ from COSMOS2020.  In addition, we also display the results obtained using {\tt LePhare} on the same seven CSST-like bands. In the {\tt LePhare} fit, in addition to the 31 COSMOS SEDs offered alongside   {\tt LePhare}, we also use 12 SEDs from \citet{BruzualCharlot2003}. Following the extinction treatment in \cite{Weaver_etal2022}, we utilize the four extinction laws considered in \cite{Weaver_etal2022} with the extinction amplitude as a free parameter.


Fig.~\ref{fig:zp_zs_COSMOS2020_Farmer_LP} demonstrates that the nflow-$z$ results from cINN and cNSF are similar, and they are much better than the {\tt LePhare} fit.  In terms of $\eta$  and $\snmad$, these metrics from the  nflow-$z$  algorithms are generally lower by a factor of a few relative to the {\tt LePhare} results.  As expected, the metrics get worse for fainter galaxies because of poor photometric measurements and reduction in training data size.

In Table \ref{tab:b_sigma_eta_COSMOS2020}, we show the  photo-$z$ metrics  $b$, $\sigma_{\rm NMAD}$, and $\eta$ for different galaxy samples built from the COSMOS catalogs. Different combinations of the galaxy photometry (FARMER or CLASSIC) and reference redshift (high-quality  {\tt LePhare} or {\tt EAZY} photo-$z$ from COSMOS2020  or spec-$z$) are considered. Again, we consider four $i$-band magnitude groups. For each galaxy sample, the trends are similar to those found in Fig.~\ref{fig:zp_zs_COSMOS2020_Farmer_LP}. That is, cINN and cNSF give comparable results, while the {\tt LePhare} fit is worse by a factor of a few. In greater detail, cNSF tends to yield more accurate photo-$z$ estimations across different samples. This is apparent for the FARMER-spec-$z$ sample, especially in the faint groups ($24<i<25$ and $25<i<27$), where the available training sample is scarce. Thus, we conclude that cNSF is more robust than cINN.

We note that Table \ref{tab:b_sigma_eta_COSMOS2020} also reveals that the resultant photo-$z$ accuracy is dependent on the photometry and photo-$z$ estimators.  However, this trend is largely driven by the fact that there are varying numbers of galaxies in different galaxy samples. Indeed, we generally find that the overall photo-$z$ metrics are worse for a sample with a larger number of galaxies. This is particularly obvious for the faint groups.  It is conceivable that the extra galaxies are the culprit for the poor performance. This also means that some of these algorithms, such as FARMER and {\tt LePhare}, are better tuned to get rid of the low-quality galaxies. We can verify this using the same sample of galaxies. For the spec-$z$ sample, there are unique IDs allowing us to identify galaxies in the FARMER and CLASSIC photometry. For the same spec-$z$ sub-sample, we find that the variations in the photo-$z$ metrics are much reduced for the choice of FARMER or CLASSIC photometry.    We stress that our goal here is to compare the performance of different photo-$z$ estimation algorithms; as long as we concentrate on the same sample, this sample dependence does not affect our conclusions.


Finally, we comment on the missing photometry data issue and its mitigation strategies. 
When the reference redshift is photo-$z$, there are relatively abundant galaxies available. Thus, we can afford to select only those galaxies with data in all bands intact.  However, for the spec-$z$ case, the available spec-$z$ galaxies are scarce, especially in the faint end. We have to consider galaxies with some flux data in certain bands unavailable.  However, for the machine learning method, the data in all channels must be used, and so the missing data should be somehow filled in. We test a few different methods to impute the missing data, including replacing the missing data with the mean, median, minimum of the data array, or assigning an arbitrary value such as -99.9.  Besides these commonly used simple methods, we have also considered the imputation algorithm GAIN \citep{Yoon2018_gain}, which is based on the GAN algorithm (see Sec.~\ref{sec:cGAN_confrontation}).   \citet{Luo_etal2024_GAIN} demonstrate that GAIN is effective for photometric data imputation.  Among these methods, GAIN stands out in our tests, giving the lowest $\eta$ and $\snmad$ in all the cases considered, consistent with \citet{Luo_etal2024_GAIN}. Therefore, for the spec-$z$ case in Table \ref{tab:b_sigma_eta_COSMOS2020}, we have employed the GAIN to fill in the missing data.




\subsection{ Comparison with cGAN and MDN }
\label{sec:cGAN_confrontation}

\begin{figure}[!tb]
\centering
\includegraphics[width=\linewidth]{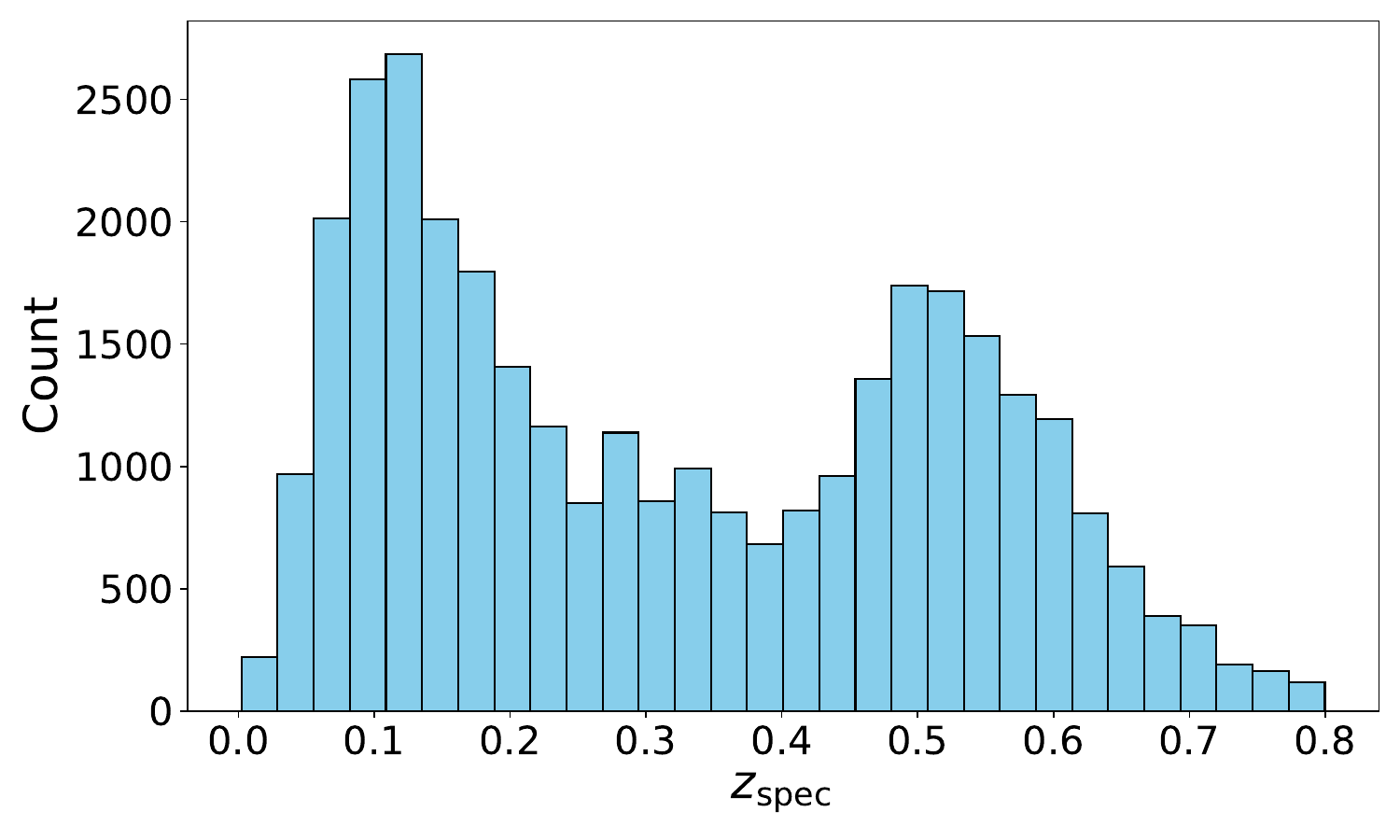}
\caption{ Redshift distribution of the galaxy sample used in the cGAN and MDN comparison test.   }
\label{fig:CGAN_specz_dist}
\end{figure}

\begin{figure*}[!htb]
\centering
\includegraphics[width=\linewidth]{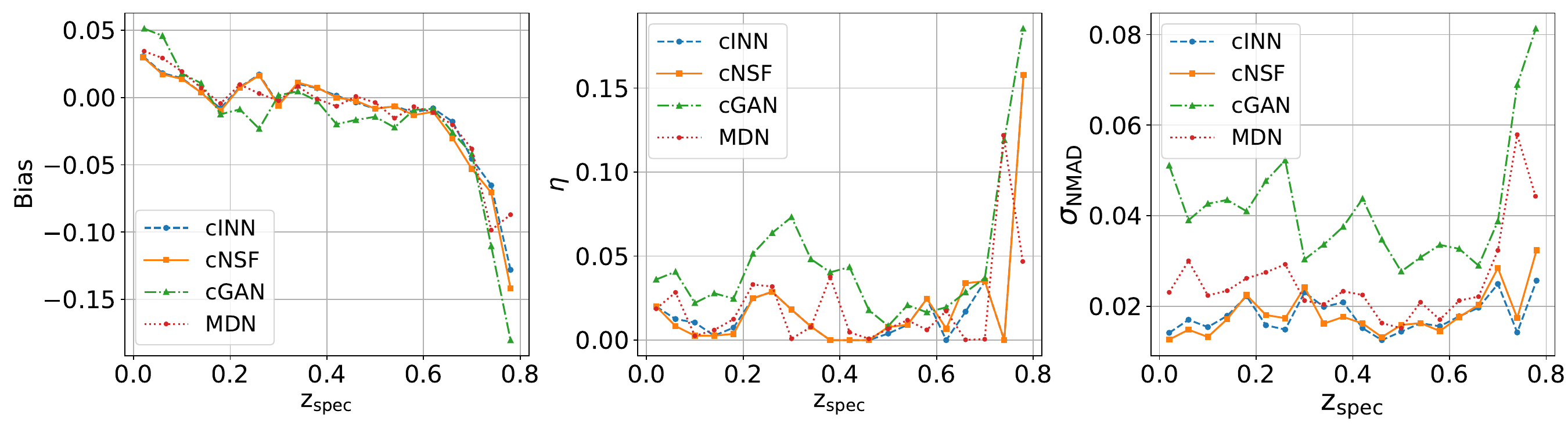}
\caption{ Bias, $\eta$, and $\snmad$ (from left to right) results for a DES Y1 sample shown in Fig.~\ref{fig:CGAN_specz_dist}. The cGAN (green, dotted-dashed) and MDN (red, dotted) results are extracted from \citet{Garcia-Fernandez2025}. The NF algorithms (cINN (blue, dashed) and cNSF (orange, solid)) perform similarly well, and their overall results are better than cGAN or MDN.   }
\label{fig:cGAN_bias_eta_sigma}
\end{figure*}

GAN \citep{Goodfellow_etal2014} is another well-known generative modeling network in deep learning. In GAN, rather than training the network with the maximum likelihood, a novel min-max game framework is introduced, where two adversarial models compete against each other.  An adversarial model, called generative model $G$ learns the underlying data distribution, while another adversarial model, dubbed discriminative model $D$ evaluates the probability that a sample originates from the training data rather than from $G$. Both $G$ and $D$ can be implemented as non-linear mapping functions, such as Multi Layer Perceptrons.   \citet{MirzaOsindero2014} extends GAN to a conditional generalization, conditional Generative Adversarial Network (cGAN) by feeding the extra conditional information to both adversarial models $G$ and $D$. Thus the treatment of the condition is similar to our nflow-$z$ models.  \citet{Garcia-Fernandez2025} applied the cGAN model to photo-$z$ estimation. As both NF and GAN are important members in generative modeling, it is interesting to check how nflow-$z$ fairs relative to cGAN.

In addition, \citet{Garcia-Fernandez2025} also compares their results against those from Mixture Density Network (MDN, \citet{Bishop_1994}).  In MDN, a sum of base distributions is used to parametrize the target probability distribution, and the weights of each distribution and the parameters for each base distribution are optimized by neural networks. The base distribution is commonly taken to be the normal distribution, and in this case, the base distribution parameters are the mean and variance of the normal distribution. 
\citet{Garcia-Fernandez2025} used the MDN implementation ported from \citet{Ansari_etal2021}.

Here, we compare our results against the cGAN and MDN results from \citet{Garcia-Fernandez2025}. To do so, we apply our method to the same dataset provided by \citet{Garcia-Fernandez2025}, which can be downloaded alongside the cGAN code.  The dataset is a subset of the  DES Y1 galaxies \citep{Drlica-Wagner_etal2018}. Only galaxies in Stripe 82 region with the associated spec-$z$ from SDSS falling in  $0<z_{\rm spec} < 0.8$ are selected. The redshift distribution of this galaxy sample is shown in Fig.~\ref{fig:CGAN_specz_dist}.  This dataset contains $griz$ bands magnitude only, and in particular, no error bars for the magnitudes are provided. Thus, in our NF runs, we only use the fluxes and colors computed from these magnitudes.

In Fig.~\ref{fig:cGAN_bias_eta_sigma}, we plot the bias, $\eta$, and $\snmad $ from different algorithms. The cGAN and MDN results are extracted from \citet{Garcia-Fernandez2025} directly. Note that, following \citet{Garcia-Fernandez2025},  the bias is defined as the mean of $ \Delta z $ rather than its median.  cINN and cNSF perform similarly well, and are better than cGAN and MDN when all the metrics are taken into account.  The cGAN fares the worst in all the metrics. Although MDN yields competitive bias and $ \eta $, its $\snmad$ is substantially higher than the NF results.

While both NF and GAN are members of generative modeling in deep learning, NF directly models the probability distribution of the target space, and GAN does not. Our studies suggest  that direct probability distribution modeling is instrumental in enhancing the precision of photo-$z$ estimation.  MDN is similar to NF in spirit, as it uses a mixture of Gaussian distributions to model the target distribution.  However, it is less flexible than NF because it still has a fixed function form, while NF allows for more degrees of freedom in constructing the coordinate transformation.



\subsection{Comparison with ANNz2 }

ANNz2 \citep{Sadeh:2015lsa} is a  widely used  photo-$z$ estimation code, and  improves upon the previous version ANNz \citep{CollisterLahav_2004}.  This implementation employs multiple machine learning methods (MLMs) from the {\tt TMVA} package \citep{Hoecker:2007gu}. Among the available methods, artificial neural networks (ANNs) and boosted decision trees (BDTs) are found to be the most useful in photo-$z$ estimation.

 There are two principal modes of operation in ANNz2: Single Regression and Randomized Regression. The former outputs a single photo-$z$ estimate, while the latter is capable of generating PDFs. We focus on the Randomized Regression here.  An ensemble of 100 MLMs is trained, which can be obtained by setting the initialization random seed or the parameters in the configuration.  For instance, in ANNs, we can set the number of hidden layers and the number of neurons thereof, while the number of trees and the type of forest construction algorithm can be varied in BDTs. Both ANN and BDT are used in our runs. The performance of the MLMs is ranked based on metrics such as the bias and spread of the estimation, and the best-performing MLM is chosen to produce the  photo-$z$ prediction.  Random weights are assigned to the MLMs, and the combined PDF yielding the best PIT test is adopted for the final PDF.

\begin{figure*}[!ht]
\centering
\begin{minipage}{0.45\textwidth}
\centering
\includegraphics[width=\textwidth]{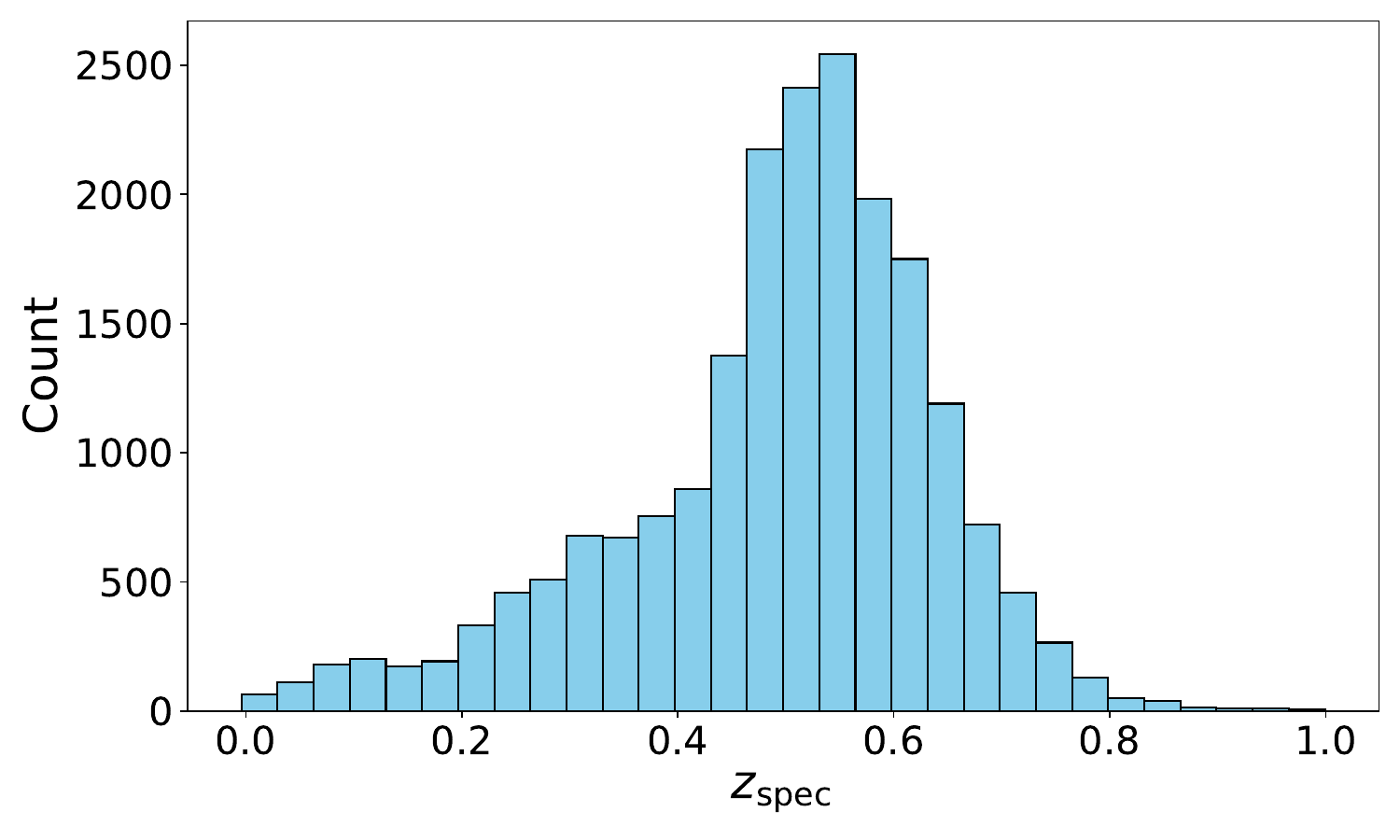}
\end{minipage}%
\hspace{0.5cm}
\begin{minipage}{0.45\textwidth}
\centering
\includegraphics[width=\textwidth]{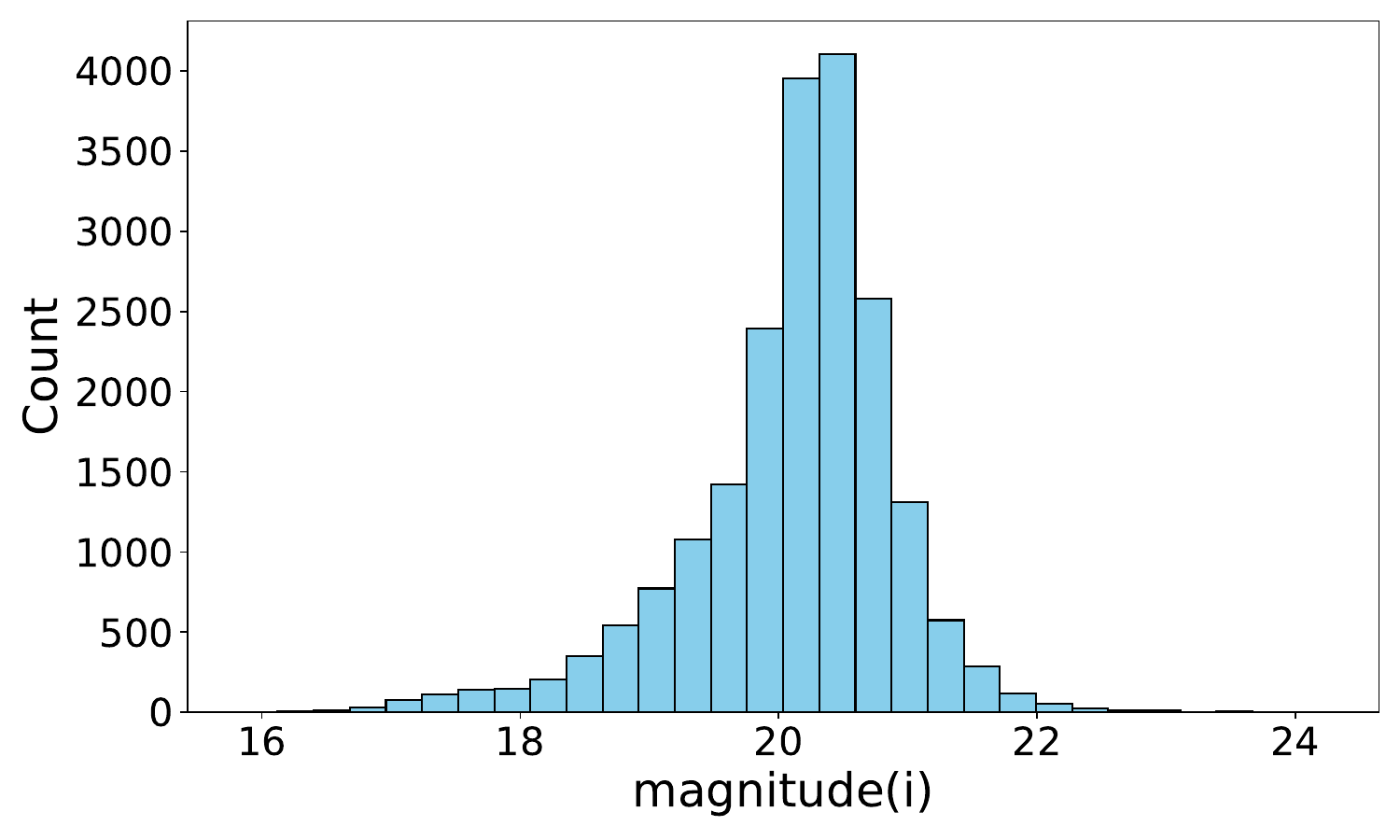}
\end{minipage}
\caption{  Redshift (left) and $i$-band magnitude (right) distribution for the galaxy sample used for ANNz2 comparison. }
\label{fig:z_magi_dist_ANNZ2}
\end{figure*}

\begin{figure*}[!htb]
\centering
\includegraphics[width=\linewidth]{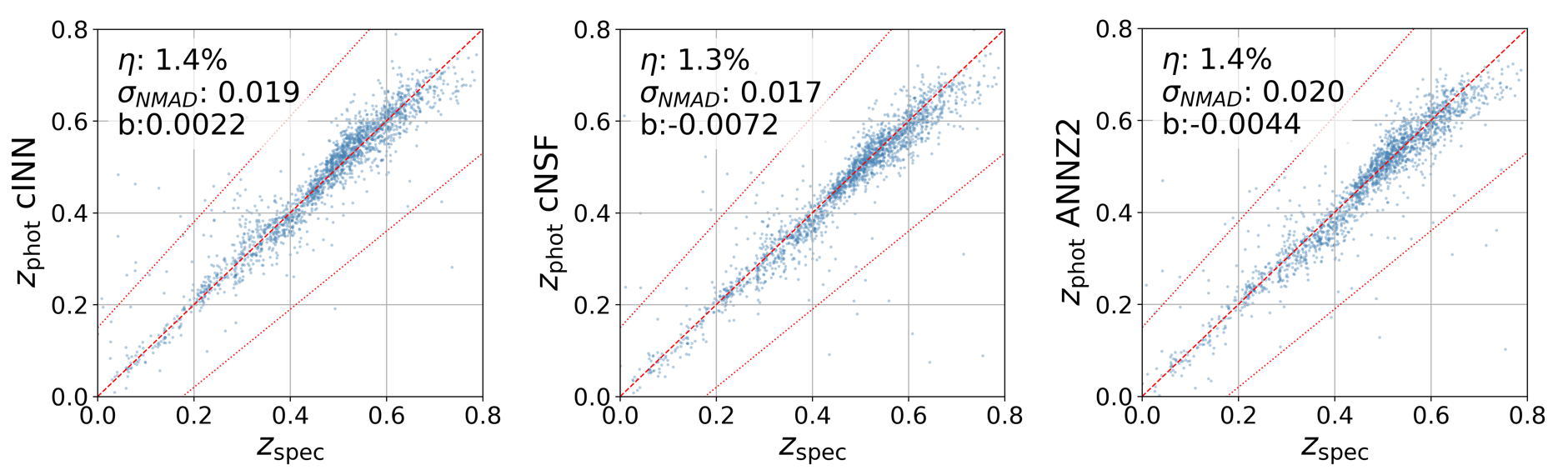}
\caption{ Scatter plot between the spec-$z$ and photo-$z$ estimated by cINN, cNSF, and ANNz2 (left to right). While the performance of these methods is similar, cINN and cNSF take significantly less time to run.     } 
\label{fig:scatter_zp_zs_ANNz2}
\end{figure*}

 We conduct the tests on the dataset offered alongside ANNz2. This dataset consists of the $ugriz$ band magnitudes and their associated errors from the SDSS DR10 dataset \citep{Ahn_etal2014}. The redshift and magnitude distribution of this sample are shown in Fig.~\ref{fig:z_magi_dist_ANNZ2}. Fig.~\ref{fig:scatter_zp_zs_ANNz2} is a scatter plot between the spec-$z$ and photo-$z$ estimates from cINN, cNSF, and ANNz2.  The results from these methods are not significantly different, with cNSF giving the smallest  $\eta $ and $ \snmad$  and cINN the smallest bias.

Because ANNz2 requires running many MLMs, it is much more time-consuming.  In our test, the ANNz2 run takes nearly 1.5 days, while cINN and cNSF take about 15 and 3 minutes, respectively.


\subsection{ Application to DESCaLS data  }

\begin{table*}[!htb]
  \caption{
    The photo-$z$ quality metrics, $\eta$  and $\snmad$, from the cNSF algorithm on the DESCaLS data. The dataset is split into $z<21$ and $z>21$ based on the $z$-band magnitude. The results obtained using various combinations of the flux (P), flux error (Perr) and morphological (M) information  are compared.    We have also included the random forest results from \cite{ZhouNewman_etal2021}, which makes use of both photometric and morphological information.  } 
\label{tab:DESCaLS_test}
  \begin{tabular}{|l|c|c|c|c|c||c|c|c|c|c||}
    \hline
    \multirow{1}{*}{} &
      \multicolumn{5}{c}{  $z<21$ } &
      \multicolumn{5}{c}{  $z>21$ }  \\
      \hline 
           & P      & P+Perr & P+M    &  P+Perr+M  & Zhou+2021 &  P       & P+Perr    & P+M     &  P+Perr+M  & Zhou+2021  \\  
      \hline
$\eta$   & 2.17\%  &  1.81\% & 1.45\% &  1.41\%   & 1.51\%     & 17.85\%  & 16.65\%   & 17.04\% & 16.29\%    & 24.6\%     \\  
$\snmad$ & 0.0162  & 0.0151  & 0.0140 &  0.0139   & 0.0133     & 0.0538   & 0.0524    & 0.0527  &  0.0517    & 0.0725     \\
\hline
  \end{tabular}
\end{table*}

In this section, we apply our algorithm to the photometric data from the Dark Energy Camera Legacy Survey (DESCaLS), which is part of the DESI Legacy Imaging Surveys \citep{Dey_etal2019}.   Besides its original purpose for DESI target selection, the DESCaLS catalog has been used for numerous cosmological studies, e.g.,~\citet{Song_etal2024,Qin_etal2015}.

As a value-added product, Photometric Redshifts for the Legacy Surveys (PRLS) catalog \citep{ZhouNewman_etal2021} offers photo-$z$ estimates for the  Legacy Surveys sample.  The photo-$z$s are derived from the random forest algorithm, based on the DR8 dataset of the Legacy Surveys.  In addition to the photometric data in $g$, $r$, and $z$ bands from DESCaLS, the near infrared imaging data in $W1$ and $W2$ bands from Wide-field Infrared Survey Explorer (WISE, \citet{Wright_etal2010}) are also used to assist photo-$z$ estimation.  In the random forest analysis of \citet{ZhouNewman_etal2021}, their input takes the $r$-band magnitude, $g-r$, $r-z$, $z-W1$, and $W1 - W2$ colors, and additional morphological information by means of  the half-light radius, axis ratio, and shape probability.  The usage of the shape information was inspired by the study of \citet{Soo_etal2018}, which finds that the morphological information can add substantial gain in photo-$z$ accuracy if there are few bands available, such as $grz$ only.

In Table \ref{tab:DESCaLS_test}, we present the photo-$z$ estimate results on this DESCaLS catalog. The results are derived from the cNSF algorithm. We have isolated the impact of flux errors by utilizing flux information alone and  both flux and flux error. Motivated by  \cite{ZhouNewman_etal2021}, we also consider the case when both  photometry and morphological information are employed. We use the three types of morphological information as  \cite{ZhouNewman_etal2021}. In order to directly compare with  \cite{ZhouNewman_etal2021}, we divide the samples into two groups based on $z$-band magnitude: $z<21$ and $z>21$. Note that for these samples, we impute the missing entries using the GAIN algorithm.

For the bright group ($z<21$), while adding the photometry error gives some gain on the photo-$z$ accuracy (17\% on $\eta$ and 8\% on $\snmad$), including the morphological information significantly improves the photo-$z$ accuracy (33\% on $\eta$ and 14\% on $\snmad$). Further adding the flux error on top of the flux plus morphology case only marginally improves the results.  For the faint group ($z>21$), adding the flux error or morphological information only mildly improves the accuracy, with flux error contributing slightly larger gain. By including both, the accuracy is enhanced by (14\% on $\eta$ and 9\% on $\snmad$).  The finding that the morphological information is only significant for the bright group is consistent with \cite{ZhouNewman_etal2021}.


Table \ref{tab:DESCaLS_test} also shows the PRLS results, which are extracted from  Appendix B in \cite{ZhouNewman_etal2021}. The PRLS results are derived from both photometric and morphological information. For the bright group ($z<21$), our results are similar to PRLS. However, for the faint group ($z>21$), our algorithm gives substantial improvement over the PRLS results, with  $\eta $ reduced by 34\% and $\snmad$ by 29\%.

We have demonstrated that our algorithm can bring substantial improvement in the photo-$z$ estimates in the faint end. In turn, this is expected to improve the cosmological application results.  This exercise demonstrates that our algorithm is particularly advantageous for the faint galaxy samples, where the training samples are scarce and the measurements are noisy.



\section{ Conclusions} 

\label{sec:conclusions}

Many large-scale cosmological surveys are on-going or upcoming, e.g.,~those mentioned in the Introduction.  These surveys are expected to bring about an enormous amount of photometric data. In particular, the observed sample is expected to extend to high redshift and much deeper magnitude.  Getting an accurate photo-$z$ estimate for the photometric samples is a prerequisite to achieving exquisite scientific results from these data samples.   Although many works have been devoted to photo-$z$ estimation, in particular, various machine learning algorithms have been applied for photo-$z$ estimation, there is still room for improvement, especially in the regime where the spec-$z$ training data are sparse.

In this work, we apply normalizing flow (NF), a powerful machine learning method, to develop a new photo-$z$ estimation pipeline. We call the NF-based photo-$z$ estimation framework, nflow-$z$.  nflow-$z$ directly models the redshift probability distribution, conditional on the observables such as fluxes and colors.  It learns the coordinate transformation required to transform the complex target distribution to a simple base distribution, such as normal.  We explore two types of architectures to implement nflow-$z$: Conditional Invertible Neural Network (cINN) and  Conditional Neural Spline Flow (cNSF), and contrast the performance of these implementations.

We apply nflow-$z$ to several datasets and compare the results against other state-of-the-art algorithms.  The datasets tested include  a CSST mock, COSMOS2020 catalog, a DES Y1 sample, an SDSS sample, and a DESCaLS catalog.  These datasets cover diverse redshift ranges, magnitude spans, and various training sample sizes, and thus, they can help to check the performance of the NF methods under different situations. Among others, we employ the normalized median absolute deviation $\snmad$, the outlier fraction $\eta$, and the bias $b$ to quantify the photo-$z$ accuracy.  Our testing results generally indicate that the NF methods compare favorably to other algorithms. In the CSST mock test, the cNSF algorithm yields the lowest $\eta$ and $\snmad$, outperforming others, including random forest, MLP, CNN, hybrid transfer network, and recurrent neural network \change{(the list in Sec.~\ref{sec:CSST_quantification})}. Moreover, the NF network is generally much more lightweight than many of these networks, and so it requires substantially fewer resources to run.  For the tests on COSMOS2020 catalogs, we find that our results generally improve over the SED fit by {\tt LePhare} by a factor of few \change{(Fig.~\ref{fig:zp_zs_COSMOS2020_Farmer_LP} and Table \ref{tab:b_sigma_eta_COSMOS2020})}.  In the DESCaLS dataset test, while for the bright sample ($z<21$) our results are similar to the official PRLS results from random forest, ours improve over the official ones by about 30\% for the faint sample ($z>21$) \change{in terms of $\eta$ or $\snmad$ (Table \ref{tab:DESCaLS_test})}.   We have compared nflow-$z$ against the results from cGAN and MDN on a DES Y1 sample.  We find that nflow-$z$ surpasses \change{ cGAN (in terms of $\eta$ or $\snmad$) and beats MDN (in terms of $\snmad$) (Fig.~\ref{fig:cGAN_bias_eta_sigma})}.   Even though our test results are similar to the ANNz2 in performance on an SDSS sample, the NF method takes much less time to run (ten minutes versus a couple of days). Thus, we have demonstrated that the NF-based method is effective in estimating photo-$z$ on a number of datasets.  Some of these datasets have been applied to various cosmological and astrophysical applications; the improvement in photo-$z$ precision brought by the NF method should be beneficial to these studies.

Here, we summarize a few key properties of nflow-$z$. It models the redshift probability distribution directly and naturally yields a PDF for the photo-$z$ estimate.  The redshift-color degeneracy is a physical effect limiting the accuracy of the photo-$z$ estimation.   \change{ nflow-$z$ reduces the impact of this degeneracy by resolving the degeneracy into a higher-dimensional surface in latent space, thanks to the assignment of a latent variable to each training sample in the NF network structure. }  \change{ Moreover, the nflow-$z$ network is relatively simple. It is a 1-dimensional network for redshift and this enables the network to concentrate on the redshift information in the data. }   In practice, the simple network structure means that it takes little training and running time, so it is much less resource demanding than other competing algorithms and is easy to tune its parameters to optimize its performance. \change{ For example, in the CSST mock test, running the training and testing steps takes less than half an hour. }

Our tests suggest that cNSF tends to deliver smaller $\eta$ and $\snmad$ than cINN, although it seems that cNSF often gives a larger bias (this can be corrected for). Also, we find that cNSF is more robust in the sense that it still manages to deliver good results in the case of poor photometry measurements or a sparse training sample. Consequently, we recommend cNSF instead of cINN. The advantages of nflow-$z$ are especially apparent in the faint end regime where the training sample is relatively sparse, and the data measurements are noisy. Thus, it can be particularly useful for deep surveys or noisy datasets.  The codes implementing nflow-$z$ are available for download on \href{https://github.com/kcc274/nflow-z.git}{GitHub}\footnote{https://github.com/kcc274/nflow-z.git} \change{ or the frozen version on Zenodo\footnote{https://doi.org/10.5281/zenodo.17595685}}.  They can be readily applied to other photometric datasets. We encourage readers to explore and experiment with them.

\section*{Acknowledgments}
We thank Jon{\'a}s Chaves-Montero, Dezi Liu, Jian Qin, and Ji Yao for useful discussions. 
YR, KCC, and RS are  supported by the National Science Foundation of China under the grant number 12273121 and 12533002.  YL is supported by the Major Key Project of Peng Cheng Laboratory and the National Key Research and Development Program of China under grant number 2023YFA1605600.  This work is also supported by the science research grants from the China Manned Space Project with NO.CMS-CSST-2021-B01.

\clearpage
\bibliographystyle{aa}
\bibliography{reference}

\begin{thebibliography}{121}
\expandafter\ifx\csname natexlab\endcsname\relax\def\natexlab#1{#1}\fi

\bibitem[{Abbott {et~al.}(2022{\natexlab{a}})Abbott, Aguena, Alarcon, Allam,
  Alves, Amon, Andrade-Oliveira, Annis, Avila, Bacon, Baxter, Bechtol, Becker,
  Bernstein, Bhargava, Birrer, Blazek, Brandao-Souza, Bridle, Brooks,
  Buckley-Geer, Burke, Camacho, Campos, Carnero~Rosell, Carrasco~Kind,
  Carretero, Castander, Cawthon, Chang, Chen, Chen, Choi, Conselice, Cordero,
  Costanzi, Crocce, da~Costa, da~Silva~Pereira, Davis, Davis, De~Vicente,
  DeRose, Desai, Di~Valentino, Diehl, Dietrich, Dodelson, Doel, Doux,
  Drlica-Wagner, Eckert, Eifler, Elsner, Elvin-Poole, Everett, Evrard, Fang,
  Farahi, Fernandez, Ferrero, Ferté, Fosalba, Friedrich, Frieman,
  García-Bellido, Gatti, Gaztanaga, Gerdes, Giannantonio, Giannini, Gruen,
  Gruendl, Gschwend, Gutierrez, Harrison, Hartley, Herner, Hinton, Hollowood,
  Honscheid, Hoyle, Huff, Huterer, Jain, James, Jarvis, Jeffrey, Jeltema,
  Kovacs, Krause, Kron, Kuehn, Kuropatkin, Lahav, Leget, Lemos, Liddle, Lidman,
  Lima, Lin, MacCrann, Maia, Marshall, Martini, McCullough, Melchior,
  Mena-Fernández, Menanteau, Miquel, Mohr, Morgan, Muir, Myles, Nadathur,
  Navarro-Alsina, Nichol, Ogando, Omori, Palmese, Pandey, Park, Paz-Chinchón,
  Petravick, Pieres, Plazas~Malagón, Porredon, Prat, Raveri, Rodriguez-Monroy,
  Rollins, Romer, Roodman, Rosenfeld, Ross, Rykoff, Samuroff, Sánchez,
  Sanchez, Sanchez, Sanchez~Cid, Scarpine, Schubnell, Scolnic, Secco, Serrano,
  Sevilla-Noarbe, Sheldon, Shin, Smith, Soares-Santos, Suchyta, Swanson,
  Tabbutt, Tarle, Thomas, To, Troja, Troxel, Tucker, Tutusaus, Varga, Walker,
  Weaverdyck, Wechsler, Weller, Yanny, Yin, Zhang, \& Zuntz}]{DESY3_3x2pt}
Abbott, T., Aguena, M., Alarcon, A., {et~al.} 2022{\natexlab{a}}, Physical
  Review D, 105

\bibitem[{Abbott {et~al.}(2019)}]{Abbott:2017wcz}
Abbott, T. {et~al.} 2019, Mon. Not. Roy. Astron. Soc., 483, 4866

\bibitem[{{Abbott} {et~al.}(2018){Abbott}, {Abdalla}, {Alarcon}, {Aleksi{\'c}},
  {Allam}, {Allen}, {Amara}, {Annis}, {Asorey}, {Avila}, {Bacon}, {Balbinot},
  {Banerji}, {Banik}, {Barkhouse}, {Baumer}, {Baxter}, {Bechtol}, {Becker},
  {Benoit-L{\'e}vy}, {Benson}, {Bernstein}, {Bertin}, {Blazek}, {Bridle},
  {Brooks}, {Brout}, {Buckley-Geer}, {Burke}, {Busha}, {Campos}, {Capozzi},
  {Carnero Rosell}, {Carrasco Kind}, {Carretero}, {Castander}, {Cawthon},
  {Chang}, {Chen}, {Childress}, {Choi}, {Conselice}, {Crittenden}, {Crocce},
  {Cunha}, {D'Andrea}, {da Costa}, {Das}, {Davis}, {Davis}, {De Vicente},
  {DePoy}, {DeRose}, {Desai}, {Diehl}, {Dietrich}, {Dodelson}, {Doel},
  {Drlica-Wagner}, {Eifler}, {Elliott}, {Elsner}, {Elvin-Poole}, {Estrada},
  {Evrard}, {Fang}, {Fernandez}, {Fert{\'e}}, {Finley}, {Flaugher}, {Fosalba},
  {Friedrich}, {Frieman}, {Garc{\'\i}a-Bellido}, {Garcia-Fernandez}, {Gatti},
  {Gaztanaga}, {Gerdes}, {Giannantonio}, {Gill}, {Glazebrook}, {Goldstein},
  {Gruen}, {Gruendl}, {Gschwend}, {Gutierrez}, {Hamilton}, {Hartley}, {Hinton},
  {Honscheid}, {Hoyle}, {Huterer}, {Jain}, {James}, {Jarvis}, {Jeltema},
  {Johnson}, {Johnson}, {Kacprzak}, {Kent}, {Kim}, {King}, {Kirk}, {Kokron},
  {Kovacs}, {Krause}, {Krawiec}, {Kremin}, {Kuehn}, {Kuhlmann}, {Kuropatkin},
  {Lacasa}, {Lahav}, {Li}, {Liddle}, {Lidman}, {Lima}, {Lin}, {MacCrann},
  {Maia}, {Makler}, {Manera}, {March}, {Marshall}, {Martini}, {McMahon},
  {Melchior}, {Menanteau}, {Miquel}, {Miranda}, {Mudd}, {Muir}, {M{\"o}ller},
  {Neilsen}, {Nichol}, {Nord}, {Nugent}, {Ogando}, {Palmese}, {Peacock},
  {Peiris}, {Peoples}, {Percival}, {Petravick}, {Plazas}, {Porredon}, {Prat},
  {Pujol}, {Rau}, {Refregier}, {Ricker}, {Roe}, {Rollins}, {Romer}, {Roodman},
  {Rosenfeld}, {Ross}, {Rozo}, {Rykoff}, {Sako}, {Salvador}, {Samuroff},
  {S{\'a}nchez}, {Sanchez}, {Santiago}, {Scarpine}, {Schindler}, {Scolnic},
  {Secco}, {Serrano}, {Sevilla-Noarbe}, {Sheldon}, {Smith}, {Smith}, {Smith},
  {Soares-Santos}, {Sobreira}, {Suchyta}, {Tarle}, {Thomas}, {Troxel},
  {Tucker}, {Tucker}, {Uddin}, {Varga}, {Vielzeuf}, {Vikram}, {Vivas},
  {Walker}, {Wang}, {Wechsler}, {Weller}, {Wester}, {Wolf}, {Yanny}, {Yuan},
  {Zenteno}, {Zhang}, {Zhang}, {Zuntz}, \& {Dark Energy Survey
  Collaboration}}]{DESY1_3x2pt}
{Abbott}, T.~M.~C., {Abdalla}, F.~B., {Alarcon}, A., {et~al.} 2018, \prd, 98,
  043526

\bibitem[{{Abbott} {et~al.}(2024){Abbott}, {Adamow}, {Aguena}, {Allam},
  {Alves}, {Amon}, {Andrade-Oliveira}, {Asorey}, {Avila}, {Bacon}, {Bechtol},
  {Bernstein}, {Bertin}, {Blazek}, {Bocquet}, {Brooks}, {Burke}, {Camacho},
  {Carnero Rosell}, {Carollo}, {Carr}, {Carretero}, {Castander}, {Cawthon},
  {Chan}, {Chang}, {Conselice}, {Costanzi}, {Crocce}, {da Costa}, {Pereira},
  {Davis}, {De Vicente}, {Deiosso}, {Desai}, {Diehl}, {Dodelson}, {Doux},
  {Drlica-Wagner}, {Elvin-Poole}, {Everett}, {Ferrero}, {Fert{\'e}},
  {Flaugher}, {Fosalba}, {Frieman}, {Garc{\'\i}a-Bellido}, {Gaztanaga},
  {Giannini}, {Glazebrook}, {Gruendl}, {Gutierrez}, {Hartley}, {Hinton},
  {Hollowood}, {Honscheid}, {Huterer}, {James}, {Kent}, {Kuehn}, {Lahav},
  {Lee}, {Lewis}, {Lidman}, {Lima}, {Lin}, {Malik}, {Maraston}, {Marshall},
  {Martini}, {Mena-Fern{\'a}ndez}, {Menanteau}, {Miquel}, {Mohr}, {Myles},
  {M{\"o}ller}, {Nichol}, {Ogando}, {Palmese}, {Percival}, {Pieres}, {Plazas
  Malag{\'o}n}, {Porredon}, {Prat}, {Rodr{\'\i}guez-Monroy}, {Romer},
  {Roodman}, {Rosenfeld}, {Ross}, {Rykoff}, {Sako}, {Samuroff}, {S{\'a}nchez},
  {Sanchez}, {Sanchez Cid}, {Santiago}, {Schubnell}, {Sevilla-Noarbe},
  {Sheldon}, {Smith}, {Suchyta}, {Swanson}, {Tarle}, {Thomas}, {To}, {Toribio
  San Cipriano}, {Troxel}, {Tucker}, {Tucker}, {Walker}, {Weaverdyck},
  {Weller}, {Wiseman}, {Yanny}, \& {DES Collaboration}}]{DESY6_BAO}
{Abbott}, T.~M.~C., {Adamow}, M., {Aguena}, M., {et~al.} 2024, \prd, 110,
  063515

\bibitem[{Abbott {et~al.}(2022{\natexlab{b}})}]{DES:2021esc}
Abbott, T. M.~C. {et~al.} 2022{\natexlab{b}}, Phys. Rev. D, 105, 043512

\bibitem[{{Ahn} {et~al.}(2014){Ahn}, {Alexandroff}, {Allende Prieto}, {Anders},
  {Anderson}, {Anderton}, {Andrews}, {Aubourg}, {Bailey}, {Bastien},
  {Bautista}, {Beers}, {Beifiori}, {Bender}, {Berlind}, {Beutler}, {Bhardwaj},
  {Bird}, {Bizyaev}, {Blake}, {Blanton}, {Blomqvist}, {Bochanski}, {Bolton},
  {Borde}, {Bovy}, {Shelden Bradley}, {Brandt}, {Brauer}, {Brinkmann},
  {Brownstein}, {Busca}, {Carithers}, {Carlberg}, {Carnero}, {Carr},
  {Chiappini}, {Chojnowski}, {Chuang}, {Comparat}, {Crepp}, {Cristiani},
  {Croft}, {Cuesta}, {Cunha}, {da Costa}, {Dawson}, {De Lee}, {Dean},
  {Delubac}, {Deshpande}, {Dhital}, {Ealet}, {Ebelke}, {Edmondson},
  {Eisenstein}, {Epstein}, {Escoffier}, {Esposito}, {Evans}, {Fabbian}, {Fan},
  {Favole}, {Femen{\'\i}a Castell{\'a}}, {Fern{\'a}ndez Alvar}, {Feuillet},
  {Filiz Ak}, {Finley}, {Fleming}, {Font-Ribera}, {Frinchaboy},
  {Galbraith-Frew}, {Garc{\'\i}a-Hern{\'a}ndez}, {Garc{\'\i}a P{\'e}rez}, {Ge},
  {G{\'e}nova-Santos}, {Gillespie}, {Girardi}, {Gonz{\'a}lez Hern{\'a}ndez},
  {Gott}, {Gunn}, {Guo}, {Halverson}, {Harding}, {Harris}, {Hasselquist},
  {Hawley}, {Hayden}, {Hearty}, {Herrero Dav{\'o}}, {Ho}, {Hogg}, {Holtzman},
  {Honscheid}, {Huehnerhoff}, {Ivans}, {Jackson}, {Jiang}, {Johnson},
  {Kinemuchi}, {Kirkby}, {Klaene}, {Kneib}, {Koesterke}, {Lan}, {Lang}, {Le
  Goff}, {Leauthaud}, {Lee}, {Lee}, {Long}, {Loomis}, {Lucatello}, {Lupton},
  {Ma}, {Mack}, {Mahadevan}, {Maia}, {Majewski}, {Malanushenko},
  {Malanushenko}, {Manchado}, {Manera}, {Maraston}, {Margala}, {Martell},
  {Masters}, {McBride}, {McGreer}, {McMahon}, {M{\'e}nard}, {M{\'e}sz{\'a}ros},
  {Miralda-Escud{\'e}}, {Miyatake}, {Montero-Dorta}, {Montesano}, {More},
  {Morrison}, {Muna}, {Munn}, {Myers}, {Nguyen}, {Nichol}, {Nidever},
  {Noterdaeme}, {Nuza}, {O'Connell}, {O'Connell}, {O'Connell}, {Olmstead},
  {Oravetz}, {Owen}, {Padmanabhan}, {Palanque-Delabrouille}, {Pan}, {Parejko},
  {Parihar}, {P{\^a}ris}, {Pepper}, {Percival}, {P{\'e}rez-R{\`a}fols}, {Dotto
  Perottoni}, {Petitjean}, {Pieri}, {Pinsonneault}, {Prada}, {Price-Whelan},
  {Raddick}, {Rahman}, {Rebolo}, {Reid}, {Richards}, {Riffel}, {Robin},
  {Rocha-Pinto}, {Rockosi}, {Roe}, {Ross}, {Ross}, {Rossi}, {Roy},
  {Rubi{\~n}o-Martin}, {Sabiu}, {S{\'a}nchez}, {Santiago}, {Sayres},
  {Schiavon}, {Schlegel}, {Schlesinger}, {Schmidt}, {Schneider}, {Schultheis},
  {Sellgren}, {Seo}, {Shen}, {Shetrone}, {Shu}, {Simmons}, {Skrutskie}, \&
  {Slosar}}]{Ahn_etal2014}
{Ahn}, C.~P., {Alexandroff}, R., {Allende Prieto}, C., {et~al.} 2014, \apjs,
  211, 17

\bibitem[{{Aihara} {et~al.}(2022){Aihara}, {AlSayyad}, {Ando}, {Armstrong},
  {Bosch}, {Egami}, {Furusawa}, {Furusawa}, {Harasawa}, {Harikane}, {Hsieh},
  {Ikeda}, {Ito}, {Iwata}, {Kodama}, {Koike}, {Kokubo}, {Komiyama}, {Li},
  {Liang}, {Lin}, {Lupton}, {Lust}, {MacArthur}, {Mawatari}, {Mineo},
  {Miyatake}, {Miyazaki}, {More}, {Morishima}, {Murayama}, {Nakajima},
  {Nakata}, {Nishizawa}, {Oguri}, {Okabe}, {Okura}, {Ono}, {Osato}, {Ouchi},
  {Pan}, {Plazas Malag{\'o}n}, {Price}, {Reed}, {Rykoff}, {Shibuya},
  {Simunovic}, {Strauss}, {Sugimori}, {Suto}, {Suzuki}, {Takada}, {Takagi},
  {Takata}, {Takita}, {Tanaka}, {Tang}, {Taranu}, {Terai}, {Toba}, {Turner},
  {Uchiyama}, {Vijarnwannaluk}, {Waters}, {Yamada}, {Yamamoto}, \&
  {Yamashita}}]{Aihara_etal2022}
{Aihara}, H., {AlSayyad}, Y., {Ando}, M., {et~al.} 2022, \pasj, 74, 247

\bibitem[{{Almosallam} {et~al.}(2016){Almosallam}, {Jarvis}, \&
  {Roberts}}]{Almosallam_etal2016}
{Almosallam}, I.~A., {Jarvis}, M.~J., \& {Roberts}, S.~J. 2016, \mnras, 462,
  726

\bibitem[{{Amon} {et~al.}(2022){Amon}, {Gruen}, {Troxel}, {MacCrann},
  {Dodelson}, {Choi}, {Doux}, {Secco}, {Samuroff}, {Krause}, {Cordero},
  {Myles}, {DeRose}, {Wechsler}, {Gatti}, {Navarro-Alsina}, {Bernstein},
  {Jain}, {Blazek}, {Alarcon}, {Fert{\'e}}, {Lemos}, {Raveri}, {Campos},
  {Prat}, {S{\'a}nchez}, {Jarvis}, {Alves}, {Andrade-Oliveira}, {Baxter},
  {Bechtol}, {Becker}, {Bridle}, {Camacho}, {Carnero Rosell}, {Carrasco Kind},
  {Cawthon}, {Chang}, {Chen}, {Chintalapati}, {Crocce}, {Davis}, {Diehl},
  {Drlica-Wagner}, {Eckert}, {Eifler}, {Elvin-Poole}, {Everett}, {Fang},
  {Fosalba}, {Friedrich}, {Gaztanaga}, {Giannini}, {Gruendl}, {Harrison},
  {Hartley}, {Herner}, {Huang}, {Huff}, {Huterer}, {Kuropatkin}, {Leget},
  {Liddle}, {McCullough}, {Muir}, {Pandey}, {Park}, {Porredon}, {Refregier},
  {Rollins}, {Roodman}, {Rosenfeld}, {Ross}, {Rykoff}, {Sanchez},
  {Sevilla-Noarbe}, {Sheldon}, {Shin}, {Troja}, {Tutusaus}, {Tutusaus},
  {Varga}, {Weaverdyck}, {Yanny}, {Yin}, {Zhang}, {Zuntz}, {Aguena}, {Allam},
  {Annis}, {Bacon}, {Bertin}, {Bhargava}, {Brooks}, {Buckley-Geer}, {Burke},
  {Carretero}, {Costanzi}, {da Costa}, {Pereira}, {De Vicente}, {Desai},
  {Dietrich}, {Doel}, {Ferrero}, {Flaugher}, {Frieman}, {Garc{\'\i}a-Bellido},
  {Gaztanaga}, {Gerdes}, {Giannantonio}, {Gschwend}, {Gutierrez}, {Hinton},
  {Hollowood}, {Honscheid}, {Hoyle}, {James}, {Kron}, {Kuehn}, {Lahav}, {Lima},
  {Lin}, {Maia}, {Marshall}, {Martini}, {Melchior}, {Menanteau}, {Miquel},
  {Mohr}, {Morgan}, {Ogando}, {Palmese}, {Paz-Chinch{\'o}n}, {Petravick},
  {Pieres}, {Romer}, {Sanchez}, {Scarpine}, {Schubnell}, {Serrano}, {Smith},
  {Soares-Santos}, {Tarle}, {Thomas}, {To}, {Weller}, \& {DES
  Collaboration}}]{Amon_etal2022}
{Amon}, A., {Gruen}, D., {Troxel}, M.~A., {et~al.} 2022, \prd, 105, 023514

\bibitem[{{Ansari} {et~al.}(2021){Ansari}, {Agnello}, \&
  {Gall}}]{Ansari_etal2021}
{Ansari}, Z., {Agnello}, A., \& {Gall}, C. 2021, \aap, 650, A90

\bibitem[{Ardizzone {et~al.}(2018-2022)Ardizzone, Bungert, Draxler, Köthe,
  Kruse, Schmier, \& Sorrenson}]{freia}
Ardizzone, L., Bungert, T., Draxler, F., {et~al.} 2018-2022, {Framework for
  Easily Invertible Architectures (FrEIA)}

\bibitem[{{Ardizzone} {et~al.}(2018){Ardizzone}, {Kruse}, {Wirkert}, {Rahner},
  {Pellegrini}, {Klessen}, {Maier-Hein}, {Rother}, \&
  {K{\"o}the}}]{Ardizzone_etal2018}
{Ardizzone}, L., {Kruse}, J., {Wirkert}, S., {et~al.} 2018, arXiv e-prints,
  arXiv:1808.04730

\bibitem[{{Ardizzone} {et~al.}(2019){Ardizzone}, {L{\"u}th}, {Kruse}, {Rother},
  \& {K{\"o}the}}]{Ardizzone_etal2019}
{Ardizzone}, L., {L{\"u}th}, C., {Kruse}, J., {Rother}, C., \& {K{\"o}the}, U.
  2019, arXiv e-prints, arXiv:1907.02392

\bibitem[{{Arnouts} {et~al.}(1999){Arnouts}, {Cristiani}, {Moscardini},
  {Matarrese}, {Lucchin}, {Fontana}, \& {Giallongo}}]{Arnouts_1999}
{Arnouts}, S., {Cristiani}, S., {Moscardini}, L., {et~al.} 1999, \mnras, 310,
  540

\bibitem[{{Arnouts} {et~al.}(2002){Arnouts}, {Moscardini}, {Vanzella},
  {Colombi}, {Cristiani}, {Fontana}, {Giallongo}, {Matarrese}, \&
  {Saracco}}]{Arnouts_etal2002}
{Arnouts}, S., {Moscardini}, L., {Vanzella}, E., {et~al.} 2002, \mnras, 329,
  355

\bibitem[{{Asgari} {et~al.}(2021){Asgari}, {Lin}, {Joachimi}, {Giblin},
  {Heymans}, {Hildebrandt}, {Kannawadi}, {St{\"o}lzner}, {Tr{\"o}ster}, {van
  den Busch}, {Wright}, {Bilicki}, {Blake}, {de Jong}, {Dvornik}, {Erben},
  {Getman}, {Hoekstra}, {K{\"o}hlinger}, {Kuijken}, {Miller}, {Radovich},
  {Schneider}, {Shan}, \& {Valentijn}}]{Asgari_etal2021}
{Asgari}, M., {Lin}, C.-A., {Joachimi}, B., {et~al.} 2021, \aap, 645, A104

\bibitem[{{Bartelmann} \& {Schneider}(2001)}]{BartelmannSchneider2001}
{Bartelmann}, M. \& {Schneider}, P. 2001, \physrep, 340, 291

\bibitem[{{Ben{\'\i}tez}(2000)}]{Benitez_2000}
{Ben{\'\i}tez}, N. 2000, \apj, 536, 571

\bibitem[{Bingham {et~al.}(2019)Bingham, Chen, Jankowiak, Obermeyer, Pradhan,
  Karaletsos, Singh, Szerlip, Horsfall, \& Goodman}]{Pyro_2019}
Bingham, E., Chen, J.~P., Jankowiak, M., {et~al.} 2019, Journal of Machine
  Learning Research, 20, 1

\bibitem[{Bishop(1994)}]{Bishop_1994}
Bishop, C. 1994, Technical Report Aston University

\bibitem[{{Bister} {et~al.}(2022){Bister}, {Erdmann}, {K{\"o}the}, \&
  {Schulte}}]{Bister_etal2022}
{Bister}, T., {Erdmann}, M., {K{\"o}the}, U., \& {Schulte}, J. 2022, European
  Physical Journal C, 82, 171

\bibitem[{{Bolzonella} {et~al.}(2000){Bolzonella}, {Miralles}, \&
  {Pell{\'o}}}]{Bolzonella_etal2000}
{Bolzonella}, M., {Miralles}, J.~M., \& {Pell{\'o}}, R. 2000, \aap, 363, 476

\bibitem[{{Bordoloi} {et~al.}(2010){Bordoloi}, {Lilly}, \&
  {Amara}}]{Bordoloi_etal2010}
{Bordoloi}, R., {Lilly}, S.~J., \& {Amara}, A. 2010, \mnras, 406, 881

\bibitem[{{Brammer} {et~al.}(2008){Brammer}, {van Dokkum}, \&
  {Coppi}}]{Brammer_etal2008}
{Brammer}, G.~B., {van Dokkum}, P.~G., \& {Coppi}, P. 2008, \apj, 686, 1503

\bibitem[{{Bruzual} \& {Charlot}(2003)}]{BruzualCharlot2003}
{Bruzual}, G. \& {Charlot}, S. 2003, \mnras, 344, 1000

\bibitem[{{Buchs} {et~al.}(2019){Buchs}, {Davis}, {Gruen}, {DeRose}, {Alarcon},
  {Bernstein}, {S{\'a}nchez}, {Myles}, {Roodman}, {Allen}, {Amon}, {Choi},
  {Masters}, {Miquel}, {Troxel}, {Wechsler}, {Abbott}, {Annis}, {Avila},
  {Bechtol}, {Bridle}, {Brooks}, {Buckley-Geer}, {Burke}, {Carnero Rosell},
  {Carrasco Kind}, {Carretero}, {Castander}, {Cawthon}, {D'Andrea}, {da Costa},
  {De Vicente}, {Desai}, {Diehl}, {Doel}, {Drlica-Wagner}, {Eifler}, {Evrard},
  {Flaugher}, {Fosalba}, {Frieman}, {Garc{\'\i}a-Bellido}, {Gaztanaga},
  {Gruendl}, {Gschwend}, {Gutierrez}, {Hartley}, {Hollowood}, {Honscheid},
  {James}, {Kuehn}, {Kuropatkin}, {Lima}, {Lin}, {Maia}, {March}, {Marshall},
  {Melchior}, {Menanteau}, {Ogando}, {Plazas}, {Rykoff}, {Sanchez}, {Scarpine},
  {Serrano}, {Sevilla-Noarbe}, {Smith}, {Soares-Santos}, {Sobreira}, {Suchyta},
  {Swanson}, {Tarle}, {Thomas}, {Vikram}, \& {DES
  Collaboration}}]{Buchs_etal2019}
{Buchs}, R., {Davis}, C., {Gruen}, D., {et~al.} 2019, \mnras, 489, 820

\bibitem[{{Cao} {et~al.}(2018){Cao}, {Gong}, {Meng}, {Xu}, {Chen}, {Guo}, {Li},
  {Liu}, {Xue}, {Cao}, {Fu}, {Zhang}, {Wang}, \& {Zhan}}]{Cao_etal2018}
{Cao}, Y., {Gong}, Y., {Meng}, X.-M., {et~al.} 2018, \mnras, 480, 2178

\bibitem[{{Capak} {et~al.}(2007){Capak}, {Aussel}, {Ajiki}, {McCracken},
  {Mobasher}, {Scoville}, {Shopbell}, {Taniguchi}, {Thompson}, {Tribiano},
  {Sasaki}, {Blain}, {Brusa}, {Carilli}, {Comastri}, {Carollo}, {Cassata},
  {Colbert}, {Ellis}, {Elvis}, {Giavalisco}, {Green}, {Guzzo}, {Hasinger},
  {Ilbert}, {Impey}, {Jahnke}, {Kartaltepe}, {Kneib}, {Koda}, {Koekemoer},
  {Komiyama}, {Leauthaud}, {Le Fevre}, {Lilly}, {Liu}, {Massey}, {Miyazaki},
  {Murayama}, {Nagao}, {Peacock}, {Pickles}, {Porciani}, {Renzini}, {Rhodes},
  {Rich}, {Salvato}, {Sanders}, {Scarlata}, {Schiminovich}, {Schinnerer},
  {Scodeggio}, {Sheth}, {Shioya}, {Tasca}, {Taylor}, {Yan}, \&
  {Zamorani}}]{Capak_etal2007}
{Capak}, P., {Aussel}, H., {Ajiki}, M., {et~al.} 2007, \apjs, 172, 99

\bibitem[{{Carliles} {et~al.}(2010){Carliles}, {Budav{\'a}ri}, {Heinis},
  {Priebe}, \& {Szalay}}]{Carliles_etal2010}
{Carliles}, S., {Budav{\'a}ri}, T., {Heinis}, S., {Priebe}, C., \& {Szalay},
  A.~S. 2010, \apj, 712, 511

\bibitem[{{Carrasco Kind} \& {Brunner}(2013)}]{CarrascoKind_Brunner2013}
{Carrasco Kind}, M. \& {Brunner}, R.~J. 2013, \mnras, 432, 1483

\bibitem[{{Carrasco Kind} \& {Brunner}(2014)}]{KindBrunner_2014}
{Carrasco Kind}, M. \& {Brunner}, R.~J. 2014, \mnras, 438, 3409

\bibitem[{{Chan} {et~al.}(2022){Chan}, {Avila}, {Carnero Rosell}, {Ferrero},
  {Elvin-Poole}, {et~al.}}]{Chan_xip2022}
{Chan}, K.~C., {Avila}, S., {Carnero Rosell}, A., {et~al.} 2022, \prd, 106,
  123502

\bibitem[{{Collister} \& {Lahav}(2004)}]{CollisterLahav_2004}
{Collister}, A.~A. \& {Lahav}, O. 2004, \pasp, 116, 345

\bibitem[{{Crenshaw} {et~al.}(2024){Crenshaw}, {Kalmbach}, {Gagliano}, {Yan},
  {Connolly}, {Malz}, {Schmidt}, \& {The LSST Dark Energy Science
  Collaboration}}]{Crenshaw_etal2024}
{Crenshaw}, J.~F., {Kalmbach}, J.~B., {Gagliano}, A., {et~al.} 2024, \aj, 168,
  80

\bibitem[{{CSST Collaboration} {et~al.}(2025){CSST Collaboration}, {Gong},
  {Miao}, {Zhan}, {Li}, {Shangguan}, {Li}, {Liu}, {Chen}, {Yuan}, {Zhou},
  {Liu}, {Yu}, {Ji}, {Qi}, {Liu}, {Dai}, {Wang}, {Zheng}, {Hao}, {Dou}, {Ao},
  {Lin}, {Zhang}, {Wang}, {Sun}, {Li}, {Li}, {Xu}, {Li}, {Li}, {Wu}, {Zhang},
  {Wang}, {Bai}, {Cai}, {Cai}, {Chan}, {Chang}, {Chen}, {Chen}, {Chen}, {Chen},
  {Cui}, {Du}, {Duan}, {Fan}, {Fan}, {Fan}, {Fan}, {Fang}, {Fu}, {Fu}, {Fu},
  {Gao}, {Gu}, {Gu}, {Guo}, {Han}, {Huang}, {Ho}, {Jiang}, {Jing}, {Kang},
  {Kong}, {Li}, {Li}, {Li}, {Li}, {Li}, {Liao}, {Lin}, {Liu}, {Liu}, {Liu},
  {Mao}, {Mao}, {Meng}, {Pang}, {Peng}, {Peng}, {Shan}, {Shen}, {Shen}, {Shen},
  {Shi}, {Shi}, {Tan}, {Tian}, {Wang}, {Wang}, {Wang}, {Wang}, {Wu}, {Wu},
  {Wu}, {Xu}, {Xue}, {Xue}, {Yang}, {Yang}, {Yao}, {Yuan}, {Yuan}, {Zhang},
  {Zhang}, {Zhang}, {Zhao}, {Zhao}, {Zhong}, {Zhong}, {Zhou}, \&
  {Zu}}]{CSST_intro2025}
{CSST Collaboration}, {Gong}, Y., {Miao}, H., {et~al.} 2025, arXiv e-prints,
  arXiv:2507.04618

\bibitem[{{Dalal} {et~al.}(2023){Dalal}, {Li}, {Nicola}, {Zuntz}, {Strauss},
  {Sugiyama}, {Zhang}, {Rau}, {Mandelbaum}, {Takada}, {More}, {Miyatake},
  {Kannawadi}, {Shirasaki}, {Taniguchi}, {Takahashi}, {Osato}, {Hamana},
  {Oguri}, {Nishizawa}, {Malag{\'o}n}, {Sunayama}, {Alonso}, {Slosar}, {Luo},
  {Armstrong}, {Bosch}, {Hsieh}, {Komiyama}, {Lupton}, {Lust}, {MacArthur},
  {Miyazaki}, {Murayama}, {Nishimichi}, {Okura}, {Price}, {Tait}, {Tanaka}, \&
  {Wang}}]{Dalal_etal2023}
{Dalal}, R., {Li}, X., {Nicola}, A., {et~al.} 2023, \prd, 108, 123519

\bibitem[{De~Vicente {et~al.}(2016)De~Vicente, S\'anchez, \&
  Sevilla-Noarbe}]{DeVicente:2015kyp}
De~Vicente, J., S\'anchez, E., \& Sevilla-Noarbe, I. 2016, Mon. Not. Roy.
  Astron. Soc., 459, 3078

\bibitem[{{Dey} {et~al.}(2019){Dey}, {Schlegel}, {Lang}, {Blum}, {Burleigh},
  {Fan}, {Findlay}, {Finkbeiner}, {Herrera}, {Juneau}, {Landriau}, {Levi},
  {McGreer}, {Meisner}, {Myers}, {Moustakas}, {Nugent}, {Patej}, {Schlafly},
  {Walker}, {Valdes}, {Weaver}, {Y{\`e}che}, {Zou}, {Zhou}, {Abareshi},
  {Abbott}, {Abolfathi}, {Aguilera}, {Alam}, {Allen}, {Alvarez}, {Annis},
  {Ansarinejad}, {Aubert}, {Beechert}, {Bell}, {BenZvi}, {Beutler}, {Bielby},
  {Bolton}, {Brice{\~n}o}, {Buckley-Geer}, {Butler}, {Calamida}, {Carlberg},
  {Carter}, {Casas}, {Castander}, {Choi}, {Comparat}, {Cukanovaite}, {Delubac},
  {DeVries}, {Dey}, {Dhungana}, {Dickinson}, {Ding}, {Donaldson}, {Duan},
  {Duckworth}, {Eftekharzadeh}, {Eisenstein}, {Etourneau}, {Fagrelius},
  {Farihi}, {Fitzpatrick}, {Font-Ribera}, {Fulmer}, {G{\"a}nsicke},
  {Gaztanaga}, {George}, {Gerdes}, {Gontcho}, {Gorgoni}, {Green}, {Guy},
  {Harmer}, {Hernandez}, {Honscheid}, {Huang}, {James}, {Jannuzi}, {Jiang},
  {Joyce}, {Karcher}, {Karkar}, {Kehoe}, {Kneib}, {Kueter-Young}, {Lan},
  {Lauer}, {Le Guillou}, {Le Van Suu}, {Lee}, {Lesser}, {Perreault Levasseur},
  {Li}, {Mann}, {Marshall}, {Mart{\'\i}nez-V{\'a}zquez}, {Martini}, {du Mas des
  Bourboux}, {McManus}, {Meier}, {M{\'e}nard}, {Metcalfe},
  {Mu{\~n}oz-Guti{\'e}rrez}, {Najita}, {Napier}, {Narayan}, {Newman}, {Nie},
  {Nord}, {Norman}, {Olsen}, {Paat}, {Palanque-Delabrouille}, {Peng},
  {Poppett}, {Poremba}, {Prakash}, {Rabinowitz}, {Raichoor}, {Rezaie},
  {Robertson}, {Roe}, {Ross}, {Ross}, {Rudnick}, {Safonova}, {Saha},
  {S{\'a}nchez}, {Savary}, {Schweiker}, {Scott}, {Seo}, {Shan}, {Silva},
  {Slepian}, {Soto}, {Sprayberry}, {Staten}, {Stillman}, {Stupak}, {Summers},
  {Sien Tie}, {Tirado}, {Vargas-Maga{\~n}a}, {Vivas}, {Wechsler}, {Williams},
  {Yang}, {Yang}, {Yapici}, {Zaritsky}, {Zenteno}, {Zhang}, {Zhang}, {Zhou}, \&
  {Zhou}}]{Dey_etal2019}
{Dey}, A., {Schlegel}, D.~J., {Lang}, D., {et~al.} 2019, \aj, 157, 168

\bibitem[{Dinh {et~al.}(2016)Dinh, Sohl-Dickstein, \& Bengio}]{Dinh_etal2016}
Dinh, L., Sohl-Dickstein, J., \& Bengio, S. 2016, arXiv preprint
  arXiv:1605.08803

\bibitem[{{D'Isanto} \& {Polsterer}(2018)}]{DIsantoPolsterer2018}
{D'Isanto}, A. \& {Polsterer}, K.~L. 2018, \aap, 609, A111

\bibitem[{Dolatabadi {et~al.}(2020)Dolatabadi, Erfani, \&
  Leckie}]{Dolatabadi_etal2020}
Dolatabadi, H.~M., Erfani, S., \& Leckie, C. 2020, in International Conference
  on Artificial Intelligence and Statistics, PMLR, 4236--4246

\bibitem[{{Drlica-Wagner} {et~al.}(2018){Drlica-Wagner}, {Sevilla-Noarbe},
  {Rykoff}, {Gruendl}, {Yanny}, {Tucker}, {Hoyle}, {Carnero Rosell},
  {Bernstein}, {Bechtol}, {Becker}, {Benoit-L{\'e}vy}, {Bertin}, {Carrasco
  Kind}, {Davis}, {de Vicente}, {Diehl}, {Gruen}, {Hartley}, {Leistedt}, {Li},
  {Marshall}, {Neilsen}, {Rau}, {Sheldon}, {Smith}, {Troxel}, {Wyatt}, {Zhang},
  {Abbott}, {Abdalla}, {Allam}, {Banerji}, {Brooks}, {Buckley-Geer}, {Burke},
  {Capozzi}, {Carretero}, {Cunha}, {D'Andrea}, {da Costa}, {DePoy}, {Desai},
  {Dietrich}, {Doel}, {Evrard}, {Fausti Neto}, {Flaugher}, {Fosalba},
  {Frieman}, {Garc{\'\i}a-Bellido}, {Gerdes}, {Giannantonio}, {Gschwend},
  {Gutierrez}, {Honscheid}, {James}, {Jeltema}, {Kuehn}, {Kuhlmann},
  {Kuropatkin}, {Lahav}, {Lima}, {Lin}, {Maia}, {Martini}, {McMahon},
  {Melchior}, {Menanteau}, {Miquel}, {Nichol}, {Ogando}, {Plazas}, {Romer},
  {Roodman}, {Sanchez}, {Scarpine}, {Schindler}, {Schubnell}, {Smith}, {Smith},
  {Soares-Santos}, {Sobreira}, {Suchyta}, {Tarle}, {Vikram}, {Walker},
  {Wechsler}, {Zuntz}, \& {DES Collaboration}}]{Drlica-Wagner_etal2018}
{Drlica-Wagner}, A., {Sevilla-Noarbe}, I., {Rykoff}, E.~S., {et~al.} 2018,
  \apjs, 235, 33

\bibitem[{Durkan {et~al.}(2019)Durkan, Bekasov, Murray, \&
  Papamakarios}]{Durkan_etal2019}
Durkan, C., Bekasov, A., Murray, I., \& Papamakarios, G. 2019, Advances in
  neural information processing systems, 32

\bibitem[{{Eifler} {et~al.}(2021){Eifler}, {Miyatake}, {Krause}, {Heinrich},
  {Miranda}, {Hirata}, {Xu}, {Hemmati}, {Simet}, {Capak}, {Choi}, {Dor{\'e}},
  {Doux}, {Fang}, {Hounsell}, {Huff}, {Huang}, {Jarvis}, {Kruk}, {Masters},
  {Rozo}, {Scolnic}, {Spergel}, {Troxel}, {von der Linden}, {Wang}, {Weinberg},
  {Wenzl}, \& {Wu}}]{WFIRST_2021}
{Eifler}, T., {Miyatake}, H., {Krause}, E., {et~al.} 2021, \mnras, 507, 1746

\bibitem[{{Garcia-Fernandez}(2025)}]{Garcia-Fernandez2025}
{Garcia-Fernandez}, M. 2025, arXiv e-prints, arXiv:2501.06532

\bibitem[{{Gerdes} {et~al.}(2010){Gerdes}, {Sypniewski}, {McKay}, {Hao},
  {Weis}, {Wechsler}, \& {Busha}}]{Gerdes_etal2010}
{Gerdes}, D.~W., {Sypniewski}, A.~J., {McKay}, T.~A., {et~al.} 2010, \apj, 715,
  823

\bibitem[{{Gong} {et~al.}(2019){Gong}, {Liu}, {Cao}, {Chen}, {Fan}, {Li}, {Li},
  {Li}, {Zhang}, \& {Zhan}}]{Gong_etal2019}
{Gong}, Y., {Liu}, X., {Cao}, Y., {et~al.} 2019, \apj, 883, 203

\bibitem[{Goodfellow {et~al.}(2014)Goodfellow, Pouget-Abadie, Mirza, Xu,
  Warde-Farley, Ozair, Courville, \& Bengio}]{Goodfellow_etal2014}
Goodfellow, I.~J., Pouget-Abadie, J., Mirza, M., {et~al.} 2014, Advances in
  neural information processing systems, 27

\bibitem[{{Haldemann} {et~al.}(2023){Haldemann}, {Ksoll}, {Walter}, {Alibert},
  {Klessen}, {Benz}, {Koethe}, {Ardizzone}, \& {Rother}}]{Haldemann_etal2023}
{Haldemann}, J., {Ksoll}, V., {Walter}, D., {et~al.} 2023, \aap, 672, A180

\bibitem[{{Heymans} {et~al.}(2013){Heymans}, {Grocutt}, {Heavens}, {Kilbinger},
  {Kitching}, {Simpson}, {Benjamin}, {Erben}, {Hildebrandt}, {Hoekstra},
  {Mellier}, {Miller}, {Van Waerbeke}, {Brown}, {Coupon}, {Fu},
  {Harnois-D{\'e}raps}, {Hudson}, {Kuijken}, {Rowe}, {Schrabback}, {Semboloni},
  {Vafaei}, \& {Velander}}]{Heymans_etal2013}
{Heymans}, C., {Grocutt}, E., {Heavens}, A., {et~al.} 2013, \mnras, 432, 2433

\bibitem[{{Heymans} {et~al.}(2021){Heymans}, {Tr{\"o}ster}, {Asgari}, {Blake},
  {Hildebrandt}, {Joachimi}, {Kuijken}, {Lin}, {S{\'a}nchez}, {van den Busch},
  {Wright}, {Amon}, {Bilicki}, {de Jong}, {Crocce}, {Dvornik}, {Erben},
  {Fortuna}, {Getman}, {Giblin}, {Glazebrook}, {Hoekstra}, {Joudaki},
  {Kannawadi}, {K{\"o}hlinger}, {Lidman}, {Miller}, {Napolitano}, {Parkinson},
  {Schneider}, {Shan}, {Valentijn}, {Verdoes Kleijn}, \&
  {Wolf}}]{Heymans_etal2021}
{Heymans}, C., {Tr{\"o}ster}, T., {Asgari}, M., {et~al.} 2021, \aap, 646, A140

\bibitem[{{Hikage} {et~al.}(2019){Hikage}, {Oguri}, {Hamana}, {More},
  {Mandelbaum}, {Takada}, {K{\"o}hlinger}, {Miyatake}, {Nishizawa}, {Aihara},
  {Armstrong}, {Bosch}, {Coupon}, {Ducout}, {Ho}, {Hsieh}, {Komiyama},
  {Lanusse}, {Leauthaud}, {Lupton}, {Medezinski}, {Mineo}, {Miyama},
  {Miyazaki}, {Murata}, {Murayama}, {Shirasaki}, {Sif{\'o}n}, {Simet},
  {Speagle}, {Spergel}, {Strauss}, {Sugiyama}, {Tanaka}, {Utsumi}, {Wang}, \&
  {Yamada}}]{Hikage_etal2019}
{Hikage}, C., {Oguri}, M., {Hamana}, T., {et~al.} 2019, \pasj, 71, 43

\bibitem[{{Hildebrandt} {et~al.}(2017){Hildebrandt}, {Viola}, {Heymans},
  {Joudaki}, {Kuijken}, {Blake}, {Erben}, {Joachimi}, {Klaes}, {Miller},
  {Morrison}, {Nakajima}, {Verdoes Kleijn}, {Amon}, {Choi}, {Covone}, {de
  Jong}, {Dvornik}, {Fenech Conti}, {Grado}, {Harnois-D{\'e}raps}, {Herbonnet},
  {Hoekstra}, {K{\"o}hlinger}, {McFarland}, {Mead}, {Merten}, {Napolitano},
  {Peacock}, {Radovich}, {Schneider}, {Simon}, {Valentijn}, {van den Busch},
  {van Uitert}, \& {Van Waerbeke}}]{Hildebrandt_etal2017}
{Hildebrandt}, H., {Viola}, M., {Heymans}, C., {et~al.} 2017, \mnras, 465, 1454

\bibitem[{Hocker {et~al.}(2007)}]{Hoecker:2007gu}
Hocker, A. {et~al.} 2007, arXiv: physics/070303 [\eprint{physics/0703039}]

\bibitem[{{Hoyle}(2016)}]{Hoyle_2016}
{Hoyle}, B. 2016, Astronomy and Computing, 16, 34

\bibitem[{{Ilbert} {et~al.}(2006){Ilbert}, {Arnouts}, {McCracken},
  {Bolzonella}, {Bertin}, {Le F{\`e}vre}, {Mellier}, {Zamorani}, {Pell{\`o}},
  {Iovino}, {Tresse}, {Le Brun}, {Bottini}, {Garilli}, {Maccagni}, {Picat},
  {Scaramella}, {Scodeggio}, {Vettolani}, {Zanichelli}, {Adami}, {Bardelli},
  {Cappi}, {Charlot}, {Ciliegi}, {Contini}, {Cucciati}, {Foucaud}, {Franzetti},
  {Gavignaud}, {Guzzo}, {Marano}, {Marinoni}, {Mazure}, {Meneux}, {Merighi},
  {Paltani}, {Pollo}, {Pozzetti}, {Radovich}, {Zucca}, {Bondi}, {Bongiorno},
  {Busarello}, {de La Torre}, {Gregorini}, {Lamareille}, {Mathez}, {Merluzzi},
  {Ripepi}, {Rizzo}, \& {Vergani}}]{Ilbert_etal2006}
{Ilbert}, O., {Arnouts}, S., {McCracken}, H.~J., {et~al.} 2006, \aap, 457, 841

\bibitem[{{Ilbert} {et~al.}(2009){Ilbert}, {Capak}, {Salvato}, {Aussel},
  {McCracken}, {Sanders}, {Scoville}, {Kartaltepe}, {Arnouts}, {Le Floc'h},
  {Mobasher}, {Taniguchi}, {Lamareille}, {Leauthaud}, {Sasaki}, {Thompson},
  {Zamojski}, {Zamorani}, {Bardelli}, {Bolzonella}, {Bongiorno}, {Brusa},
  {Caputi}, {Carollo}, {Contini}, {Cook}, {Coppa}, {Cucciati}, {de la Torre},
  {de Ravel}, {Franzetti}, {Garilli}, {Hasinger}, {Iovino}, {Kampczyk},
  {Kneib}, {Knobel}, {Kovac}, {Le Borgne}, {Le Brun}, {Le F{\`e}vre}, {Lilly},
  {Looper}, {Maier}, {Mainieri}, {Mellier}, {Mignoli}, {Murayama}, {Pell{\`o}},
  {Peng}, {P{\'e}rez-Montero}, {Renzini}, {Ricciardelli}, {Schiminovich},
  {Scodeggio}, {Shioya}, {Silverman}, {Surace}, {Tanaka}, {Tasca}, {Tresse},
  {Vergani}, \& {Zucca}}]{Ilbert_etal2009}
{Ilbert}, O., {Capak}, P., {Salvato}, M., {et~al.} 2009, \apj, 690, 1236

\bibitem[{{Ivezi{\'c}} {et~al.}(2019){Ivezi{\'c}}, {Kahn}, {Tyson}, {Abel},
  {Acosta}, {Allsman}, {Alonso}, {AlSayyad}, {Anderson}, {Andrew}, {Angel},
  {Angeli}, {Ansari}, {Antilogus}, {Araujo}, {Armstrong}, {Arndt}, {Astier},
  {Aubourg}, {Auza}, {Axelrod}, {Bard}, {Barr}, {Barrau}, {Bartlett}, {Bauer},
  {Bauman}, {Baumont}, {Bechtol}, {Bechtol}, {Becker}, {Becla}, {Beldica},
  {Bellavia}, {Bianco}, {Biswas}, {Blanc}, {Blazek}, {Blandford}, {Bloom},
  {Bogart}, {Bond}, {Booth}, {Borgland}, {Borne}, {Bosch}, {Boutigny},
  {Brackett}, {Bradshaw}, {Brandt}, {Brown}, {Bullock}, {Burchat}, {Burke},
  {Cagnoli}, {Calabrese}, {Callahan}, {Callen}, {Carlin}, {Carlson},
  {Chandrasekharan}, {Charles-Emerson}, {Chesley}, {Cheu}, {Chiang}, {Chiang},
  {Chirino}, {Chow}, {Ciardi}, {Claver}, {Cohen-Tanugi}, {Cockrum}, {Coles},
  {Connolly}, {Cook}, {Cooray}, {Covey}, {Cribbs}, {Cui}, {Cutri}, {Daly},
  {Daniel}, {Daruich}, {Daubard}, {Daues}, {Dawson}, {Delgado}, {Dellapenna},
  {de Peyster}, {de Val-Borro}, {Digel}, {Doherty}, {Dubois},
  {Dubois-Felsmann}, {Durech}, {Economou}, {Eifler}, {Eracleous}, {Emmons},
  {Fausti Neto}, {Ferguson}, {Figueroa}, {Fisher-Levine}, {Focke}, {Foss},
  {Frank}, {Freemon}, {Gangler}, {Gawiser}, {Geary}, {Gee}, {Geha}, {Gessner},
  {Gibson}, {Gilmore}, {Glanzman}, {Glick}, {Goldina}, {Goldstein}, {Goodenow},
  {Graham}, {Gressler}, {Gris}, {Guy}, {Guyonnet}, {Haller}, {Harris},
  {Hascall}, {Haupt}, {Hernandez}, {Herrmann}, {Hileman}, {Hoblitt}, {Hodgson},
  {Hogan}, {Howard}, {Huang}, {Huffer}, {Ingraham}, {Innes}, {Jacoby}, {Jain},
  {Jammes}, {Jee}, {Jenness}, {Jernigan}, {Jevremovi{\'c}}, {Johns}, {Johnson},
  {Johnson}, {Jones}, {Juramy-Gilles}, {Juri{\'c}}, {Kalirai}, {Kallivayalil},
  {Kalmbach}, {Kantor}, {Karst}, {Kasliwal}, {Kelly}, {Kessler}, {Kinnison},
  {Kirkby}, {Knox}, {Kotov}, {Krabbendam}, {Krughoff}, {Kub{\'a}nek},
  {Kuczewski}, {Kulkarni}, {Ku}, {Kurita}, {Lage}, {Lambert}, {Lange},
  {Langton}, {Le Guillou}, {Levine}, {Liang}, {Lim}, {Lintott}, {Long},
  {Lopez}, {Lotz}, {Lupton}, {Lust}, {MacArthur}, {Mahabal}, {Mandelbaum},
  {Markiewicz}, {Marsh}, {Marshall}, {Marshall}, {May}, {McKercher}, {McQueen},
  {Meyers}, {Migliore}, {Miller}, {Mills}, {Miraval}, {Moeyens}, {Moolekamp},
  {Monet}, {Moniez}, {Monkewitz}, {Montgomery}, {Morrison}, {Mueller},
  {Muller}, {Mu{\~n}oz Arancibia}, {Neill}, {Newbry}, {Nief}, {Nomerotski},
  {Nordby}, {O'Connor}, {Oliver}, {Olivier}, {Olsen}, {O'Mullane}, {Ortiz},
  {Osier}, {Owen}, {Pain}, {Palecek}, {Parejko}, {Parsons}, {Pease},
  {Peterson}, {Peterson}, {Petravick}, {Libby Petrick}, {Petry},
  {Pierfederici}, {Pietrowicz}, {Pike}, {Pinto}, {Plante}, {Plate}, {Plutchak},
  {Price}, {Prouza}, {Radeka}, {Rajagopal}, {Rasmussen}, {Regnault}, {Reil},
  {Reiss}, {Reuter}, {Ridgway}, {Riot}, {Ritz}, {Robinson}, {Roby}, {Roodman},
  {Rosing}, {Roucelle}, {Rumore}, {Russo}, {Saha}, {Sassolas}, {Schalk},
  {Schellart}, {Schindler}, {Schmidt}, {Schneider}, {Schneider}, {Schoening},
  {Schumacher}, {Schwamb}, {Sebag}, {Selvy}, {Sembroski}, {Seppala}, {Serio},
  {Serrano}, {Shaw}, {Shipsey}, {Sick}, {Silvestri}, {Slater}, {Smith},
  {Smith}, {Sobhani}, {Soldahl}, {Storrie-Lombardi}, {Stover}, {Strauss},
  {Street}, {Stubbs}, {Sullivan}, {Sweeney}, {Swinbank}, {Szalay}, {Takacs},
  {Tether}, {Thaler}, {Thayer}, {Thomas}, {Thornton}, {Thukral}, {Tice},
  {Trilling}, {Turri}, {Van Berg}, {Vanden Berk}, {Vetter}, {Virieux},
  {Vucina}, {Wahl}, {Walkowicz}, {Walsh}, {Walter}, {Wang}, {Wang}, {Warner},
  {Wiecha}, {Willman}, {Winters}, {Wittman}, {Wolff}, {Wood-Vasey}, {Wu},
  {Xin}, {Yoachim}, \& {Zhan}}]{LSST_2019}
{Ivezi{\'c}}, {\v{Z}}., {Kahn}, S.~M., {Tyson}, J.~A., {et~al.} 2019, \apj,
  873, 111

\bibitem[{{Kang} {et~al.}(2022){Kang}, {Pellegrini}, {Ardizzone}, {Klessen},
  {Koethe}, {Glover}, \& {Ksoll}}]{Kang_etal2022}
{Kang}, D.~E., {Pellegrini}, E.~W., {Ardizzone}, L., {et~al.} 2022, \mnras,
  512, 617

\bibitem[{{Khostovan} {et~al.}(2025){Khostovan}, {Kartaltepe}, {Salvato},
  {Ilbert}, {Casey}, {Algera}, {Antwi-Danso}, {Battisti}, {Brinch}, {Brusa},
  {Calabro}, {Capak}, {Chartab}, {Cooper}, {Cox}, {Darvish}, {Drakos},
  {Faisst}, {George}, {Gozaliasl}, {Harish}, {Hasinger}, {Hatamnia}, {Iovino},
  {Jin}, {Kashino}, {Koekemoer}, {Laishram}, {Lee}, {Lertprasertpong}, {Lilly},
  {Masters}, {Mobasher}, {Nagao}, {Onodera}, {Peng}, {Sanders}, {Sanders},
  {Sattari}, {Scoville}, {Shah}, {Silverman}, {Suzuki}, {Tanaka}, {Toft},
  {Trakhtenbrot}, {Trump}, {Vaccari}, {Valentino}, {Vanderhoof}, {Weaver},
  {Yun}, \& {Zavala}}]{Khostovan_etal2025}
{Khostovan}, A.~A., {Kartaltepe}, J.~S., {Salvato}, M., {et~al.} 2025, arXiv
  e-prints, arXiv:2503.00120

\bibitem[{Kingma \& Ba(2015)}]{KingmaBa2015}
Kingma, D.~P. \& Ba, J. 2015, arXiv preprint arXiv:1412.6980

\bibitem[{Kingma \& Dhariwal(2018)}]{KingmaDhariwal_2018}
Kingma, D.~P. \& Dhariwal, P. 2018, Advances in neural information processing
  systems, 31

\bibitem[{Kobyzev {et~al.}(2020)Kobyzev, Prince, \&
  Brubaker}]{Kobyzev_etal2019}
Kobyzev, I., Prince, S.~J., \& Brubaker, M.~A. 2020, IEEE transactions on
  pattern analysis and machine intelligence, 43, 3964

\bibitem[{{Ksoll} {et~al.}(2020){Ksoll}, {Ardizzone}, {Klessen}, {Koethe},
  {Sabbi}, {Robberto}, {Gouliermis}, {Rother}, {Zeidler}, \&
  {Gennaro}}]{Ksoll_etal2020}
{Ksoll}, V.~F., {Ardizzone}, L., {Klessen}, R., {et~al.} 2020, \mnras, 499,
  5447

\bibitem[{{Laigle} {et~al.}(2016){Laigle}, {McCracken}, {Ilbert}, {Hsieh},
  {Davidzon}, {Capak}, {Hasinger}, {Silverman}, {Pichon}, {Coupon}, {Aussel},
  {Le Borgne}, {Caputi}, {Cassata}, {Chang}, {Civano}, {Dunlop}, {Fynbo},
  {Kartaltepe}, {Koekemoer}, {Le F{\`e}vre}, {Le Floc'h}, {Leauthaud}, {Lilly},
  {Lin}, {Marchesi}, {Milvang-Jensen}, {Salvato}, {Sanders}, {Scoville},
  {Smolcic}, {Stockmann}, {Taniguchi}, {Tasca}, {Toft}, {Vaccari}, \&
  {Zabl}}]{Laigle_etal2016}
{Laigle}, C., {McCracken}, H.~J., {Ilbert}, O., {et~al.} 2016, \apjs, 224, 24

\bibitem[{{Lang} {et~al.}(2016){Lang}, {Hogg}, \& {Mykytyn}}]{Lang_etal2016}
{Lang}, D., {Hogg}, D.~W., \& {Mykytyn}, D. 2016, {The Tractor: Probabilistic
  astronomical source detection and measurement}, Astrophysics Source Code
  Library, record ascl:1604.008

\bibitem[{{Laureijs} {et~al.}(2011){Laureijs}, {Amiaux}, {Arduini},
  {Augu{\`e}res}, {Brinchmann}, {Cole}, {Cropper}, {Dabin}, {Duvet}, {Ealet},
  {Garilli}, {Gondoin}, {Guzzo}, {Hoar}, {Hoekstra}, {Holmes}, {Kitching},
  {Maciaszek}, {Mellier}, {Pasian}, {Percival}, {Rhodes}, {Saavedra Criado},
  {Sauvage}, {Scaramella}, {Valenziano}, {Warren}, {Bender}, {Castander},
  {Cimatti}, {Le F{\`e}vre}, {Kurki-Suonio}, {Levi}, {Lilje}, {Meylan},
  {Nichol}, {Pedersen}, {Popa}, {Rebolo Lopez}, {Rix}, {Rottgering},
  {Zeilinger}, {Grupp}, {Hudelot}, {Massey}, {Meneghetti}, {Miller}, {Paltani},
  {Paulin-Henriksson}, {Pires}, {Saxton}, {Schrabback}, {Seidel}, {Walsh},
  {Aghanim}, {Amendola}, {Bartlett}, {Baccigalupi}, {Beaulieu}, {Benabed},
  {Cuby}, {Elbaz}, {Fosalba}, {Gavazzi}, {Helmi}, {Hook}, {Irwin}, {Kneib},
  {Kunz}, {Mannucci}, {Moscardini}, {Tao}, {Teyssier}, {Weller}, {Zamorani},
  {Zapatero Osorio}, {Boulade}, {Foumond}, {Di Giorgio}, {Guttridge}, {James},
  {Kemp}, {Martignac}, {Spencer}, {Walton}, {Bl{\"u}mchen}, {Bonoli},
  {Bortoletto}, {Cerna}, {Corcione}, {Fabron}, {Jahnke}, {Ligori}, {Madrid},
  {Martin}, {Morgante}, {Pamplona}, {Prieto}, {Riva}, {Toledo}, {Trifoglio},
  {Zerbi}, {Abdalla}, {Douspis}, {Grenet}, {Borgani}, {Bouwens}, {Courbin},
  {Delouis}, {Dubath}, {Fontana}, {Frailis}, {Grazian}, {Koppenh{\"o}fer},
  {Mansutti}, {Melchior}, {Mignoli}, {Mohr}, {Neissner}, {Noddle}, {Poncet},
  {Scodeggio}, {Serrano}, {Shane}, {Starck}, {Surace}, {Taylor},
  {Verdoes-Kleijn}, {Vuerli}, {Williams}, {Zacchei}, {Altieri}, {Escudero
  Sanz}, {Kohley}, {Oosterbroek}, {Astier}, {Bacon}, {Bardelli}, {Baugh},
  {Bellagamba}, {Benoist}, {Bianchi}, {Biviano}, {Branchini}, {Carbone},
  {Cardone}, {Clements}, {Colombi}, {Conselice}, {Cresci}, {Deacon}, {Dunlop},
  {Fedeli}, {Fontanot}, {Franzetti}, {Giocoli}, {Garcia-Bellido}, {Gow},
  {Heavens}, {Hewett}, {Heymans}, {Holland}, {Huang}, {Ilbert}, {Joachimi},
  {Jennins}, {Kerins}, {Kiessling}, {Kirk}, {Kotak}, {Krause}, {Lahav}, {van
  Leeuwen}, {Lesgourgues}, {Lombardi}, {Magliocchetti}, {Maguire}, {Majerotto},
  {Maoli}, {Marulli}, {Maurogordato}, {McCracken}, {McLure}, {Melchiorri},
  {Merson}, {Moresco}, {Nonino}, {Norberg}, {Peacock}, {Pello}, {Penny},
  {Pettorino}, {Di Porto}, {Pozzetti}, {Quercellini}, {Radovich}, {Rassat},
  {Roche}, {Ronayette}, {Rossetti}, {Sartoris}, {Schneider}, {Semboloni},
  {Serjeant}, {Simpson}, {Skordis}, {Smadja}, {Smartt}, {Spano}, {Spiro},
  {Sullivan}, {Tilquin}, {Trotta}, {Verde}, {Wang}, {Williger}, {Zhao},
  {Zoubian}, \& {Zucca}}]{Euclid_2011}
{Laureijs}, R., {Amiaux}, J., {Arduini}, S., {et~al.} 2011, arXiv e-prints,
  arXiv:1110.3193

\bibitem[{{Li} {et~al.}(2022){Li}, {Napolitano}, {Roy}, {Tortora}, {La
  Barbera}, {Sonnenfeld}, {Qiu}, \& {Liu}}]{LiNapolitano_etal2022}
{Li}, R., {Napolitano}, N.~R., {Roy}, N., {et~al.} 2022, \apj, 929, 152

\bibitem[{{Li} {et~al.}(2023){Li}, {Zhang}, {Sugiyama}, {Dalal}, {Terasawa},
  {Rau}, {Mandelbaum}, {Takada}, {More}, {Strauss}, {Miyatake}, {Shirasaki},
  {Hamana}, {Oguri}, {Luo}, {Nishizawa}, {Takahashi}, {Nicola}, {Osato},
  {Kannawadi}, {Sunayama}, {Armstrong}, {Bosch}, {Komiyama}, {Lupton}, {Lust},
  {MacArthur}, {Miyazaki}, {Murayama}, {Nishimichi}, {Okura}, {Price}, {Tait},
  {Tanaka}, \& {Wang}}]{Li_etal2023}
{Li}, X., {Zhang}, T., {Sugiyama}, S., {et~al.} 2023, \prd, 108, 123518

\bibitem[{{Lima} {et~al.}(2008){Lima}, {Cunha}, {Oyaizu}, {Frieman}, {Lin}, \&
  {Sheldon}}]{Lima_etal2008}
{Lima}, M., {Cunha}, C.~E., {Oyaizu}, H., {et~al.} 2008, \mnras, 390, 118

\bibitem[{{Lu} {et~al.}(2024){Lu}, {Luo}, {Chen}, {Fu}, {Du}, {Gong}, {Li},
  {Meng}, {Tang}, {Zhang}, {Shu}, {Zhou}, \& {Fan}}]{Lu_etal2024}
{Lu}, J., {Luo}, Z., {Chen}, Z., {et~al.} 2024, \mnras, 527, 12140

\bibitem[{{Luo} {et~al.}(2024{\natexlab{a}}){Luo}, {Li}, {Lu}, {Chen}, {Fu},
  {Zhang}, {Xiao}, {Du}, {Gong}, {Shu}, {Ma}, {Meng}, {Zhou}, \&
  {Fan}}]{Luo_etal2024}
{Luo}, Z., {Li}, Y., {Lu}, J., {et~al.} 2024{\natexlab{a}}, \mnras, 535, 1844

\bibitem[{{Luo} {et~al.}(2024{\natexlab{b}}){Luo}, {Tang}, {Chen}, {Fu}, {Du},
  {Zhang}, {Gong}, {Shu}, {Lu}, {Li}, {Meng}, {Zhou}, \&
  {Fan}}]{Luo_etal2024_GAIN}
{Luo}, Z., {Tang}, Z., {Chen}, Z., {et~al.} 2024{\natexlab{b}}, \mnras, 531,
  3539

\bibitem[{{Lupton} {et~al.}(1999){Lupton}, {Gunn}, \&
  {Szalay}}]{Lupton_etal1999}
{Lupton}, R.~H., {Gunn}, J.~E., \& {Szalay}, A.~S. 1999, \aj, 118, 1406

\bibitem[{{Mandelbaum} {et~al.}(2018){Mandelbaum}, {Eifler}, {Hlo{\v{z}}ek},
  {Collett}, {Gawiser}, {Scolnic}, {Alonso}, {Awan}, {Biswas}, {Blazek},
  {Burchat}, {Chisari}, {Dell'Antonio}, {Digel}, {Frieman}, {Goldstein},
  {Hook}, {Ivezi{\'c}}, {Kahn}, {Kamath}, {Kirkby}, {Kitching}, {Krause},
  {Leget}, {Marshall}, {Meyers}, {Miyatake}, {Newman}, {Nichol}, {Rykoff},
  {Sanchez}, {Slosar}, {Sullivan}, \& {Troxel}}]{LSST_DESC_2018}
{Mandelbaum}, R., {Eifler}, T., {Hlo{\v{z}}ek}, R., {et~al.} 2018, arXiv
  e-prints, arXiv:1809.01669

\bibitem[{{Masters} {et~al.}(2015){Masters}, {Capak}, {Stern}, {Ilbert},
  {Salvato}, {Schmidt}, {Longo}, {Rhodes}, {Paltani}, {Mobasher}, {Hoekstra},
  {Hildebrandt}, {Coupon}, {Steinhardt}, {Speagle}, {Faisst}, {Kalinich},
  {Brodwin}, {Brescia}, \& {Cavuoti}}]{Masters_etal2015}
{Masters}, D., {Capak}, P., {Stern}, D., {et~al.} 2015, \apj, 813, 53

\bibitem[{{McQuinn} \& {White}(2013)}]{McQuinnWhite_2013}
{McQuinn}, M. \& {White}, M. 2013, \mnras, 433, 2857

\bibitem[{{M{\'e}nard} {et~al.}(2013){M{\'e}nard}, {Scranton}, {Schmidt},
  {Morrison}, {Jeong}, {Budavari}, \& {Rahman}}]{Menard_etal2013}
{M{\'e}nard}, B., {Scranton}, R., {Schmidt}, S., {et~al.} 2013, arXiv e-prints,
  arXiv:1303.4722

\bibitem[{Mirza \& Osindero(2014)}]{MirzaOsindero2014}
Mirza, M. \& Osindero, S. 2014, arXiv preprint arXiv:1411.1784

\bibitem[{{Miyatake} {et~al.}(2023){Miyatake}, {Sugiyama}, {Takada},
  {Nishimichi}, {Li}, {Shirasaki}, {More}, {Kobayashi}, {Nishizawa}, {Rau},
  {Zhang}, {Takahashi}, {Dalal}, {Mandelbaum}, {Strauss}, {Hamana}, {Oguri},
  {Osato}, {Luo}, {Kannawadi}, {Hsieh}, {Armstrong}, {Bosch}, {Komiyama},
  {Lupton}, {Lust}, {MacArthur}, {Miyazaki}, {Murayama}, {Okura}, {Price},
  {Sunayama}, {Tait}, {Tanaka}, \& {Wang}}]{Miyatake_etal2023}
{Miyatake}, H., {Sugiyama}, S., {Takada}, M., {et~al.} 2023, \prd, 108, 123517

\bibitem[{{Moskowitz} {et~al.}(2024){Moskowitz}, {Gawiser}, {Crenshaw},
  {Andrews}, {Malz}, {Schmidt}, \& {LSST Dark Energy Science
  Collaboration}}]{Moskowitz_etal2024}
{Moskowitz}, I., {Gawiser}, E., {Crenshaw}, J.~F., {et~al.} 2024, \apjl, 967,
  L6

\bibitem[{{Newman}(2008)}]{Newman_2008}
{Newman}, J.~A. 2008, \apj, 684, 88

\bibitem[{{Newman} \& {Gruen}(2022)}]{NewmanGruen_2022}
{Newman}, J.~A. \& {Gruen}, D. 2022, \araa, 60, 363

\bibitem[{{Padmanabhan} {et~al.}(2007){Padmanabhan}, {Schlegel}, {Seljak},
  {Makarov}, {Bahcall}, {et~al.}}]{Padmanabhan_etal2007}
{Padmanabhan}, N., {Schlegel}, D.~J., {Seljak}, U., {et~al.} 2007, MNRAS, 378,
  852

\bibitem[{Papamakarios {et~al.}(2021)Papamakarios, Nalisnick, Rezende, Mohamed,
  \& Lakshminarayanan}]{Papamakarios_etal2019}
Papamakarios, G., Nalisnick, E., Rezende, D.~J., Mohamed, S., \&
  Lakshminarayanan, B. 2021, Journal of Machine Learning Research, 22, 1

\bibitem[{{Peng} {et~al.}(2022){Peng}, {Xu}, {Zhang}, {Chen}, \&
  {Yu}}]{Peng_etal2022}
{Peng}, H., {Xu}, H., {Zhang}, L., {Chen}, Z., \& {Yu}, Y. 2022, \mnras, 516,
  6210

\bibitem[{Phan {et~al.}(2019)Phan, Pradhan, \& Jankowiak}]{Pyro_2019composable}
Phan, D., Pradhan, N., \& Jankowiak, M. 2019, arXiv preprint arXiv:1912.11554

\bibitem[{Prince(2023)}]{Prince2023}
Prince, S.~J. 2023, Understanding Deep Learning (MIT Press)

\bibitem[{{Qin} {et~al.}(2025){Qin}, {Zhang}, {Chen}, {Fu}, {Yu}, {Xu}, {Yao},
  {Shi}, \& {Shan}}]{Qin_etal2015}
{Qin}, J., {Zhang}, P., {Chen}, Z., {et~al.} 2025, arXiv e-prints,
  arXiv:2505.16420

\bibitem[{{Ramachandra} {et~al.}(2022){Ramachandra}, {Chaves-Montero},
  {Alarcon}, {Fadikar}, {Habib}, \& {Heitmann}}]{Ramachandra_etal2022}
{Ramachandra}, N., {Chaves-Montero}, J., {Alarcon}, A., {et~al.} 2022, \mnras,
  515, 1927

\bibitem[{Rezende \& Mohamed(2015)}]{RezendeMohamed2015}
Rezende, D. \& Mohamed, S. 2015, in International conference on machine
  learning, PMLR, 1530--1538

\bibitem[{Sadeh {et~al.}(2016)Sadeh, Abdalla, \& Lahav}]{Sadeh:2015lsa}
Sadeh, I., Abdalla, F.~B., \& Lahav, O. 2016, Publ. Astron. Soc. Pac., 128,
  104502

\bibitem[{{Salvato} {et~al.}(2019){Salvato}, {Ilbert}, \&
  {Hoyle}}]{Salvato_etal2019}
{Salvato}, M., {Ilbert}, O., \& {Hoyle}, B. 2019, Nature Astronomy, 3, 212

\bibitem[{{Schmidt} {et~al.}(2013){Schmidt}, {M{\'e}nard}, {Scranton},
  {Morrison}, \& {McBride}}]{Schmidt_etal2013}
{Schmidt}, S.~J., {M{\'e}nard}, B., {Scranton}, R., {Morrison}, C., \&
  {McBride}, C.~K. 2013, \mnras, 431, 3307

\bibitem[{{Schneider} {et~al.}(2006){Schneider}, {Knox}, {Zhan}, \&
  {Connolly}}]{Schneider_etal2006}
{Schneider}, M., {Knox}, L., {Zhan}, H., \& {Connolly}, A. 2006, \apj, 651, 14

\bibitem[{{Schuldt} {et~al.}(2021){Schuldt}, {Suyu}, {Ca{\~n}ameras},
  {Taubenberger}, {Meinhardt}, {Leal-Taix{\'e}}, \& {Hsieh}}]{Schuldt_etal2021}
{Schuldt}, S., {Suyu}, S.~H., {Ca{\~n}ameras}, R., {et~al.} 2021, \aap, 651,
  A55

\bibitem[{{Secco} {et~al.}(2022){Secco}, {Samuroff}, {Krause}, {Jain},
  {Blazek}, {Raveri}, {Campos}, {Amon}, {Chen}, {Doux}, {Choi}, {Gruen},
  {Bernstein}, {Chang}, {DeRose}, {Myles}, {Fert{\'e}}, {Lemos}, {Huterer},
  {Prat}, {Troxel}, {MacCrann}, {Liddle}, {Kacprzak}, {Fang}, {S{\'a}nchez},
  {Pandey}, {Dodelson}, {Chintalapati}, {Hoffmann}, {Alarcon}, {Alves},
  {Andrade-Oliveira}, {Baxter}, {Bechtol}, {Becker}, {Brandao-Souza},
  {Camacho}, {Carnero Rosell}, {Carrasco Kind}, {Cawthon}, {Cordero}, {Crocce},
  {Davis}, {Di Valentino}, {Drlica-Wagner}, {Eckert}, {Eifler}, {Elidaiana},
  {Elsner}, {Elvin-Poole}, {Everett}, {Fosalba}, {Friedrich}, {Gatti},
  {Giannini}, {Gruendl}, {Harrison}, {Hartley}, {Herner}, {Huang}, {Huff},
  {Jarvis}, {Jeffrey}, {Kuropatkin}, {Leget}, {Muir}, {Mccullough}, {Navarro
  Alsina}, {Omori}, {Park}, {Porredon}, {Rollins}, {Roodman}, {Rosenfeld},
  {Ross}, {Rykoff}, {Sanchez}, {Sevilla-Noarbe}, {Sheldon}, {Shin}, {Troja},
  {Tutusaus}, {Varga}, {Weaverdyck}, {Wechsler}, {Yanny}, {Yin}, {Zhang},
  {Zuntz}, {Abbott}, {Aguena}, {Allam}, {Annis}, {Bacon}, {Bertin}, {Bhargava},
  {Bridle}, {Brooks}, {Buckley-Geer}, {Burke}, {Carretero}, {Costanzi}, {da
  Costa}, {De Vicente}, {Diehl}, {Dietrich}, {Doel}, {Ferrero}, {Flaugher},
  {Frieman}, {Garc{\'\i}a-Bellido}, {Gaztanaga}, {Gerdes}, {Giannantonio},
  {Gschwend}, {Gutierrez}, {Hinton}, {Hollowood}, {Honscheid}, {Hoyle},
  {James}, {Jeltema}, {Kuehn}, {Lahav}, {Lima}, {Lin}, {Maia}, {Marshall},
  {Martini}, {Melchior}, {Menanteau}, {Miquel}, {Mohr}, {Morgan}, {Ogando},
  {Palmese}, {Paz-Chinch{\'o}n}, {Petravick}, {Pieres}, {Plazas Malag{\'o}n},
  {Rodriguez-Monroy}, {Romer}, {Sanchez}, {Scarpine}, {Schubnell}, {Scolnic},
  {Serrano}, {Smith}, {Soares-Santos}, {Suchyta}, {Swanson}, {Tarle}, {Thomas},
  {To}, \& {DES Collaboration}}]{Secco_etal2022}
{Secco}, L.~F., {Samuroff}, S., {Krause}, E., {et~al.} 2022, \prd, 105, 023515

\bibitem[{{Seo} {et~al.}(2012){Seo}, {Ho}, {White}, {Cuesta}, {Ross},
  {et~al.}}]{Seo_etal2012}
{Seo}, H.-J., {Ho}, S., {White}, M., {et~al.} 2012, \apj, 761, 13

\bibitem[{{Song} {et~al.}(2024){Song}, {Chan}, {Xu}, \&
  {Zheng}}]{Song_etal2024}
{Song}, R., {Chan}, K.~C., {Xu}, H., \& {Zheng}, W. 2024, \mnras, 530, 881

\bibitem[{{Soo} {et~al.}(2018){Soo}, {Moraes}, {Joachimi}, {Hartley}, {Lahav},
  {Charbonnier}, {Makler}, {Pereira}, {Comparat}, {Erben}, {Leauthaud}, {Shan},
  \& {Van Waerbeke}}]{Soo_etal2018}
{Soo}, J. Y.~H., {Moraes}, B., {Joachimi}, B., {et~al.} 2018, \mnras, 475, 3613

\bibitem[{{Spergel} {et~al.}(2015){Spergel}, {Gehrels}, {Baltay}, {Bennett},
  {Breckinridge}, {Donahue}, {Dressler}, {Gaudi}, {Greene}, {Guyon}, {Hirata},
  {Kalirai}, {Kasdin}, {Macintosh}, {Moos}, {Perlmutter}, {Postman},
  {Rauscher}, {Rhodes}, {Wang}, {Weinberg}, {Benford}, {Hudson}, {Jeong},
  {Mellier}, {Traub}, {Yamada}, {Capak}, {Colbert}, {Masters}, {Penny},
  {Savransky}, {Stern}, {Zimmerman}, {Barry}, {Bartusek}, {Carpenter}, {Cheng},
  {Content}, {Dekens}, {Demers}, {Grady}, {Jackson}, {Kuan}, {Kruk}, {Melton},
  {Nemati}, {Parvin}, {Poberezhskiy}, {Peddie}, {Ruffa}, {Wallace}, {Whipple},
  {Wollack}, \& {Zhao}}]{WFIRST_2015}
{Spergel}, D., {Gehrels}, N., {Baltay}, C., {et~al.} 2015, arXiv e-prints,
  arXiv:1503.03757

\bibitem[{{Sugiyama} {et~al.}(2023){Sugiyama}, {Miyatake}, {More}, {Li},
  {Shirasaki}, {Takada}, {Kobayashi}, {Takahashi}, {Nishimichi}, {Nishizawa},
  {Rau}, {Zhang}, {Dalal}, {Mandelbaum}, {Strauss}, {Hamana}, {Oguri}, {Osato},
  {Kannawadi}, {Hsieh}, {Luo}, {Armstrong}, {Bosch}, {Komiyama}, {Lupton},
  {Lust}, {Miyazaki}, {Murayama}, {Okura}, {Price}, {Tait}, {Tanaka}, \&
  {Wang}}]{Sugiyama_etal2023}
{Sugiyama}, S., {Miyatake}, H., {More}, S., {et~al.} 2023, \prd, 108, 123521

\bibitem[{{Sun} {et~al.}(2023){Sun}, {Speagle}, {Huang}, {Ting}, \&
  {Cai}}]{Sun_etal2023}
{Sun}, Z., {Speagle}, J.~S., {Huang}, S., {Ting}, Y.-S., \& {Cai}, Z. 2023,
  arXiv e-prints, arXiv:2310.20125

\bibitem[{Tabak \& Turner(2013)}]{Tabak:2013cnz}
Tabak, E.~G. \& Turner, C.~V. 2013, Commun. Pure Appl. Math., 66, 145

\bibitem[{{Troxel} {et~al.}(2018){Troxel}, {MacCrann}, {Zuntz}, {Eifler},
  {Krause}, {Dodelson}, {Gruen}, {Blazek}, {Friedrich}, {Samuroff}, {Prat},
  {Secco}, {Davis}, {Fert{\'e}}, {DeRose}, {Alarcon}, {Amara}, {Baxter},
  {Becker}, {Bernstein}, {Bridle}, {Cawthon}, {Chang}, {Choi}, {De Vicente},
  {Drlica-Wagner}, {Elvin-Poole}, {Frieman}, {Gatti}, {Hartley}, {Honscheid},
  {Hoyle}, {Huff}, {Huterer}, {Jain}, {Jarvis}, {Kacprzak}, {Kirk}, {Kokron},
  {Krawiec}, {Lahav}, {Liddle}, {Peacock}, {Rau}, {Refregier}, {Rollins},
  {Rozo}, {Rykoff}, {S{\'a}nchez}, {Sevilla-Noarbe}, {Sheldon}, {Stebbins},
  {Varga}, {Vielzeuf}, {Wang}, {Wechsler}, {Yanny}, {Abbott}, {Abdalla},
  {Allam}, {Annis}, {Bechtol}, {Benoit-L{\'e}vy}, {Bertin}, {Brooks},
  {Buckley-Geer}, {Burke}, {Carnero Rosell}, {Carrasco Kind}, {Carretero},
  {Castander}, {Crocce}, {Cunha}, {D'Andrea}, {da Costa}, {DePoy}, {Desai},
  {Diehl}, {Dietrich}, {Doel}, {Fernandez}, {Flaugher}, {Fosalba},
  {Garc{\'\i}a-Bellido}, {Gaztanaga}, {Gerdes}, {Giannantonio}, {Goldstein},
  {Gruendl}, {Gschwend}, {Gutierrez}, {James}, {Jeltema}, {Johnson}, {Johnson},
  {Kent}, {Kuehn}, {Kuhlmann}, {Kuropatkin}, {Li}, {Lima}, {Lin}, {Maia},
  {March}, {Marshall}, {Martini}, {Melchior}, {Menanteau}, {Miquel}, {Mohr},
  {Neilsen}, {Nichol}, {Nord}, {Petravick}, {Plazas}, {Romer}, {Roodman},
  {Sako}, {Sanchez}, {Scarpine}, {Schindler}, {Schubnell}, {Smith}, {Smith},
  {Soares-Santos}, {Sobreira}, {Suchyta}, {Swanson}, {Tarle}, {Thomas},
  {Tucker}, {Vikram}, {Walker}, {Weller}, {Zhang}, \& {DES
  Collaboration}}]{Troxel_eltal2018}
{Troxel}, M.~A., {MacCrann}, N., {Zuntz}, J., {et~al.} 2018, \prd, 98, 043528

\bibitem[{{van den Busch} {et~al.}(2020){van den Busch}, {Hildebrandt},
  {Wright}, {Morrison}, {Blake}, {Joachimi}, {Erben}, {Heymans}, {Kuijken}, \&
  {Taylor}}]{vandenBusch_etal2020}
{van den Busch}, J.~L., {Hildebrandt}, H., {Wright}, A.~H., {et~al.} 2020,
  \aap, 642, A200

\bibitem[{{Weaver} {et~al.}(2022){Weaver}, {Kauffmann}, {Ilbert}, {McCracken},
  {Moneti}, {Toft}, {Brammer}, {Shuntov}, {Davidzon}, {Hsieh}, {Laigle},
  {Anastasiou}, {Jespersen}, {Vinther}, {Capak}, {Casey}, {McPartland},
  {Milvang-Jensen}, {Mobasher}, {Sanders}, {Zalesky}, {Arnouts}, {Aussel},
  {Dunlop}, {Faisst}, {Franx}, {Furtak}, {Fynbo}, {Gould}, {Greve}, {Gwyn},
  {Kartaltepe}, {Kashino}, {Koekemoer}, {Kokorev}, {Le F{\`e}vre}, {Lilly},
  {Masters}, {Magdis}, {Mehta}, {Peng}, {Riechers}, {Salvato}, {Sawicki},
  {Scarlata}, {Scoville}, {Shirley}, {Silverman}, {Sneppen}, {Smolc̆i{\'c}},
  {Steinhardt}, {Stern}, {Tanaka}, {Taniguchi}, {Teplitz}, {Vaccari}, {Wang},
  \& {Zamorani}}]{Weaver_etal2022}
{Weaver}, J.~R., {Kauffmann}, O.~B., {Ilbert}, O., {et~al.} 2022, \apjs, 258,
  11

\bibitem[{{Wright} {et~al.}(2020){Wright}, {Hildebrandt}, {van den Busch}, \&
  {Heymans}}]{Wright_etal2020}
{Wright}, A.~H., {Hildebrandt}, H., {van den Busch}, J.~L., \& {Heymans}, C.
  2020, \aap, 637, A100

\bibitem[{{Wright} {et~al.}(2025){Wright}, {St{\"o}lzner}, {Asgari}, {Bilicki},
  {Giblin}, {Heymans}, {Hildebrandt}, {Hoekstra}, {Joachimi}, {Kuijken}, {Li},
  {Reischke}, {von Wietersheim-Kramsta}, {Yoon}, {Burger}, {Chisari}, {de
  Jong}, {Dvornik}, {Georgiou}, {Harnois-D{\'e}raps}, {Jalan}, {William},
  {Joudaki}, {Lesci}, {Linke}, {Loureiro}, {Mahony}, {Maturi}, {Miller},
  {Moscardini}, {Napolitano}, {Porth}, {Radovich}, {Schneider}, {Tr{\"o}ster},
  {Wittje}, {Yan}, \& {Zhang}}]{Wright_etal2025}
{Wright}, A.~H., {St{\"o}lzner}, B., {Asgari}, M., {et~al.} 2025, arXiv
  e-prints, arXiv:2503.19441

\bibitem[{{Wright} {et~al.}(2010){Wright}, {Eisenhardt}, {Mainzer}, {Ressler},
  {Cutri}, {Jarrett}, {Kirkpatrick}, {Padgett}, {McMillan}, {Skrutskie},
  {Stanford}, {Cohen}, {Walker}, {Mather}, {Leisawitz}, {Gautier}, {McLean},
  {Benford}, {Lonsdale}, {Blain}, {Mendez}, {Irace}, {Duval}, {Liu}, {Royer},
  {Heinrichsen}, {Howard}, {Shannon}, {Kendall}, {Walsh}, {Larsen}, {Cardon},
  {Schick}, {Schwalm}, {Abid}, {Fabinsky}, {Naes}, \& {Tsai}}]{Wright_etal2010}
{Wright}, E.~L., {Eisenhardt}, P. R.~M., {Mainzer}, A.~K., {et~al.} 2010, \aj,
  140, 1868

\bibitem[{{Xu} {et~al.}(2023){Xu}, {Zhang}, {Peng}, {Yu}, {Zhang}, {Yao},
  {Qin}, {Sun}, {He}, \& {Yang}}]{XU2023}
{Xu}, H., {Zhang}, P., {Peng}, H., {et~al.} 2023, MNRAS, 520, 161

\bibitem[{Yoon {et~al.}(2018)Yoon, Jordon, \& Schaar}]{Yoon2018_gain}
Yoon, J., Jordon, J., \& Schaar, M. 2018, International conference on machine
  learning, 80, 5689

\bibitem[{{Zhan}(2011)}]{Zhan_2011}
{Zhan}, H. 2011, Scientia Sinica Physica, Mechanica \& Astronomica, 41, 1441

\bibitem[{{Zhang} {et~al.}(2023){Zhang}, {Zuo}, \& {Zhang}}]{Zhang_etal2023}
{Zhang}, H., {Zuo}, S., \& {Zhang}, L. 2023, Research in Astronomy and
  Astrophysics, 23, 075011

\bibitem[{{Zhang} {et~al.}(2017){Zhang}, {Yu}, \& {Zhang}}]{Zhang_etal2017}
{Zhang}, L., {Yu}, Y., \& {Zhang}, P. 2017, \apj, 848, 44

\bibitem[{{Zhang} {et~al.}(2010){Zhang}, {Pen}, \&
  {Bernstein}}]{Zhang_etal2010}
{Zhang}, P., {Pen}, U.-L., \& {Bernstein}, G. 2010, \mnras, 405, 359

\bibitem[{{Zheng} {et~al.}(2024){Zheng}, {Chan}, {Xu}, {Zhang}, \&
  {Song}}]{Zheng_etal2024}
{Zheng}, W., {Chan}, K.~C., {Xu}, H., {Zhang}, L., \& {Song}, R. 2024, \aap,
  692, A186

\bibitem[{{Zhou} {et~al.}(2021{\natexlab{a}}){Zhou}, {Newman}, {Mao},
  {Meisner}, {Moustakas}, {Myers}, {Prakash}, {Zentner}, {Brooks}, {Duan},
  {Landriau}, {Levi}, {Prada}, \& {Tarle}}]{ZhouNewman_etal2021}
{Zhou}, R., {Newman}, J.~A., {Mao}, Y.-Y., {et~al.} 2021{\natexlab{a}}, \mnras,
  501, 3309

\bibitem[{{Zhou} {et~al.}(2022{\natexlab{a}}){Zhou}, {Gong}, {Meng}, {Cao},
  {Chen}, {Chen}, {Du}, {Fu}, \& {Luo}}]{Zhou_etal2022}
{Zhou}, X., {Gong}, Y., {Meng}, X.-M., {et~al.} 2022{\natexlab{a}}, \mnras,
  512, 4593

\bibitem[{{Zhou} {et~al.}(2022{\natexlab{b}}){Zhou}, {Gong}, {Meng}, {Chen},
  {Chen}, {Du}, {Fu}, \& {Luo}}]{Zhou_etal2022RAA}
{Zhou}, X., {Gong}, Y., {Meng}, X.-M., {et~al.} 2022{\natexlab{b}}, Research in
  Astronomy and Astrophysics, 22, 115017

\bibitem[{{Zhou} {et~al.}(2021{\natexlab{b}}){Zhou}, {Gong}, {Meng}, {Zhang},
  {Cao}, {Chen}, {Amaro}, {Fan}, \& {Fu}}]{Zhou_etal2021}
{Zhou}, X., {Gong}, Y., {Meng}, X.-M., {et~al.} 2021{\natexlab{b}}, \apj, 909,
  53

\end{thebibliography}



\end{document}